\documentclass[twocolumn]{aastex62}

\sfcode`A=1000 \sfcode`B=1000 \sfcode`C=1000 \sfcode`D=1000
\sfcode`E=1000 \sfcode`F=1000 \sfcode`G=1000 \sfcode`H=1000
\sfcode`I=1000 \sfcode`J=1000 \sfcode`K=1000 \sfcode`L=1000
\sfcode`M=1000 \sfcode`N=1000 \sfcode`O=1000 \sfcode`P=1000
\sfcode`Q=1000 \sfcode`R=1000 \sfcode`S=1000 \sfcode`T=1000
\sfcode`U=1000 \sfcode`V=1000 \sfcode`W=1000 \sfcode`X=1000
\sfcode`Y=1000 \sfcode`Z=1000

\received{March 9, 2018}
\accepted{May 30, 2018}
\submitjournal{ApJ}

\shorttitle{Near Infrared Variability of Sgr~A*}
\shortauthors{Witzel et~al.}

\usepackage{graphicx}
\usepackage{natbib}
\usepackage{mathrsfs}
\usepackage[lined]{algorithm2e}
\usepackage{amsmath}
\usepackage{placeins}

\usepackage{epsfig}
\usepackage{verbatim, amsmath, amsfonts, amssymb,amssymb}
\usepackage[utf8x]{inputenc}
\usepackage{float}
\usepackage{animate}
\usepackage{enumerate}
\usepackage{url}
\usepackage{bm}
\usepackage{color}
\usepackage[lined]{algorithm2e}
\usepackage{listings}
\usepackage{animate}
\usepackage{textcomp}
\usepackage{xfrac}

\newcommand{\Sp}{{\it Spitzer\/}}
\newcommand{\SST}{{\it Spitzer Space Telescope\/}}
\newcommand{\CXO}{{\it Chandra X-ray Observatory}}
\newcommand{\Ch}{{\it Chandra}}
\newcommand{\Sg}{Sgr~A*}

\newcommand{\epi}{\epsilon}
\newcommand\logn{\operatorname{logn}}
\newcommand{\Msun}{\mbox{M$_\sun$}}

\begin{document}

\title{Variability Timescale and Spectral Index of Sgr A* in the Near Infrared:\\
Approximate Bayesian Computation Analysis of the \\ 
Variability of the Closest Supermassive Black Hole}
\author[0000-0003-2618-797X]{G.\ Witzel}
\affiliation{University of California, Los Angeles, CA USA}
\author[0000-0002-7476-2521]{G.\ Martinez}
\affiliation{University of California, Los Angeles, CA USA}
\author[0000-0002-5599-4650]{J.\ Hora}
\affiliation{Harvard-Smithsonian Center for Astrophysics, 60 Garden St., Cambridge, MA 02138 USA}
\author[0000-0002-9895-5758]{S.\ P.\ Willner}
\affiliation{Harvard-Smithsonian Center for Astrophysics, 60 Garden St., Cambridge, MA 02138 USA}
\author[0000-0002-6753-2066]{M.\ R.\ Morris}
\affiliation{University of California, Los Angeles, CA USA}
\author[0000-0001-7451-8935]{C.\ Gammie}
\affiliation{Department of Astronomy, University of Illinois, 1002 West Green Street, Urbana, IL 61801}
\author{E.\ E.\ Becklin}
\affiliation{University of California, Los Angeles, CA USA}
\affiliation{SOFIA Science Center, Moffett Field, CA USA}
\author[0000-0002-3993-0745]{M.\ L.\ N.\ Ashby}
\affiliation{Harvard-Smithsonian Center for Astrophysics, 60 Garden St., Cambridge, MA 02138 USA}
\author[0000-0003-3852-6545]{F.\ Baganoff}
\affiliation{MIT Kavli Institute for Astrophysics and Space Research, Cambridge, MA 02139, USA}
\author{S.\ Carey}
\affiliation{Spitzer Science Center, California Institute of Technology, Pasadena, CA 91125 USA}
\author[0000-0001-9554-6062]{T.\ Do}
\affiliation{University of California, Los Angeles, CA USA}
\author[0000-0002-0670-0708]{G.\ G.\ Fazio}
\affiliation{Harvard-Smithsonian Center for Astrophysics, 60 Garden St., Cambridge, MA 02138 USA}
\author[0000-0003-3230-5055]{A.\ Ghez}
\affiliation{University of California, Los Angeles, CA USA}
\author{W.\ J.\  Glaccum}
\affiliation{Spitzer Science Center, California Institute of Technology, Pasadena, CA 91125 USA}
\author[0000-0001-6803-2138]{D.\ Haggard}
\affiliation{Department of Physics, McGill University, 3600
University St., Montreal, QC H3A 2T8, Canada}
\affiliation{McGill Space Institute, McGill University, Montreal, QC H3A 2A7, Canada}
\author[0000-0002-7758-8717]{R.\ Herrero-Illana}
\affiliation{European Southern Observatory (ESO), Alonso de C\'ordova 3107, Vitacura, Casilla 19001, Santiago de Chile, Chile}
\author[0000-0003-4714-1364]{J.\ Ingalls}
\affiliation{Spitzer Science Center, California Institute of Technology, Pasadena, CA 91125 USA}
\author[0000-0002-1919-2730]{R.\ Narayan}
\affiliation{Harvard-Smithsonian Center for Astrophysics, 60 Garden St., Cambridge, MA 02138 USA}
\author{H.\ A.\ Smith}
\affiliation{Harvard-Smithsonian Center for Astrophysics, 60 Garden St., Cambridge, MA 02138 USA}

\begin{abstract}

Sagittarius A* (Sgr~A*) is the variable radio, near-infrared (NIR), and X-ray source associated with accretion onto the  Galactic center black hole. We present an analysis of the most comprehensive NIR variability dataset of Sgr~A* to date: eight 24 hr epochs of continuous monitoring of Sgr~A* at 4.5~\micron\ with the IRAC instrument on the \SST, 93 epochs of 2.18~\micron\ data from Naos Conica at the Very Large Telescope, and 30 epochs of 2.12~\micron\ data from the NIRC2 camera at the Keck Observatory, in total 94,929 measurements. \deleted{We make these consistently reduced and calibrated light curves publicly available.} 
%This unprecedented dataset allows an in-depth analysis of the power spectrum of this unique source and its relation to the innermost accretion zone of the black hole. 
A new approximate Bayesian computation method for fitting the first-order structure function extracts information beyond current fast Fourier transformation (FFT) methods of power spectral density \added{(PSD)} estimation \deleted{used in earlier studies}. With a combined fit of the data of all three observatories, the characteristic coherence timescale of Sgr~A* is $\tau_{b} = 243^{+82}_{-57}$~minutes ($90\%$ \replaced{Bayesian confidence level}{credible} interval). 
%The power spectral density is featureless on timescales down to 8.8~minutes ($95\%$ Bayesian confidence level).  If the emission process creates features that correspond to the orbital frequency at the innermost stable circular orbit, the black hole dimensionless spin parameter must be $\ga$0.91. 
The PSD has no detectable features on timescales down
to 8.5 minutes ($95\%$ \replaced{Bayesian confidence level}{credible} level), which is the ISCO orbital frequency \replaced{if the black hole has
an}{for a} dimensionless spin parameter $a = 0.92$. One light curve measured simultaneously at 2.12 and 4.5~\micron\ during a low flux-density phase gave a spectral index $\alpha_s = 1.6 \pm 0.1$ ($F_\nu \propto \nu^{-\alpha_s}$). This value
%is the most precise determination yet of $\alpha_s$ during a dim phase. It
implies that the \Sg\ \added{NIR} color \deleted{at these wavelengths} becomes bluer during higher flux-density phases. The probability densities of flux densities of the combined datasets are best fit by log-normal distributions. Based on these distributions, the \Sg\ spectral energy distribution is consistent with synchrotron radiation from a non-thermal electron population from below 20~GHz through the NIR.

%[T.Do comments: The last two sentences are somewhat confusing. Are you concluding that there are separate outburst and quiescent states of Sgr A* variability? I think you analysis actually encompasses all the data, not a 'quiescent' state.]
\end{abstract}


\section{Introduction}\label{intro}
The broadband radiation source Sgr~A* is located at the heart of the so-called S-star cluster \citep{2012A&A...545A..70S} at the center of the Milky Way. \Sg's position is coincident with the dynamical center of the S-stars and therefore coincident with the dynamically derived location (to within $\sim$2~mas) of the central supermassive black hole (SMBH) of our Galaxy (e.g., \citealt{2010ApJ...725..331Y}). That makes \Sg\  more than 100 times closer than any other supermassive black hole (SMBH), and it can therefore be studied in far greater detail.

Sgr~A* is visible as a compact, moderately variable radio source having flux densities between 0.5 and 4~Jy in the range  0.1 to 360~GHz (\citealt{1974ApJ...194..265B,1998ApJ...499..731F,2000A&A...362..113F,2001ApJ...547L..29Z,2004AJ....127.3399H,2004ApJ...611L..97M,2005ApJ...623L..25M,2006ApJ...650..189Y,2008ApJ...682..373M,2009ApJ...706..348Y,2009ApJ...700..417L,2010A&A...517A..46K,2011ApJ...738..158G,2015ApJ...802...69B,2016A&A...587A..37R,2017ApJ...845...35C}). \Sg\ has much dimmer NIR and X-ray counterparts that are  variable by up to 30 times the mean flux density in the NIR and up to a factor 500 in the X-rays (\citealt{2001Natur.413...45B,2002ApJ...577L...9H,2003Natur.425..934G,2004ApJ...601L.159G,2005ApJ...628..246E,2007ApJ...667..900H,2008ApJ...688L..17M,2008A&A...488..549P,2009ApJ...691.1021D,2009ApJ...698..676D,2010A&A...512A...2S,2011ApJ...728...37D,2012ApJS..203...18W,2013ApJ...774...42N,2015ApJ...799..199N,2017MNRAS.468.2447P,2017ApJ...843...96Z}). The X-ray energy output can become comparable to the submm level during the brightest flares. This strong, rapid variability may be associated with accretion processes close to the supermassive black hole's event horizon. The connection of the variability to regions close to the event horizon is based on:
(1) the observed timescales of the variability, with common changes of a factor $\gtrsim$10 within $\sim$10~minutes in the NIR (\citealt{2003Natur.425..934G,2004ApJ...601L.159G}); 
(2) the spectral index\footnote{The spectral index is defined here as 
$F_\nu \propto \nu^{-\alpha_s}$.}
$\alpha_s \approx 0.6$ 
(\citealt{2005ApJ...635.1087G,2007ApJ...667..900H,2011AA...532A..26B,2014IAUS..303..274W}); 
(3) linear polarization in the NIR and submm (\citealt{2006A&A...455....1E,2006ApJ...640..308M,2006A&A...458L..25M,2007MNRAS.375..764T,2007ApJ...654L..57M,2008A&A...479..625E,2007ApJ...668L..47Y,2009ApJ...702L..56N,2011A&A...525A.130W,2015A&A...576A..20S}); and
(4) temporal correlations between the submm, NIR, and X-ray regimes. 
All of these observational results point to a population of relativistic electrons in a region that is smaller than $\sim$10 light minutes (the distance associated with the light crossing time, $<$15 Schwarzschild radii) emitting synchrotron radiation at NIR wavelengths. The variable submm and X-ray radiation may be synchrotron emission or may be linked by radiative transfer processes such as adiabatic expansion and inverse Compton or synchrotron self-Compton scattering, respectively 
(\citealt{2001Natur.413...45B,2004A&A...427....1E,2006A&A...450..535E,2006ApJ...640L.163G,2006ApJ...650..189Y,2006ApJ...644..198Y,2008A&A...492..337E,2008A&A...479..625E,2008ApJ...682..373M,2008ApJ...682..361Y,2009ApJ...698..676D,2009ApJ...706..348Y,2011AA...528A.140T,2012A&A...537A..52E,2012AJ....144....1Y,2012A&A...540A..41H,2016A&A...589A.116M,2016A&A...587A..37R,2016MNRAS.461..552D,2017MNRAS.468.2447P}).

In order to shed light on the physical and radiative mechanisms at work and on the interrelation between wavelengths, many studies have attempted to find and categorize recurring patterns and regularities in the behavior of Sgr~A*, both statistically for individual wavelength regimes as well as in the form of correlations between bands (\citealt{2007A&A...473..707M,2006A&A...460...15M,2006A&A...458L..25M,2006ApJ...640L.163G,2007ApJ...667..900H,2009ApJ...691.1021D,2009ApJ...694L..87M,2010A&A...510A...3Z,2011ApJ...728...37D,2012ApJS..203...18W,2013ApJ...774...42N,2014ApJ...791...24M,2014ApJ...793..120H,2014MNRAS.442.2797D,2015ApJ...799..199N,2017A&A...601A..80S}). In recent years, the preponderance of studies has arrived at the following set of phenomenological but statistically rigorous results:
\begin{itemize}
\item
Sgr~A* is a continuously variable NIR source that emits above the 2.12~\micron\ detection level (0.05~mJy observed or 0.5~mJy dereddened, 3$\sigma$ above the noise level of the NIRC2 camera at the Keck~II telescope)  $\sim$90\% of the time (\citealt{2012ApJS..203...18W,2014ApJ...791...24M}).  Its probability density function (PDF) of flux densities\footnote{
 {The PDF of flux densities is the probability that an {\em independent} observation will yield a flux density in a particular interval.}} 
  at 2.18~\micron\ is highly skewed (\citealt{2011ApJ...728...37D}) and can be described by a power law with a slope $\beta_{\rm{IR}} \approx 4$ (\citealt{2012ApJS..203...18W}).
%  \replaced{.  The }{, in which case the} 
The first three moments of the PDF are well defined with mean $\approx$5.8~{mJy} dereddened ($\approx$0.6~{mJy} observed), variance $\approx$9.4~{mJy}$^{2}$ dereddened, and skewness $\approx$52.3~{mJy}$^{3}$ dereddened. The brightest observed NIR peak reached $\sim$30~mJy (dereddened, \citealt{2009ApJ...698..676D}). Peaks with $F(2.18~\micron) > 10$~mJy (dereddened)  occur about four times a day (\citealt{2009ApJ...691.1021D,2009ApJ...694L..87M,2014ApJ...791...24M,2014ApJ...793..120H}). %Occasionally, large, rapid fluctuations similar to a step function of 1-2 mJy (dereddened) and faster than the typical sampling at 8-10~m class AO-supported telescopes ($\sim 30$~seconds) have been observed.
\item
The X-ray emission comes from a steady, extended ($\sim$1\arcsec) source plus outbursts from an unresolved source. Outburst flux densities can be several hundred times the level of the quiescent state. Outbursts (frequently called "flares" in the literature) have the character of distinct events and occur about once per day. The unresolved source is detectable only during its outbursts.  At other times, fluctuations are sufficiently described by the Poisson distribution expected for the steady source (\citealt{2015ApJ...799..199N}). The flux-density PDF, as for the NIR, is well described by a power-law distribution but with $\beta_X \approx 2$. X-ray flares seem always to be accompanied by NIR peaks (\citealt{2012RAA....12..995M} and references therein). However, the reverse is not true, and only about one in four $F(2.18~\micron) > 10$~mJy (dereddened) NIR peaks has an X-ray counterpart (\citealt{2001Natur.413...45B,2004A&A...427....1E,2008ApJ...682..373M,2008A&A...488..549P,2009ApJ...691.1021D,2013ApJ...774...42N,2015ApJ...799..199N}). 
There is no obvious relationship between X-ray and NIR flux-density levels.
\item
The spectral energy distribution of Sgr A* peaks in the submm (\citealt{1992A&A...261..119Z,1995A&A...297...83Z,1998ApJ...499..731F,2001ARA&A..39..309M}), where it is visible as a synchrotron source powered by the dominant thermal electron population (\citealt{2003ApJ...598..301Y}). An analysis by \cite{2014MNRAS.442.2797D} of $\sim$10~years of 1.3, 0.87, and 0.43 mm observations with CARMA and SMA shows a steady flux-density level of $\sim$3~Jy with Gaussian fluctuations about that mean. \replaced{Submm flux-density enhancements rising 25\% or more above the mean occur approximately four times per day (\citealt{2008ApJ...682..373M,2014MNRAS.442.2797D}).}{Submm flux-density enhancements rising $\sim 1$~Jy above the mean occur approximately 1.2 times per day (\citealt{2008ApJ...682..373M}).}  A time-series analysis of submm light curves gave a mean reversion time scale of $\sim$8~hr (\citealt{2014MNRAS.442.2797D}). 
\end{itemize}

{The patterns of correlation between wavelengths are still unclear.} Several authors have suggested that the submm peaks often follow bright NIR peaks by 1--3~hr (\citealt{2008ApJ...682..373M,2006A&A...450..535E,2006ApJ...650..189Y,2008A&A...492..337E,2009ApJ...706..348Y,2009A&A...500..935E,2011ApJ...729...44Y,2012A&A...537A..52E}), but most observations remain inconclusive in this regard because of the lack of simultaneous multi-wavelength data of sufficient length and overlap. Indeed, there are counterexamples.  Recent observations obtained with the \SST, the \CXO, the SMA, and the W.~M.\ Keck Observatory suggest that the phenomenology of these correlations is not simple \citep{Fazio2018}. In particular, SMA and \Sp\ observed the first example of an effectively synchronous sequence of variations in the submm and NIR. Another example obtained with SMA, \Ch, and Keck  showed an even more surprising sequence in which a submm peak precedes an X-ray flare, which in turn was followed by a NIR peak. Albeit not  conclusive due to the limitations of ground-based observations, such a sequence of peaks contradicts the canonical phenomenology of simultaneous X-ray and NIR followed by delayed submm variations.

%[T. Do comments: These series of points begin by stating "phenomenological but statistically rigorous results, but in this last paragraph, your point is that the statistics for the lags in submm are problematic. Since your paper is not about submm observations, can this point be shorten?]

There are many previous studies of {the statistical properties of} Sgr~A*'s variability. Initially, these studies focused on putative quasi-periodicity (QPO) at time scales between 10 and 20 minutes and its relation to the innermost stable orbit of the $4 \times 10^6~\rm{M}_{\odot}$ SMBH (\citealt{2003Natur.425..934G,2006A&A...458L..25M,2006A&A...460...15M,2007A&A...473..707M,2007MNRAS.375..764T,2010A&A...510A...3Z,2017MNRAS.472.4422K}). \cite{2009ApJ...691.1021D} found no evidence for such a QPO based on available data at the time. Consequently, the scope of the statistical analysis was broadened with a determination of the red-noise correlation timescale ($128^{+329}_{−77}$~minutes) in the NIR (\citealt{2009ApJ...694L..87M}) that allowed for a comparison of Sgr~A* with black holes of different mass regimes. This comparison revealed that the mass and characteristic timescale of Sgr~A* are consistent with a linear mass--timescale relation without a luminosity correction term as proposed by, for example,  \cite{2006Natur.444..730M}, who discussed characteristic timescales of AGN and black hole X-ray binaries (BHXRB). In this context, \cite{2009ApJ...694L..87M} pointed out the \added{exceptional} value of Sgr~A* because it is the SMBH with the most precise mass determination so far: ${M}_{\rm{bh}} = (4.02 \pm 0.16 \pm 0.04) \times 10^6$~\Msun\ \citep{2016ApJ...830...17B}, where the error bar terms give the statistical and systematic uncertainties, respectively.

Another line of inquiry has considered the possibility of a dichotomy 
of the NIR variability into statistically different processes {(or `states')} with either
different flux-density PDFs or different timing behavior or both.
These inquiries have been motivated 
by some NIR flares having X-ray counterparts while others do
not (\citealt{2011ApJ...728...37D}). The statistics of the variations have been shown to be consistent with a single variability state without evidence 
for multiple superimposed or interleaved variability processes (\citealt{2012ApJS..203...18W,2014ApJ...791...24M}).

A variety of NIR spectral index values have been reported. While some authors \replaced{claimed}{found} a strong dependence of the spectral index on the 
flux-density level,  \added{other} high-cadence and high-signal-to-noise studies \added{at $K$-band-equivalent flux densities  $>$0.2~mJy} showed only minor intrinsic fluctuations around an $H$- (1.65~\micron) to $L$-band (3.8~\micron) spectral index $\alpha_s = 0.6$
(\citealt{2005ApJ...628..246E,2005ApJ...635.1087G,2006ApJ...640L.163G,2006ApJ...642L.145K,2007ApJ...667..900H,2011AA...532A..26B,2014IAUS..303..274W}). 
%\added{One limitation is that these studies only considered flux densities $F_K\ga1$~mJy.}
%[T. Do comments: I suggesting using something other than "contradictory" describe this. The papers all used different data and arrived a different spectral indices. The values from the measurements are different, but not necessarily contradictory.]

\Sg\ is linearly polarized in the NIR. \cite{2015A&A...576A..20S} statistically analyzed time series and found typical polarization of $(20\pm10)\%$ and a preferred position angle of $(13\pm15)\arcdeg$.

In summary, the NIR variability is well characterized as a red-noise process -- that is, it has a power spectral density (PSD)\footnote{
 The PSD is the Fourier transform of the autocorrelation function 
 (e.g., \citealt[][Eq.~3]{1995A&A...300..707T}).  In other words, the PSD measures how much the flux densities of two measurements separated in time are likely to differ, but the independent variable is spectral frequency, i.e., 1/(time difference).}
that is a power law with a slope $\gamma_1 \approx 2$ for timescales in the range $\sim$20 to $\sim$150 minutes. The process is a damped random walk -- that is, it has a correlation (or characteristic) timescale. For timescales longer than the correlation timescale ($128^{+329}_{−77}$~minutes at the 90\% \replaced{confidence}{credible} level{\footnote{\added{The terms ``credible interval'' and ``credible limit'' refer to  intervals and their upper and lower limits that have a specified probability of containing the true value. In the Bayesian context, these intervals are directly derived from the posteriors.}}}---\citealt{2009ApJ...694L..87M}), the process is uncorrelated white noise, and the PSD becomes flat for the corresponding lowest frequencies (see \ref{howtosim}). Based on the available datasets, no evidence for periodicity, quasi-periodicity, or changes in its statistical behavior (e.g., a two-state variability model) \replaced{can}{could} be found. In fact, the existing knowledge of the NIR variability of Sgr~A* can be described statistically by as few as five parameters: the PSD slope and break timescale, the slope and normalization of the power-law flux-density PDF, and the NIR spectral index. Two more parameters are needed to describe the linear polarization: the fixed polarization fraction and position angle. Considering the large amplitudes of  flux-density fluctuations, such constancy of statistical and physical parameters over the period of existing data is surprising.

%[T. Do comments: you refer to Figure 18 here. Are the figures out of order?]

%Theoretical efforts to model the turbulent accretion flow and the variability caused by the accretion, however, cannot fully describe all observations. In particular, the flux-density levels in the NIR are a challenge for the radiative models in the context of full 3D GRMHD simulations. The observed NIR and submm variability timescales could correspond to a thermal or viscous timescale associated with the inner radius of the accretion disk. Many emission mechanisms remain viable, e.g., magnetic reconnection events, violent disk instabilities, adiabatic expansion of plasma blobs, unsteady jet emission, or strong or weak magnetic field accretion (\citealt{2007ApJ...667..714S,2010MNRAS.408.1051Y,2010ApJ...725..450D,2012A&A...537A..52E}). Gravitational lensing is likely to be an additional factor modulating any of these mechanisms (e.g., \citealt{2015ApJ...812..103C}). A promising class of simulations using 3D GRMHD models has resulted in valid prediction of light curves in terms of variability and even in flux-density levels in either NIR, X-rays and submm. \cite{2012ApJ...746L..10D}, in particular, derive characteristic PSDs with short timescale structure below 20 minutes by modeling Sgr A* as a hot, optically thin, geometrically thick accretion flow around a spinning black hole with aligned disk orbital and black hole spin angular momenta.

Specific scenarios for producing NIR variability have invoked magnetic
reconnection events, disk instabilities, ejection and expansion of plasma
blobs, unsteady jet emission, or accretion of magnetic fields (\citealt{2007ApJ...667..714S,2010MNRAS.408.1051Y,2010ApJ...725..450D,2012A&A...537A..52E}). However, these theoretical efforts to model the turbulent accretion flow and the variability caused by the accretion cannot fully explain all observations to date. In particular, the peak NIR flux densities  are  higher than predicted by radiative transfer models of three-dimensional general relativistic
magnetohydrodynamic (GRMHD) simulations with a thermal electron distribution
function that matches millimeter flux densities.   The observed NIR variability
may therefore be due to the acceleration of electrons out of the dominant
thermal component of the distribution function into a non-thermal tail (e.g., \citealt{2010ApJ...725..450D}).

GRMHD models with a thermal electron distribution,
while producing only relatively weak variability
(\citealt{2012ApJ...746L..10D}), have an interesting feature in their power spectrum
near $f_{\rm{ISCO}}$, the orbital
frequency of the innermost stable circular orbit (ISCO).
In particular, these models show an approximately $f^{-2}$ power spectrum
at $f < f_{\rm{ISCO}}$, a bump in power close to $f_{\rm{ISCO}}$, and a break in the
spectrum to approximately $f^{-4}$ at $f > f_{\rm{ISCO}}$.
This is consistent with the notion that variability in the disk at frequencies
above the orbital frequency is associated with disk turbulence, which is known
from simulations to have a steeply declining spatial power spectrum
(e.g., \citealt{2009ApJ...694.1010G}).  This would naturally give rise to a steeply declining
temporal power spectrum as well.
With \Sp\ (\citealt{2014ApJ...793..120H}), in combination with ground-based 8--10~m  telescopes, the predicted PSD short-timescale structure is testable.

This work presents the first analysis of the NIR PSD of Sgr~A* that includes continuous datasets for all relevant timescales from 24~hr down to the sub-minute level. We use an unprecedented dataset from three different observatories: the W.~M.\ Keck Observatory, the European Southern Observatory Very Large Telescope (ESO/VLT), and the \SST. The observatories contribute complementary information about the PSD: the Keck data have the best signal-to-noise and can detect Sgr~A* variations at timescales below 1 minute.  A limitation is that most of the Keck datasets have a duration of $\leq$2~hr. The VLT data cover timescales between 4 minutes and 6 hr, and the \Sp\ data timescales from $\sim$7 minutes to 24~hr, much longer than the previously derived correlation timescale. Together, these data enable the most precise estimate possible today of the correlation timescale and a test for PSD features at timescales below 50 minutes.

To exploit the combined data sets, we have developed an entirely new algorithm. 
It  uses the first-order structure function as the central tool for analyzing the timing of Sgr~A* and a  customized population Monte Carlo approximate Bayesian computation (PMC-ABC) sampler to derive parameter values.
The goals of this paper are to
\begin{itemize}
\item provide this extensive dataset to the community with a full statistical characterization;
\item introduce the new PMC-ABC algorithm that will have wide application to variable sources; 
\item determine the PSD of the variability process of Sgr~A*, including a new determination of the correlation timescale; 
\item determine the \Sg\ flux-density PDF in both $K$- and $M$-band (4.5~\micron); 
\item characterize the \Sg\ spectral index between these two bands; and
\item characterize the instrumental performance of this kind of space-based variability study in comparison to ground-based AO telescopes.  
\end{itemize}
Section~\ref{obs_section} describes the observations and datasets used in this work. Section~\ref{meth} and \ref{ABC} present the newly developed algorithm for analyzing non-deterministic stationary linear time series and the results of our analysis of the Sgr~A* light curves. Sections~\ref{disc} and~\ref{concl}  discuss the results and present our conclusions. \added{Readers mainly interested in the mathematical foundation of our methodology are referred to \ref{ABC}, \ref{FFT}, and \ref{ratioform}. Readers only interested in our main results are refereed to Figures~\ref{corner}, \ref{conf}, \ref{alpha_f}, and \ref{sed} and Tables~\ref{results} and \ref{percentiles}.}

Different authors (e.g., \citealt{2003Natur.425..934G,2009ApJ...691.1021D,2011ApJ...728...37D}) 
have used different values for interstellar extinction to \Sg, making
it difficult to compare studies. To avoid ambiguity and simplify comparisons, 
data are given here without correction for interstellar extinction, contrary to prior practice (e.g., \citealt{2012ApJS..203...18W}). 
Where extinction is needed, for example to compare with models or discuss an
intrinsic spectral index, we adopt a 2.12 and 2.18~$\mu$m extinction $A_K = 2.46
\pm 0.10$~mag (\citealt{2010A&A...511A..18S,2011A&A...532A..83S}) and a 4.5~$\mu$m
extinction value of $A_M = 1.00 \pm 0.14$~mag.\footnote{Error bars include both statistical and estimated systematic uncertainties. The $A_M$ error bar is a corrected value from R.~Sch{\"o}del (2018, private communication).}
{To place our $K$-band flux densities on the same scale as \cite{2011ApJ...728...37D} or \cite{2012ApJS..203...18W}, multiply by 9.64 ($A_K = 2.46$). To compare with \cite{2003Natur.425..934G} or \cite{2006A&A...450..535E}, multiply by 13.18 ($A_K = 2.8$). To compare with \cite{2009ApJ...691.1021D}, multiply by 20.89 ($A_K = 3.3$), and to compare with \cite{2007ApJ...667..900H}, multiply by 19.23 ($A_K = 3.2$).}

\section{Observations and Data Reduction}
\label{obs_section}

\subsection{\Sp/IRAC observations}\label{IRAC_obs}
All observations in this \SST\ program (Program IDs 10060, 12034, and 13027) used IRAC 
subarray mode, which reads a 32$\times$32-pixel region
of the IRAC 4.5~\micron\ detector array 10 times per second.
Each subarray data collection event obtains 64 consecutive images (a ``frame set'') of these pixels, and there is typically 2~s idle time between images.
The subarray pixel area starts at pixel (9,9)
of the full 256$\times$256-pixel array, and the angular scale is 1\farcs21 per pixel.

Each of eight \Sp\ observing epochs used the same basic observing procedure. This comprised an initial peakup from a reference star to place \Sg\ at the center of pixel (16,16), making a small map, during which time the telescope temperature settled down, a second peakup, a staring mode observation lasting $\sim$12~hr, a third peakup, and a second stare.
The staring observations in 2013 and 2014 used custom Instrument Engineering Requests (IERs) to obtain two 11.6~hr monitoring periods at each epoch.  The 2016 observations used standard Astronomical Observation Requests (AORs) to do the same, but with $2\times 12$~hr of monitoring.  
The 2017 observations used new IERs to decrease the effective data rate by truncating the lowest four bits of each 0.1~s pixel value.  Because of the
high source brightness in the Galactic center, these bits contain random noise and therefore do not compress. Removing them reduced the data volume to 65\% of what it would have been without truncating.  Prior to making the 2017 observations, we used our earlier Sgr A* measurements to verify that truncating these bits would increase the noise by only $\sim$1.3\%, which does not affect our ability to measure flux density fluctuations at expected levels.   
Further details of the observations are given by \citet{2014ApJ...793..120H},
and all AORKEYs  are in Table~\ref{AORs}.

\begin{table}[b!]
\begin{center}

\caption{IRAC Observation Log}\label{startstop}
\begin{tabular}{lcrl}
\tableline
\tableline
\null\\[-1ex]
& AOR Start & Frame\\
AORKEY & Time (UTC)\tablenotemark{a} & Sets\tablenotemark{b} & Type\\[1ex]
\tableline
\null\\[-1ex]
50123264 & 2013-12-10 03:48:56 &   92 & Map\\
50123520 & 2013-12-10 04:20:24 & 5000 & Stare part 1\\
50123776 & 2013-12-10 16:04:21 & 5000 & Stare part 2\\
51040768 & 2014-06-02 22:32:00 &  126 & Map\\
51041024 & 2014-06-02 22:59:37 & 5000 & Stare part 1\\
51041280 & 2014-06-03 10:43:22 & 5000 & Stare part 2\\
51087616 & 2014-06-17 18:29:35 &  126 & Map\\
51087872 & 2014-06-17 18:57:17 & 5000 & Stare part 1\\
51088128 & 2014-06-18 06:41:01 & 5000 & Stare part 2\\
51344128 & 2014-07-04 13:21:59 &  126 & Map\\
51344384 & 2014-07-04 13:49:41 & 4999 & Stare part 1\\
51344640 & 2014-07-05 01:33:25 & 5000 & Stare part 2\\
58115840 & 2016-07-12 18:04:23 &  156 & Map\\
58116352 & 2016-07-12 18:37:45 & 5142 & Stare part 1\\
58116608 & 2016-07-13 06:41:14 & 5142 & Stare part 2\\
58116096 & 2016-07-18 11:44:02 &  156 & Map\\
58116864 & 2016-07-18 12:17:25 & 5142 & Stare part 1\\
58117120 & 2016-07-19 00:20:54 & 5142 & Stare part 2\\
60651008 & 2017-07-15 22:28:54 &  156 & Map\\
63303680 & 2017-07-15 23:02:17 & 5142 & Stare part 1\\
63303936 & 2017-07-16 11:05:46 & 5142 & Stare part 2\\
60651264 & 2017-07-25 22:39:33 &  156 & Map\\
63304192 & 2017-07-25 23:12:57 & 5142 & Stare part 1\\
63304448 & 2017-07-26 11:16:26 & 5141 & Stare part 2\\
\tableline
\end{tabular}
\end{center}
\label{AORs}
\tablenotetext{a}{Start times are UTC at the \Sp\ observatory.  Corresponding times at Earth are a few minutes earlier.  Light curves given in Table~\ref{LightcurveData} have heliocentric times.}
\tablenotetext{b}{Frame set numbers include only frame sets with 0.1~s frame times.  As explained by \citet{2014ApJ...793..120H}, 2013--2014 observations also included images with 0.02~s frame times.  These are not included in the counts.}
\end{table}
\begin{table*}[t!]
\begin{center}
\caption{IRAC Flux Correction Coefficients}
\begin{tabular}{crrrrrrrr}
\tableline
\tableline
\null\\[-1ex]
Coefficient & 2013& 2014&2014&2014&2016&2016&2017&2017\\
Name & Dec 10 & June 2 & June 17 & July 4 & July 12 & July 18 & July 15 & July 25\\
\tableline
&&\\[-1ex]
$a $&   7537.1& 5358.9&  3693.5   &6684.6 & $-$877.31 &4748.0 & 4555.274 & 6938.5\\
$b $&  $-$17216&  $-$1336.0& 3173.4     &$-$9461.1 & $-$22340  &27.372&$-$4669.305 & 15146.3\\
$c $&  3730.1& $-$2696.0&  1679.9    & $-$12401 & $-$12402  &$-$2525.2&$-$4963.178 & 13822.1\\
$d$ &  $$-$11716$& $$-$6104.9$&  9248.1    &9199.4 & $-$6054.6  &$-$9.3296& $-$102.5009&  14876.4\\
$e$ &  3396.5 &$$-$1074.1$&   7629.8  & 5026.4 & 9508.4  &$$-$5798.4$& $-$3444.561& 2254.8\\
$f$ &$   $-$3025.0$&$ $-$760.80$& 16120    & 15750 &  10070  &391.47& 5618.444& 23499.7\\
$g_1$ & 0.0054&  0.3611  & $-$0.0049   & 0.1732 &0.3128   &$-$0.0004& 0.05112&0.05934 \\
$g_2$ & $-$0.0451& $-0.2359$  & 0.1518   & $-0.2194 $&  0.3267  &0.4427& 0.03646&$-$0.1593 \\
$g_3$ & $-$0.0007 & 0.1785  & 0.0344   &$-$0.2423 &  0.1124   &$-0.1267$ &0.06231 &0.1074\\
$g_4$ &  0.04063& $-$0.2422  & 0.1685   & 0.1462&  0.0672   &$-0.2089$& 0.05674&$-$0.02738 \\
$h_1$ &  0.2897& $-$0.7424  & $-$0.2675   &$-$0.5568 & 1.3735   &$-0.2927$& 0.4828& 0.1101\\
$h_2$ &1.338 &0.5666  & $-$0.4242   & 0.9940  &   1.1245  &$-0.7833$& 0.8234&0.8551 \\
$h_3$ & 0.1075& $-0.6004$ &  0.4291  & 0.9720  &  1.0763   &1.5937& 0.1941& $-$1.3753\\
$h_4$ & 0.4762& 1.1260  &$-0.6078  $  & $-0.2426$ & 0.9778   &1.1396&0.6042 &$-$0.9479 \\
$k_1$ & 0.0679&$-$0.7136    &0.6420    & $-0.5358 $&  0.0466  &$-$1.0620& $-$0.0834&$-$0.0098 \\
$k_2$ & 0.1133&  0.8440    & $-0.7912$  &0.9836 &   $-$0.1290  &$-$1.0258& $-$1.3781&$-$1.0708 \\
$k_3$ &$-$0.3432   & $-$0.2633   &$-$0.1390& 1.6669  &  $-$0.0074  &$-$0.3891& 0.1371&0.3808 \\
$k_4$ &$-$0.0905&0.5196   &$-$0.3543    & $-$0.3662 & $-$0.0981  &0.9668 & 1.0882& 0.6798\\

\tableline
\end{tabular}
\end{center}
\tablecomments{This table refers to the coefficients defined in 
Equation~\ref{polyfunc}.  For coefficients $g$, $h$, and $k$, the 
subscripts $n=1$ to~4
refer to neighboring pixels in the order (15,16),  (17,16), 
(16,15), and (16,17).} 
\label{Coefficients}
\end{table*}

The data reduction used an improved version of the technique described by \citet{2014ApJ...793..120H}. The first image of every frame set was removed because of calibration difficulties, and the remaining 63 frames were averaged.
The major remaining problem is that telescope pointing jitter introduces 
fluctuations into the flux measured by pixel (16,16).  Those can be largely
removed by fitting the measured flux as a function of the $(X,Y)$ coordinates of
Sgr~A*\ in each frame set with $(X,Y)$ being determined by cross-correlating
each frame set with a standard one having Sgr~A*\ centered on pixel (16,16).
However, this basic scheme does not work as well for the epoch 2--8 observations
(2014 June--July) as it did for the first epoch (2013 December).
This may be due to the observations being performed at a different rotation
angle on the array than the first epoch. The new angle did not allow the same
simple correction to yield similar quality as in the first epoch, probably due
to the inherent structure of the source and the details of how it falls on the
pixel array.
For some reason, the $(X,Y)$ coordinates do not capture all of the apparent 
background variability.
Several methods were tried to improve the fit. We 
found that the dependence of the pixel output $F(X_i,Y_i)$ on the $X,Y$ position on the array for the object and reference pixels could be well-modeled by using the second-degree polynomial
\begin{equation}\label{polyfunc}
\begin{split}
F(X_i,Y_i)=a+bX_i+cY_i+dX_iY_i+eX_i^2+fY_i^2+\\
\sum_{n=1}^{4}{P_{i,n} (g_n+h_n X_i+k_n Y_i)}\quad,
\end{split}
\end{equation}
where $a, b, c, d, e, f, g_n, h_n$, and $k_n$ are constant coefficients to be derived; $i$ is the sample number in the time sequence; $X_i$ and $Y_i$ represent the position of Sgr~A*\ on the array for sample $i$ in units of pixels (relative to the center of pixel (16,16)); and $P_{i,n}$ are the data values of the four pixels that are direct neighbors to the pixel output being analyzed. For example, for the analysis of pixel (16, 16), these neighbor pixels were (15,16),  (17,16), (16,15), and (16,17). The values of the coefficients were determined by least-squares fitting, minimizing the residuals between $F(X_i,Y_i)$ and the pixel (16,16) values in the
monitoring data. The fit was done iteratively, removing frame sets in which Sgr~A*\ showed detectable flux. That typically left about 7000 frame sets to fit out of an initial $>$10,000 available in each epoch. 
Coefficients derived for each epoch are given in Table~\ref{Coefficients}.

As a test of our method, we also extracted and modeled the output of a reference pixel in the same way as for pixel (16,16). The reference pixel was  
at an image location with a significant gradient and not on a local
maximum, similar to pixel (16,16) but far enough away from it that
the pixel will not see the variability from Sgr~A*.  For the 2013 December epoch, we used pixel (18,19) as a reference, as did \citet{2014ApJ...793..120H}.
Because of the different rotation angle in all subsequent epochs, we used pixel (14,14). 

One limitation of the reduction technique is that it cannot provide an absolute zero point for the
\Sg\ flux density.  Instead, $F=0$ corresponds to the average flux
density in the frame sets used to derive the coefficients.  The actual flux density
corresponding to $F=0$ is a parameter derived from subsequent fitting of the time-series data.

\begin{figure*}
\begin{center}
\includegraphics[scale=0.65, angle=0]{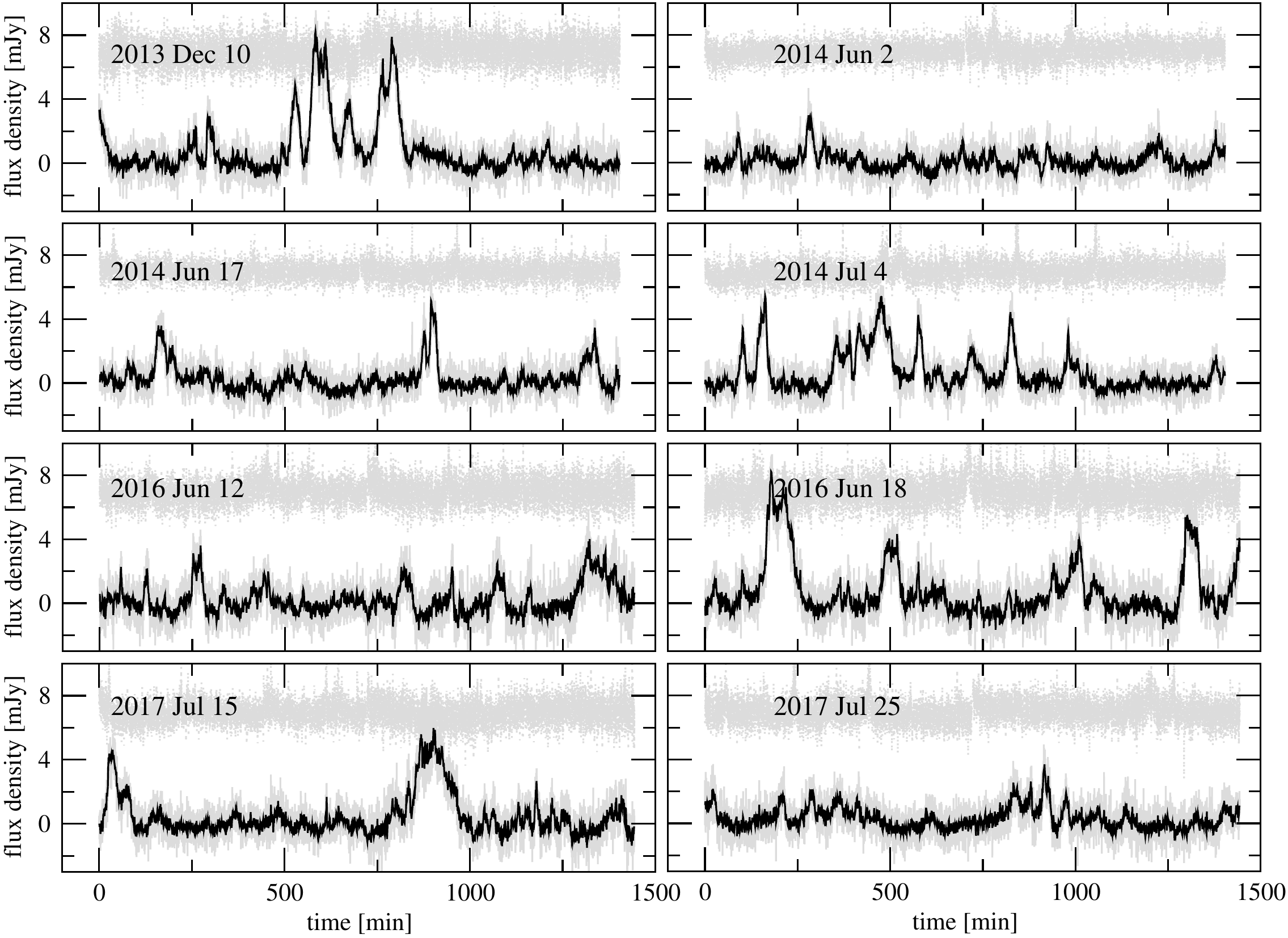}

\end{center}
\setlength{\abovecaptionskip}{0pt}
\caption{Excess 4.5~\micron\ flux density for Sgr~A* and for the reference pixel for each of the eight \Sp\ epochs. Flux densities are in
mJy with no correction for interstellar extinction. The flux density zero point
cannot be determined by the data reduction method.
In each panel, the gray lines show the flux density for each 6.4-s frame set,
and the black lines show the data binned in one minute intervals. The lower lines show the Sgr~A* flux densities, and the upper lines are for a reference pixel with $7$~mJy added to the flux density.  The
 2013 December epoch uses pixel (18,19) as the reference, and all other epochs use pixel (14,14). The values plotted are
  the difference between the observed value of the pixel in the 6.4-s
  frame set and the predicted value based on Eq.~\ref{polyfunc} and the
  measured $(X,Y)$ offset of each frame set. Flux density
  values have been corrected to total flux density for a point source by the
  position-dependent ratio of total flux density to central-pixel
  signal.  The horizontal axis shows the time in minutes relative to the start
  time (given in Table~\ref{startstop}) of the first monitoring 6.4-s frame set for that epoch. }\label{sgrAplot}
\end{figure*}

The eight light curves are plotted in Figure~\ref{sgrAplot}, and the time series data are given in Table~\ref{LightcurveData}. The new reduction of the 2013 epoch is very similar to the original result of \citeauthor{2014ApJ...793..120H}, but  the 
artifacts in the reference pixel are smaller compared to the original reduction. 
The peaks of emission from Sgr~A* in the 2013 epoch are in the same locations and very similar in amplitude and structure.

\begin{table}[ht!]
\begin{center}

\caption{Sgr A* Light Curve Data}
\begin{tabular}{crr}
\tableline
\tableline
&&\\
Observation  & Sgr A*& Reference\\
Date%\tablenotemark{a} 
& Flux Density&Flux Density\\
(HMJD)&(Jy)&(Jy)\\
\tableline
&&\\[-1ex]
\multicolumn{3}{l}{\Sp/IRAC}\\ 
\multicolumn{3}{l}{...}\\
57581.7781761 &  0.001056& 0.000016\\
57581.7782728 & 0.001576&-0.000329 \\
57581.7783700&	0.001055 & 0.000182\\
57581.7784676&	-0.000623& 0.000908\\
57581.7785647&	0.001590& -0.000180 \\
57581.7786621&	-0.000930 & -0.001045\\
57581.7787590&	0.000980& -0.000776\\
57581.7788565&	-0.000085& 0.000744\\
57581.7789539&	0.000819& -0.000939 \\
57581.7790510&	-0.000407& 0.000747\\
\multicolumn{3}{l}{...}\\
\tableline
&\\[-1ex]
\multicolumn{3}{l}{VLT/NaCo}\\
52803.1129224 & 0.0001745&\\
52803.1133356 & 0.0001585&\\
52803.1137607 & 0.0000846&\\
52803.1141797 & 0.0001671&\\
52803.1145983 & 0.0001632&\\
52803.1150173 & 0.0001849&\\
52803.1154358 & 0.0001362&\\
52803.1158572 & 0.0001725&\\
52803.1162806 & 0.0001702&\\
52803.1166930 & 0.0001488&\\
\multicolumn{3}{l}{...}\\
\tableline
&\\[-1ex]
\multicolumn{3}{l}{Keck/NIRC2}\\
53212.3510956 & 0.0004091&\\
53212.3529755 & 0.0005316&\\
53212.3532755 & 0.0004738&\\
53212.3539055 & 0.0006439&\\
53212.3550154 & 0.0005638&\\
53212.3555354 & 0.0003294&\\
53212.3796940 & 0.0000225&\\
53551.3979607 & 0.0000009&\\
53581.3320382 & 0.0001165&\\
53581.3369779 & 0.0001175&\\
\multicolumn{3}{l}{...}\\
\tableline
\end{tabular}
\end{center}
%\tablenotetext{a}{Heliocentric Modified Julian Date of the midpoint of the 6.4s BCD coadd\added{ or ground-based camera frame}.\\}
\tablecomments{The tabulated flux-density values are as observed, uncorrected for interstellar extinction.  They are plotted in Figures~\ref{sgrAplot} and \ref{vlt_keck_plot}. Times are heliocentric Modified Julian Dates.
(Table~\ref{LightcurveData}  is available in its entirety in a machine-readable form in the online journal. A portion is shown here for guidance regarding its form and content.)}\label{LightcurveData}

\end{table}

\begin{figure}
\begin{center}
\includegraphics[scale=0.65, angle=0]{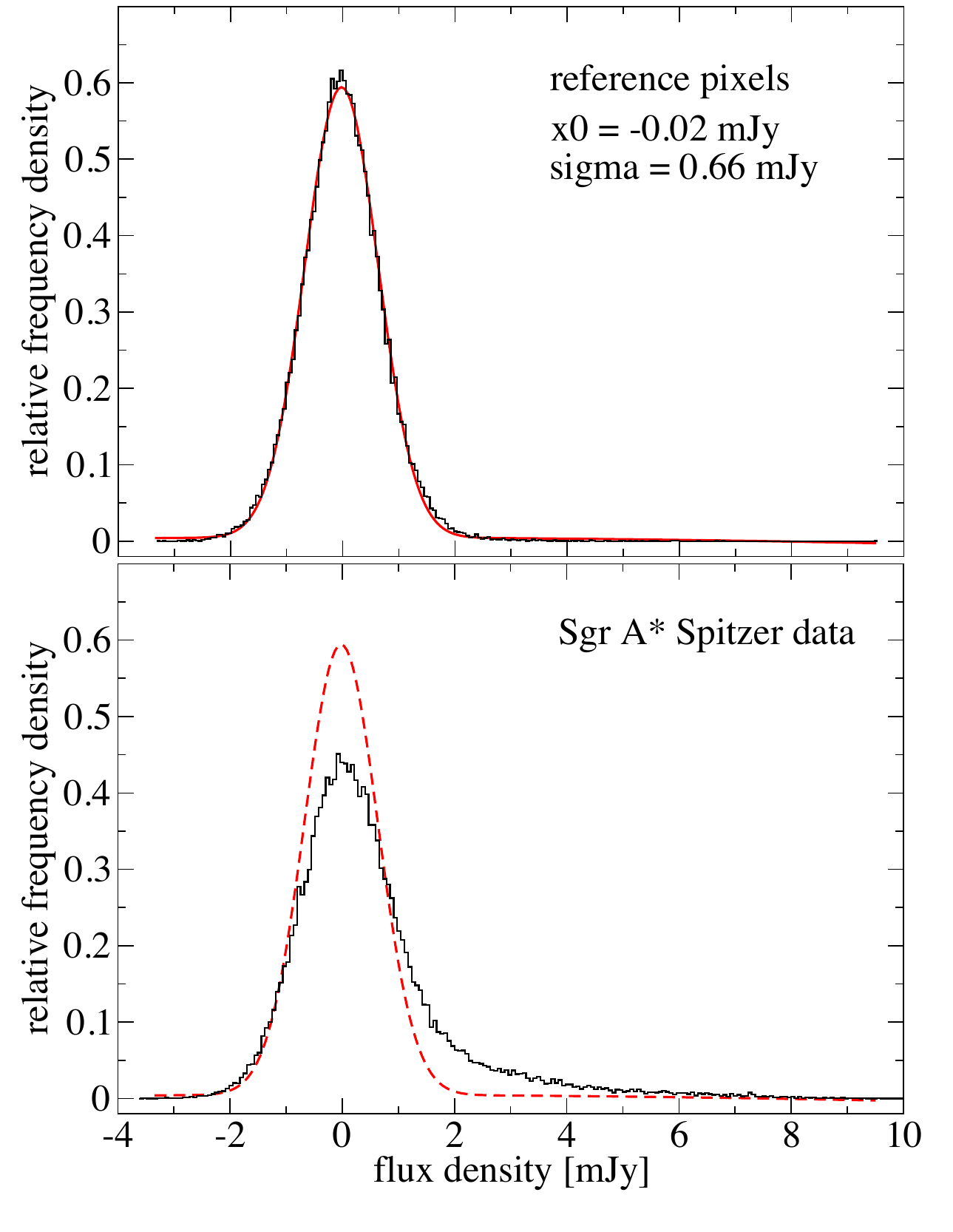}
\end{center}
\setlength{\abovecaptionskip}{-10pt}
\caption{
%\replaced
{Normalized flux density distributions}
%{Normalized distributions of excess flux densities}
for the combined IRAC data.
Black curves show the observed
distributions: the reference pixel in the upper panel and \Sg\ in the
lower panel. The red dashed curve in both panels shows a Gaussian 
distribution centered 
at $x_0 = -0.02$~mJy and with a standard deviation  $\sigma = 0.66$~mJy.
\added{As explained in \S\ref{IRAC_obs}, the zero points correspond to the average flux density during times when the flux density was small, not to an absolute zero.}
% , in excellent agreement with the value obtained by the ABC algorithm of $\sigma = 0.67 \pm 0.02$~mJy. Lower panel: flux-density PDF of Sgr~A* in the \Sp\ epochs at 4.5~$\mu$m (black histogram, not dereddened). In red we have over-plotted the resulting curve of the Gaussian fit to the reference pixel flux-density PDF of the right panel.
} \label{refpixdistr}
\setlength{\belowcaptionskip}{-100pt}
\end{figure}

All eight \Sp\ epochs showed flux-density variations intrinsic to Sgr~A* in the range of $\sim$0--8.5~mJy (not dereddened; see Figure~\ref{refpixdistr}). The first and the sixth epochs (2013 December 10, 2016 June 18) showed the highest peaks and the longest-duration excursions from zero. In contrast, the epoch of 2014 June 2 showed only minor excursions during the $>$23~hr of observations. The noise characteristics of the \Sp\ data can be estimated using the flux-density PDF of the reference pixels (shown in Figure~\ref{refpixdistr}), which has a standard deviation $\sigma_{\rm IRAC} = 0.66$~mJy for one 6.4~s frame set.

\subsection{Ground-based observations with VLT and Keck}\label{groundobs}

The VLT data (previously reported by \citealt{2012ApJS..203...18W})
were taken with the adaptive optics camera Naos Conica (NaCo; \citealt{2003SPIE.4841..944L}) in $K_s$-band 
(2.18~\micron).  The NaCo images have 68~mas resolution and integration times of
30--40~s.  Data were
taken between 2003-06-13 and 2010-06-16.
The complete data set,
after rejecting images with unstable zero points, contains  10,639 images.
The average cadence of the 
observations is one image per 1.2~minutes, the cadence being limited by 
deliberate telescope offsets (``dithering'') between frames. 
\citet{2012ApJS..203...18W} provided an observing log, and
described the data reduction and calibration.

The Keck data were obtained with the NIRC2 camera {(PI Keith Matthews)} in the $K’$-band (2.12 µm).
Images have 53~mas resolution and a fixed integration time of 28~s.  The data set contains 3157 images between 2004-07-16 and 2013-07-19. The average cadence was one image per 1.1~minutes, again limited by dithering. 
Table~\ref{datasets} lists the Keck epochs analyzed here.

\begin{table}[ht!]
\begin{center}

\caption{Keck/NIRC2 Observation Log}
\begin{tabular}{cccrr}
\tableline
\tableline
&&&\\[-1ex]
Date  & Start time    &         Stop time    &     Duration   &   Number \\
(UT) & (UT) & (UT)& (minutes) & of frames\\
\tableline
&&&\\[-1ex]
2004-07-26 & 08:18:50  &   09:00:01   &  41.18  &  7 \\  
2005-07-30 & 07:51:43  &   08:47:24   &  55.68  &  4 \\  
2006-05-03 & 11:03:03  &   13:14:12   &  131.14 &  26 \\
2006-06-20 & 08:59:22  &   11:04:45   &  125.38 &  90 \\
2006-06-21 & 08:52:27  &   11:36:53   &  164.43 &  163 \\
2006-07-17 & 06:47:50  &   09:54:03   &  186.22 &  63 \\
2007-05-17 & 11:08:23  &   13:52:39   &  164.26 &  81 \\
2007-08-10 & 06:54:19  &   08:21:05   &  86.77  &  78 \\
2007-08-12 & 06:47:09  &   07:44:37   &  57.47  &  60 \\
2008-05-15 & 10:32:40  &   13:05:16   &  152.59 &  129 \\
2008-07-24 & 06:21:14  &   09:20:04   &  178.83 &  173 \\
2009-05-01 & 11:50:04  &   14:51:44   &  181.67 &  186 \\
2009-05-02 & 11:48:28  &   12:49:31   &  61.04  &  53 \\
2009-05-04 & 12:48:42  &   13:40:32   &  51.84  &  57 \\
2009-07-24 & 07:09:43  &   09:25:34   &  135.85 &  138 \\
2009-09-09 & 05:23:34  &   06:19:27   &  55.87  &  49 \\
2010-05-04 & 11:42:12  &   14:45:44   &  183.54 &  118 \\
2010-05-05 & 13:34:16  &   14:41:24   &  67.13  &  75 \\
2010-07-06 & 07:23:03  &   09:28:04   &  125.02 &  130 \\
2010-08-15 & 05:45:35  &   08:01:03   &  135.47 &  138 \\
2011-05-27 & 10:37:31  &   13:16:23   &  158.87 &  150 \\
2011-08-23 & 05:57:35  &   07:30:44   &  93.15  &  105 \\
2011-08-24 & 05:49:56  &   07:26:34   &  96.62  &  107 \\
2012-05-15 & 10:56:28  &   14:00:01   &  183.54 &  203 \\
2012-05-18 & 10:29:53  &   12:54:26   &  144.54 &  74 \\
2012-07-24 & 06:05:04  &   09:25:28   &  200.40 &  208 \\
2013-04-26 & 12:59:28  &   14:52:09   &  112.69 &  119 \\
2013-04-27 & 12:53:26  &   15:09:22   &  135.93 &  137 \\
2013-07-20 & 06:04:26  &   09:32:51   &  208.42 &  234 \\
%\tableline
2016-07-12 & 06:59:04  & 10:08:59 & 188.21 & 204\\
\tableline
\end{tabular}
\end{center}
\tablecomments{This table lists the datasets used in this work and by \cite{2009ApJ...694L..87M}. Times are UTC at the observatory, not heliocentric.} 
\label{datasets}

\end{table}

\begin{figure}
\begin{center}
\includegraphics[scale=0.29, angle=0]{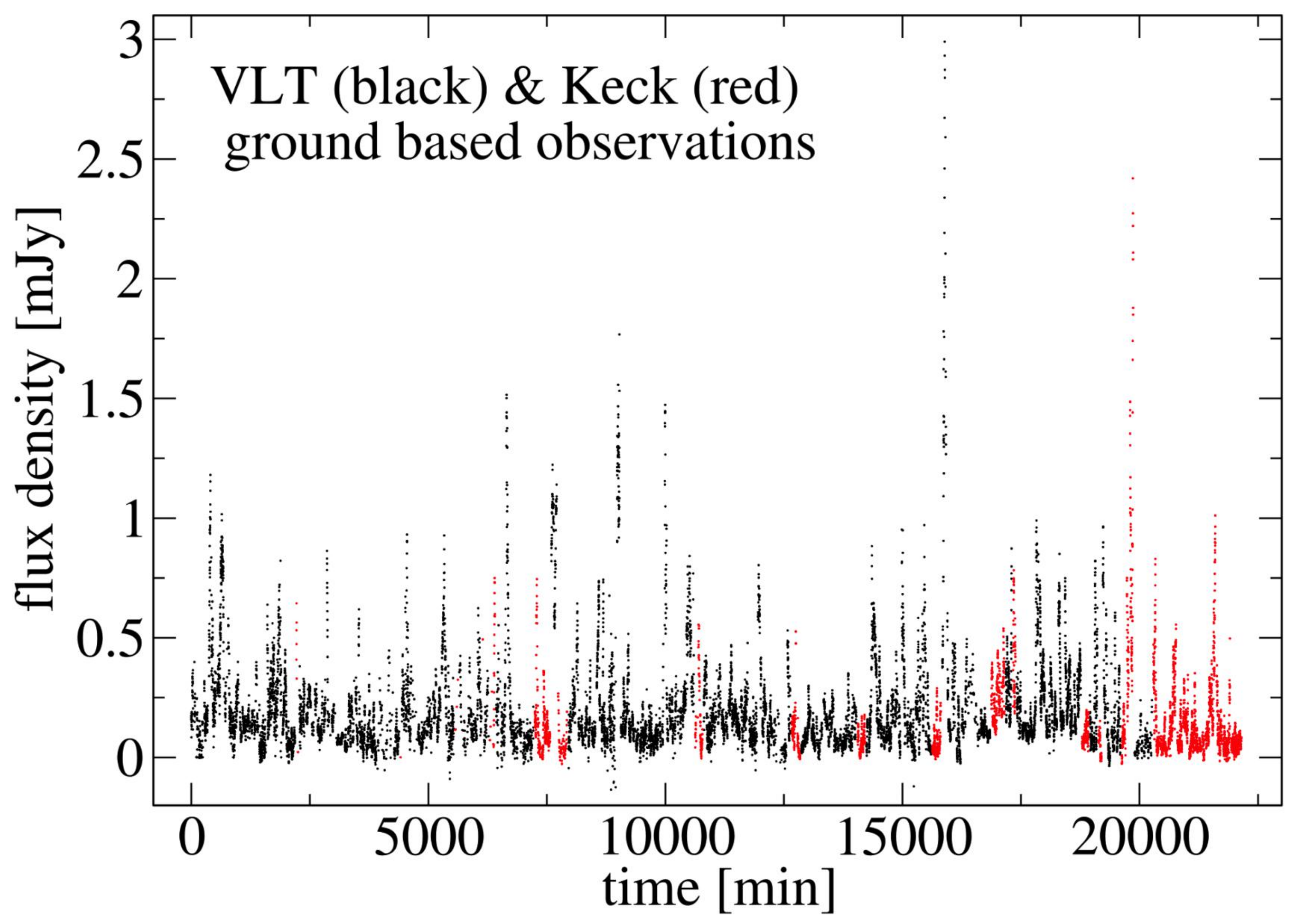}
\end{center}
\setlength{\abovecaptionskip}{-2pt}
\caption{$K$-band light curve of Sgr~A*  observed with ground-based observatories.  The data were taken in hours-long segments over more than
a decade and are
here joined together on a linear abscissa for display.
Black points show data taken with VLT/\discretionary{}{}{}NaCo at 2.18~$\mu$m (Table~2 of \citealt{2012ApJS..203...18W}).  Red points show data taken with Keck/NIRC2 (Table~\ref{datasets}) at 
2.12~$\mu$m. Flux densities are as observed with no correction
for interstellar extinction.  The combined $K$-band data have been used previously by \cite{2014ApJ...791...24M}.}\label{vlt_keck_plot}
\end{figure}

For both the NaCo and NIRC2 data sets, \Sg\ flux densities were derived from
aperture photometry on deconvolved images. Flux-density calibration used
13 non-variable stars throughout all epochs with consistent flux densities
adopted for both telescopes.  (Exact details are given by
\citealt{2012ApJS..203...18W}.) {We corrected both data sets for flux-density background levels caused by
extended point spread functions of nearby sources (source confusion) based on yearly minimums of \Sg. This procedure is justified by the fact that the mean flux density of \Sg\ is constant within the uncertainties over $\sim$20~years of observations (Chen et al. 2018, in preparation)} The (Gaussian) measurement noise
was 0.033~mJy for NaCo and 0.017~mJy for NIRC2. Typical background flux densities
estimated in the direct vicinity of Sgr~A* are 0.06~mJy (NaCo) and 0.03~mJy
(NIRC2). Observed flux densities ranged from 0 to 2.9~mJy with NaCo and from
0 to 2.3~mJy with NIRC2. We have calibrated the flux densities at the NIRC2
effective wavelength of 2.12~$\mu$m with the same magnitudes and zero point
as for NaCo with an effective wavelength of 2.18~$\mu$m. This introduces
a systematic error of $<$1\%, much smaller than the overall flux-density 
calibration uncertainty of $10\%$. The relative calibration uncertainty
is $\sim$2\%.  For a discussion of the conversion between NaCo $K_s$ and NIRC2 $K'$ photometry, see \citet[][appendix]{2013ApJ...764..154D}. Figure~\ref{vlt_keck_plot} and
Table~\ref{LightcurveData} give the $K$ light
curve data. 

\begin{figure*}
\begin{center}
\includegraphics[scale=0.18, angle=0]{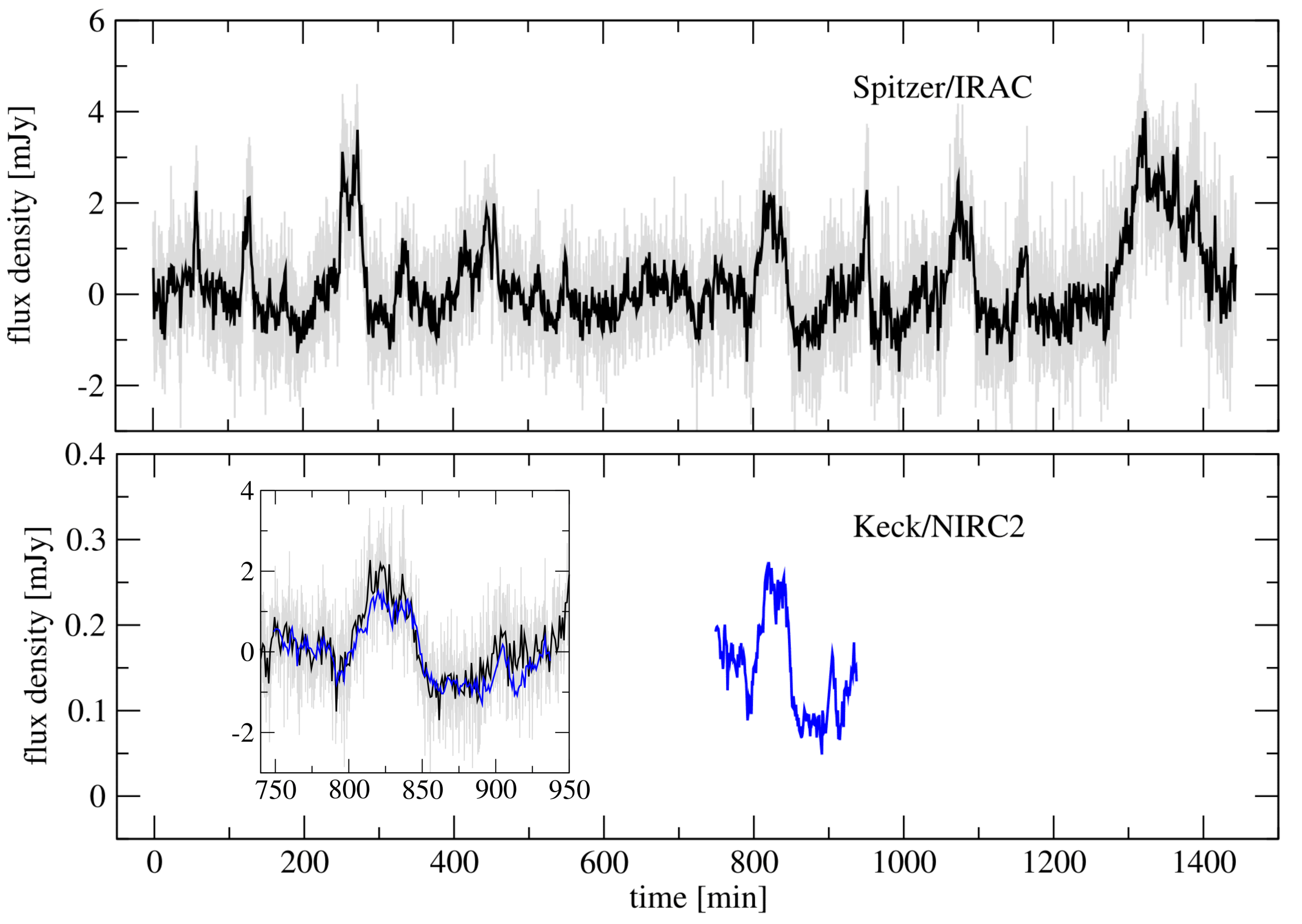}
\end{center}
\setlength{\abovecaptionskip}{-5pt}
\caption{Observations with \Sp/IRAC (black) and Keck/NIRC2 (blue) on 2016 July 12--13\deleted{, offset corrected according to the results presented in Figure~\ref{corner_mcmc}}.  \replaced{The inset shows both light curves on a scale expanded by a factor 12.4}{The inset shows both light curves on an expanded abscissa and with $K$ flux density multiplied by a factor of 12.4 and then 1.74~mJy subtracted (see \ref{bayratio} and Figure~\ref{corner_mcmc}) to match $M$}. Light curves are given in observed flux density with no interstellar extinction correction. This is the only simultaneous dataset from both observatories that shows significant variability.}\label{simdata}
\setlength{\belowcaptionskip}{0pt}
\end{figure*}

\subsection{Simultaneous observations with NIRC2 and IRAC}\label{simobs}

A key dataset was the one on 2016 July 13, when we observed \Sg\  with NIRC2 at 2.12~\micron\ during IRAC 4.5~\micron\ observations that began July 12.
The AO correction for the NIRC2 dataset was comparatively poor due to the atmospheric conditions for this night, but the frames show a significant enough flux-density excursion to be taken into account in this paper. Because of the lower data quality, the standard reduction methods described above gave poor results.  However, the UCLA Galactic center group developed a
new software package ``AIROPA'' (\citealt{2016SPIE.9909E..1OW}) based on the 
PSF-fitting code StarFinder (\citealt{2000SPIE.4007..879D}). This package was designed to take
atmospheric turbulence profiles, instrumental aberration maps, and images as
inputs, and then fit field-variable PSFs to deliver improved photometry and 
astrometry on crowded fields. AIROPA uses improved StarFinder subroutines, in particular a much improved PSF extraction that also benefits local, static (non-field-dependent) PSF-fitting as applied to these data. Running AIROPA in static PSF mode and using the resulting PSFs to
deconvolve the individual frames of 2016 July 13 improved the signal-to-noise
of the light curve by a factor of three in comparison to the standard reduction.  Figure~\ref{simdata} shows the IRAC and NIRC2 light curves.

\begin{figure}
\begin{center}
\includegraphics[scale=0.43, angle=0]{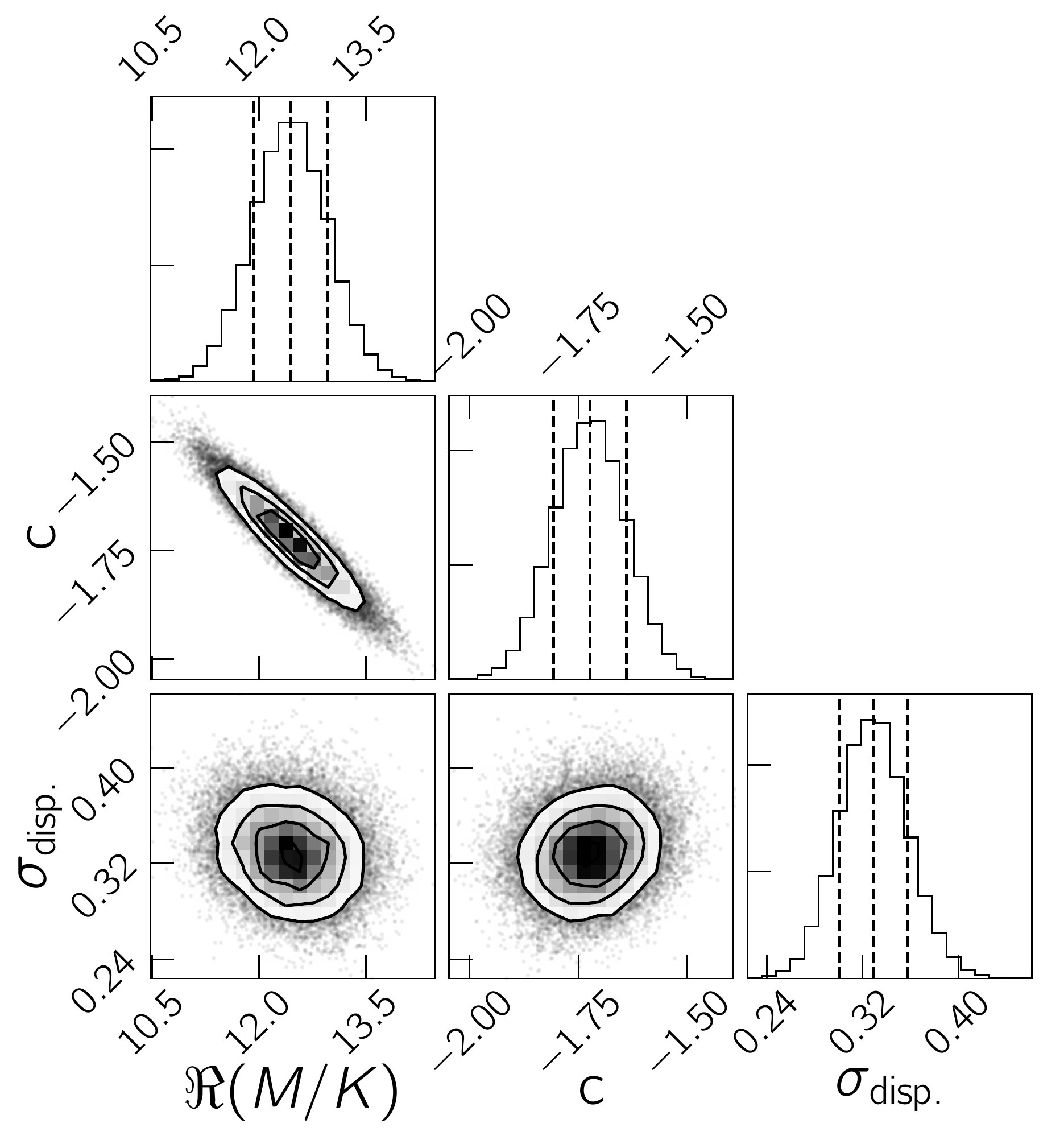}
\end{center}
\setlength{\abovecaptionskip}{-4pt}
\caption{Results of the MCMC analysis for $\Re(M/K)$ from the simultaneous IRAC and NIRC2 data (\S\ref{simobs}\added{, \ref{bayratio}}). Contours show the joint (posterior) probability density for each parameter pair, and panels along the upper right edge show histograms of the marginalized posterior of each parameter.  \added{For each histogram, the dashed lines mark the 16\%, 50\%, and 84\% quantiles.} Parameters are the ratio $\Re(M/K)$, the dispersion $\sigma_{\rm{disp}}$ in the ratio, and the constant offset $c$. }\label{corner_mcmc}
\end{figure}

It is remarkable how well the NIRC2 light curve is matched by the IRAC data. These two light curves impose strong limits on the $F(M)/F(K)$ ratio  (from here on denoted $\Re({M}/{K})$), at least for the observed flux-density levels, which have medians of \replaced{0.07}{0.15} and 0.94~mJy at $K$ and $M$ respectively \added{(but with the $M$-band zero point offset as noted in Section~\ref{IRAC_obs})}.  \replaced{These are about 5\% and 9\% of the maximum flux densities seen at these wavelengths.}{In $K$-band, this value is about 5\% of the maximum flux densities seen at this wavelength.} \replaced{However, the NIRC2 data show a significant offset due to}{Despite} confusion with the first Airy ring of the bright star S0-2 (S0-2's closest approach to Sgr~A* is anticipated for 2018), we were able to extract $K$-band fluxes at the position of \Sg\ and its vicinity with essentially zero flux density offset. \added{The remaining low-level flux density floor was determined in `empty' apertures without obvious point sources next to \Sg\ and subtracted from the $K$-band light curve.} In order to properly determine the \added{relative} offset and the flux-density ratio between the two bands, we resampled the $M$-band light curve (which has much higher cadence) to the cadence of the $K$-band light curve, and then used an MC-MC implementation in Pystan (\citealt{stan_ref}) to derive the Bayesian posteriors for the offset and the ratio while taking into account the two different measurement noise amplitudes (see \ref{bayratio}). The resulting corner plot is shown in Figure~\ref{corner_mcmc}, and the resulting uncorrected flux-density ratio $\Re({M}/{K}) = 12.4 \pm 0.5$. \added{The relative offset  $c = -1.72 \pm 0.08$, and the total dispersion $\sigma_{\rm{disp}} = 0.33 \pm 0.03$.}
{\replaced{This value is}{These values are} the integrated ratio \added{and relative offset} over the entire 204 frames and $\sim$3~hr. Instantaneous \added{ratio} values can be even higher, and around $t=820$--825~minutes, there is a significant deviation with $\Re({M}/{K}) \approx 14.7$.}

\section{Bayesian Light Curve Modeling and Results}\label{meth}

The goal of the analysis, as it was for \citet{2014ApJ...793..120H},
is to find the parameters that best describe the 
statistical variability of the observed light curves. {Compared to the earlier work, the present study uses seven additional 24~hr IRAC datasets, 123 additional epochs of ground-based observations, and a more rigorous method to explore the parameter space. 
Simple periodograms, as shown in  Figure~\ref{datapsd}, demonstrate the overall properties of the variability but do not provide the required fidelity in PSD parameter estimation.  A break near 0.01~minutes$^{-1}$ is evident, but the noise
does not permit a precise determination of the break frequency.
\begin{figure}
\begin{center}
\includegraphics[scale=0.24, angle=0]{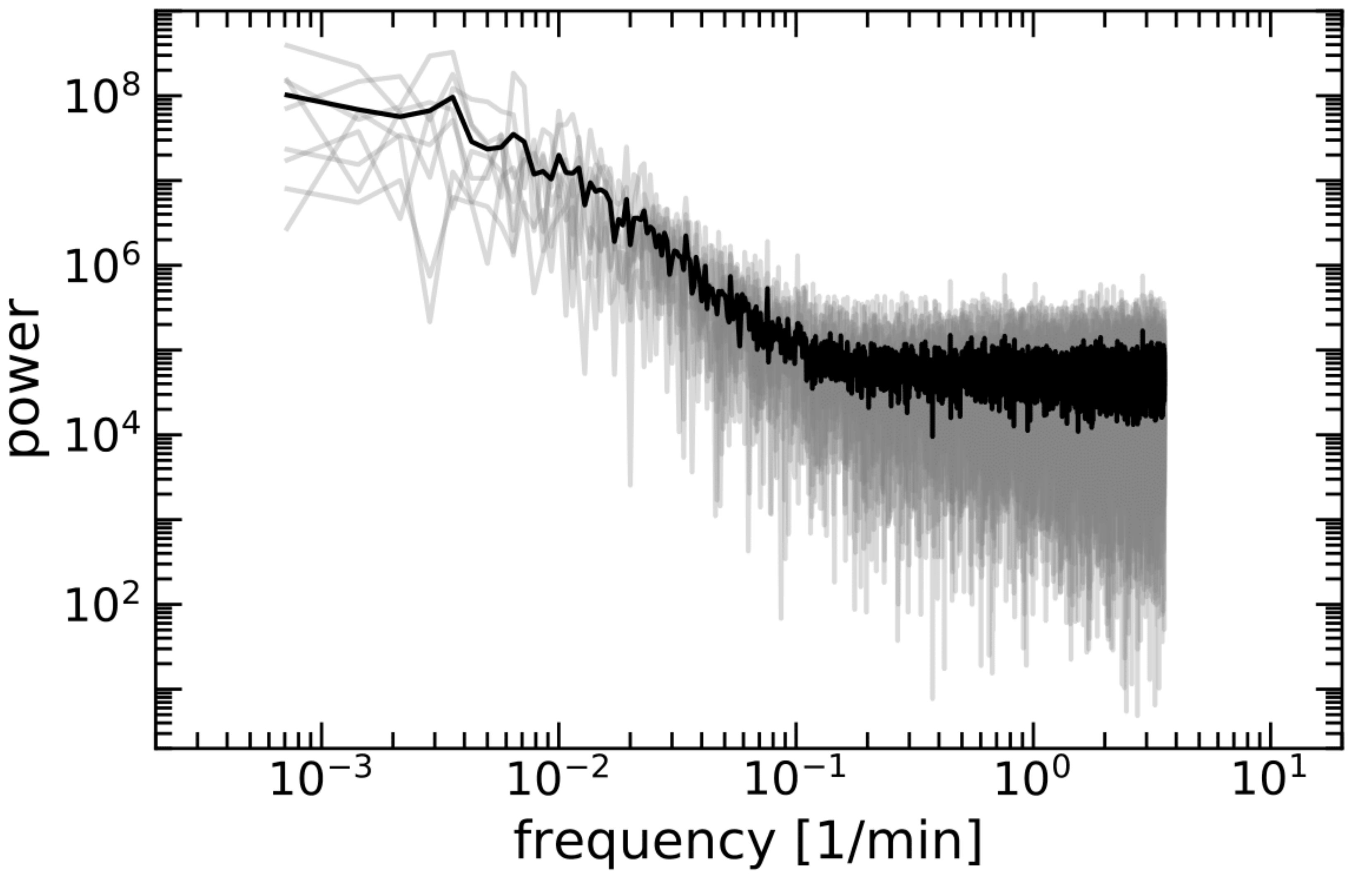}
\end{center}
\setlength{\abovecaptionskip}{-10pt}
\caption{FFT periodograms of the eight IRAC datasets.  Gray lines show the individual data sets, and the black line shows their average at each frequency.   The calculation is facilitated by the IRAC light curve points being almost equally spaced in time.
}\label{datapsd}
\end{figure}

The analysis method used here is simple
in principle but computationally expensive.} A set of statistical parameters was chosen based
on prior knowledge of the variability properties.  From each parameter
set, many mock light curves were generated and 
compared to the real ones.  
The parameters were then modified iteratively, and new sets of mock light curves
generated, seeking parameter values
that minimized the differences between the real and mock data.  Such an
approximate Bayesian computation\footnote{\added{ABC algorithms are routinely used in cosmology (see, e.g., \citealt{2015JCAP...08..043A}).}} (ABC)
gives \added{approximate} posterior distributions for the model parameters, including
proper uncertainties and correlations between the
parameters, \replaced{An ABC is needed here because there is no analytic likelihood function}{without needing an analytic likelihood function. The approximation accuracy is contingent on the selected distance function---the function that quantifies the difference between real and mock data (see \ref{dist}).}
%An ABC is needed here because there is no analytic likelihood function.  

\begin{figure}
\begin{center}
\includegraphics[scale=0.6, angle=0]{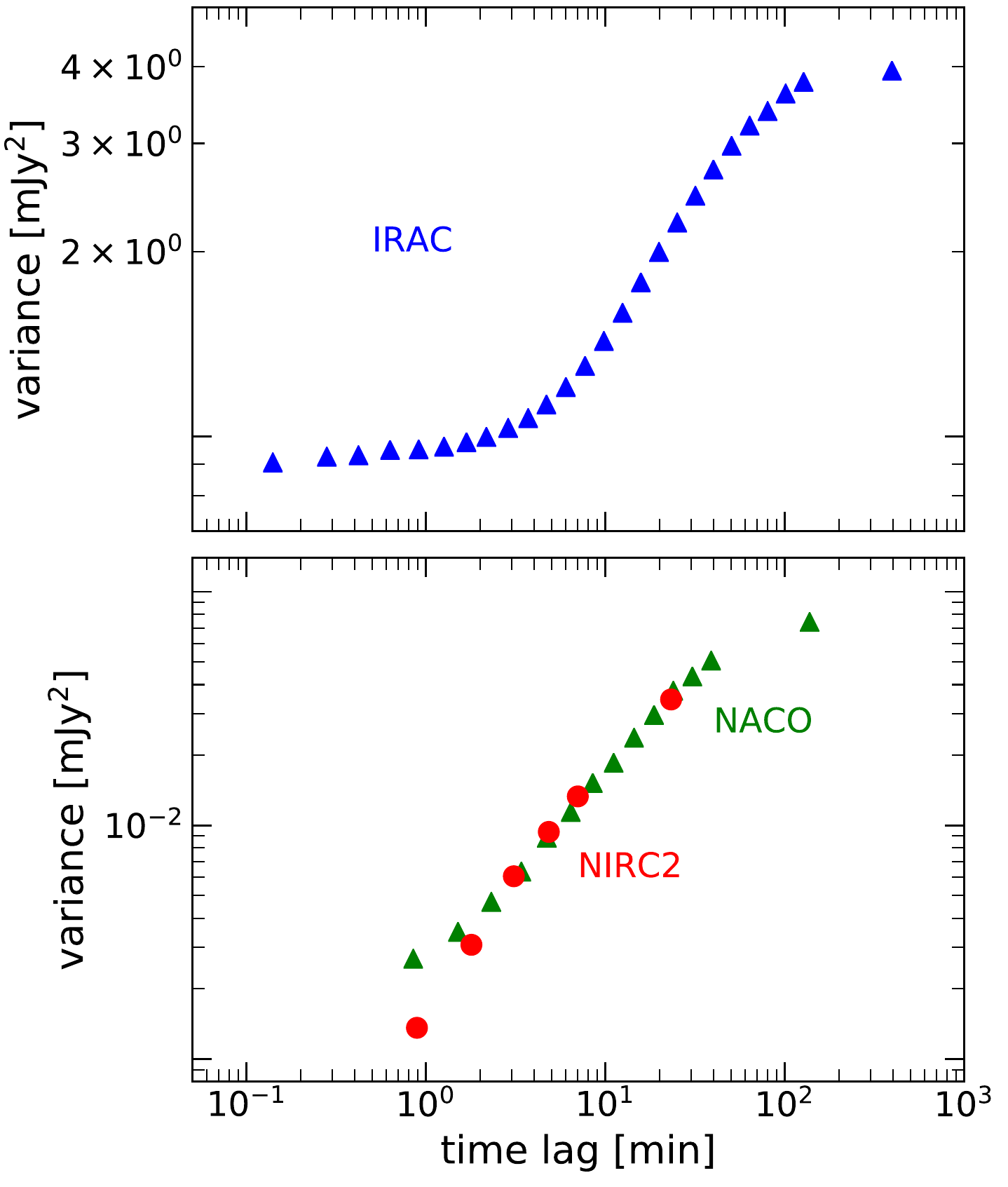}
\end{center}
\setlength{\abovecaptionskip}{-10pt}
\caption{Logarithmically binned structure functions 
(Eq.~\ref{structurefunction}) for the light curve data.  The lower panel
shows the NaCo structure function in green and the NIRC2 structure function in red.
The upper panel shows the IRAC structure function.}\label{structplot}
\end{figure}

%The actual algorithm fits structure functions rather than light curves themselves. 
\added{The variability analysis needs to model flux density differences as a function of time lag between measurements. Our analysis  is therefore based on the structure function rather than the light curves themselves.}
The first-order\footnote{In the variability
literature as followed here,
the definition of the structure function is such that a structure
function of order $M$ removes polynomials of order $M-1$ from the data -- that is, 
the {\it first}-order structure function is blind to DC offsets in the data. In the literature about turbulent media,  $V(\tau)$ as defined here is called the {\it second}-order structure function.}
structure function $V(\tau)$ of a light curve $F(t)$ is defined as
\begin{equation}
V(\tau) = \langle [F(t + \tau) - F(t)]^2 \rangle\quad,
\label{structurefunction}
\end{equation}
that is, as the variance of the process at a given time lag $\tau$ (\citealt{1985ApJ...296...46S,1992ApJ...396..469H}). The structure functions derived from the three datasets are shown in Figure~\ref{structplot}. \

{The underlying model is based on the results of earlier analyses:
\begin{itemize}
\item The long-term flux-density PDF in $K$-band is a highly skewed 
distribution, well described by either a power law with a slope 
$\beta = 4.2$ and a pole $F_0 = -0.37$~mJy (\citealt{2012ApJS..203...18W})
or by a log-normal distribution.
\item The PSD has the form of a power-law with a slope $\gamma_1 \approx 2$ and a break at a couple of hundred minutes (\citealt{2009ApJ...691.1021D,2012ApJS..203...18W,2014ApJ...791...24M,2014ApJ...793..120H}; Figure~\ref{datapsd}).
\item The noise properties of the individual datasets are well described by  Gaussians. (For the VLT and Keck data, see \citealt{2012ApJS..203...18W,2014ApJ...791...24M}; for the \Sp\ data, see \S\ref{IRAC_obs}.)
\item The uncorrected \added{average} flux-density ratio for bright phases \added{($F(K) > 0.2$~mJy)} of Sgr~A* $ \Re(M/K)  = 6_{-3}^{+5}$.  This corresponds to NIR spectral index $\alpha_s = 0.6 \pm 0.2$ (\citealt{2014ApJ...793..120H,2014IAUS..303..274W}).
\end{itemize}
}

{Two crucial parts of the ABC algorithm are (1) a method to simulate mock data from the model parameters, and (2) a distance function that describes how closely the mock data resemble the observed sample.} Our PMC-ABC implementation, which follows that of
\citet{2015A&C....13....1I}, is an iterative one that first chooses random values for each of 11 parameters (listed in Table~\ref{results})  according to the
current probability distribution for each.  (For the first iteration, the probability distribution is given by the priors.)  Each parameter set is used to generate a mock light curve for NIRC2, NaCo, and IRAC, and each light curve is transformed to its structure function.   For this step, the range and binning of time lags must match those of the real data.

Many structure functions are generated this way, each from new values of the 11 parameters but with the probability distributions fixed.  These structure functions are compared with the structure functions of the real data via a distance function (see \ref{ABC}).
The parameter sets that give structure functions closest to the real data
are used to modify the parameter probability distributions, and the cycle is repeated.

{The structure function is blind to DC offsets, which is important in the context of the arbitrary flux-density zero points of the Spitzer epochs.} It encodes information on the flux-density PDF, the measurement noise, the intrinsic correlations of the variability process, and the cadence and window function of the observations. (For detailed discussions of the structure function, see \citealt{2010MNRAS.404..931E} and \citealt{2016ApJ...826..118K}.) The intrinsic variability process and the window function are hard to disentangle, and for our analysis it is important to choose a representation that emphasizes the parts of the structure function that are dominated by the intrinsic correlations. With increasing time lag, a decreasing number of point pairs contribute to the structure function bins. For time lags longer than half the observing window (i.e., 12 hr for \Sp), not  all flux-density measurements contribute to every structure function bin, and the variance of the structure function

%\FloatBarrier

\begin{table*}[h]
\begin{center}
\caption{Priors and Posteriors of Bayesian analysis} \label{results}
\begin{tabular}{llll}
\tableline
\tableline
&&&\\[-1ex]
& & Mean of &\\
Parameter & Prior & Posterior & Description\\
\tableline
\null\\[-1ex]
\multicolumn{4}{l}{\quad Case~1: $K$ power-law/$M$ power-law model}\\
$\gamma_{1}$ & flat\tablenotemark{a} on [1.2, 3.5] & $2.21^{+0.12}_{-0.11}$ & primary PSD slope\\
$\gamma_{2}$\tablenotemark{b} & flat\tablenotemark{a} on [1.2, 10.0] & $6.0^{+2.8}_{-2.5}$ & secondary PSD slope\\
$f_{b}$ [$10^{-3} \rm{minutes}^{-1}$] & flat\tablenotemark{a} on [1.0, 600.] & $3.50^{+0.98}_{-0.89}$ & primary correlation frequency\\
$f_{b,2}$ [$\rm{minutes}^{-1}$] & flat\tablenotemark{a} on [0.001, 0.6] & $0.34^{+0.18}_{-0.16}$ & secondary break frequency\\
&& $> 0.120$ & $(95\%$ \replaced{confidence level}{credible} level) \\
$F_{0}$ [mJy] & Gaussian ($\mu = -0.36$, $\sigma=0.05$) & $-0.37^{+0.05}_{-0.05}$ &  pole of the power-law flux-density PDF ($K$- and $M$-band)\\
$\beta_{K}$ & Gaussian ($\mu = 4.22$, $\sigma=0.6$) & $4.53^{+0.34}_{-0.33}$ & slope of the power-law flux-density PDF in $K$-band \\
$\beta_{M}$ & Gaussian ($\mu = 4.22$, $\sigma=0.6$) & $4.45^{+0.56}_{-0.56}$ & slope of the power-law flux-density PDF in $M$-band\\
$s$ & flat on [0.01, 24.0] & $5.9^{+2.5}_{-1.9}$ & $M$ to $K$ flux density ratio \\
$\sigma_\mathrm{Keck}$ [mJy] &Gaussian ($\mu = 0.017$, $\sigma=0.008$)  & $0.014^{+0.004}_{-0.006}$ & measurement noise of the Keck observations\\
$\sigma_\mathrm{VLT}$ [mJy] & Gaussian ($\mu = 0.034$, $\sigma=0.008$) & $0.031^{+0.003}_{-0.003}$ & measurement noise of the VLT observations\\
$\sigma_\mathrm{IRAC}$ [mJy] & Gaussian ($\mu = 0.65$, $\sigma=0.2$) & $0.676^{+0.020}_{-0.019}$ & measurement noise of the IRAC observations\\
\tableline
% \end{tabular}

% \end{center}
% \end{table*}

% \begin{table*}
% \begin{center}
% \caption{Priors and Posteriors of our Bayesian analysis, power-law(K)/log-normal model(M).} \label{results_logn}
% \begin{tabular}{llll}
% \tableline
% \tableline
% &&&\\
% & & Mean of &\\
% Parameter & Prior & Posterior & Description\\
% \tableline
% &&&\\
\null\\[-1ex]
\multicolumn{4}{l}{\quad Case~2: $K$ power-law/$M$ log-normal model}\\
$\gamma_{1}$ & flat\tablenotemark{a} on [1.2, 3.5] & $2.21^{+0.12}_{-0.11}$ & primary PSD slope\\
$\gamma_{2}$\tablenotemark{b} & flat\tablenotemark{a} on [1.2, 10.0] & $6.0^{+2.7}_{-2.6}$ & secondary PSD slope\\
$f_{b}$ [$10^{-3} \rm{minutes}^{-1}$] & flat\tablenotemark{a} on [1.0, 600.] & $3.71^{+1.2}_{-1.1}$ & primary correlation frequency\\
$f_{b,2}$ [$\rm{minutes}^{-1}$] & flat\tablenotemark{a} on [0.001, 0.6] & $0.33^{+0.17}_{-0.16}$ & secondary break frequency\\
&& $> 0.112$ & $(95\%$ \replaced{confidence level}{credible} level) \\
$F_{0}$ [mJy] & Gaussian ($\mu = -0.37$, $\sigma=0.05$) & $-0.37^{+0.05}_{-0.04}$ &  pole of the power-law flux-density PDF in $K$-band \\
$\beta_{K}$ & Gaussian ($\mu = 4.22$, $\sigma=0.6$) & $4.59^{+0.32}_{-0.32}$ & slope of the power-law flux-density PDF in $K$-band \\
$\mu_{\logn, {M}}$ & flat on [$-$6.0, 6.0] & $-0.3^{+1.3}_{-1.3}$ & log-normal mean in $M$-band\\
$\sigma_{\logn, {M}}$ & flat on [0.001, 4.0] & $0.89^{+0.54}_{-0.50}$ & log-normal standard deviation in $M$-band \\
$\sigma_\mathrm{Keck}$ [mJy] &Gaussian ($\mu = 0.017$, $\sigma=0.008$)  & $0.015^{+0.004}_{-0.006}$ & measurement noise of the Keck observations\\
$\sigma_\mathrm{VLT}$ [mJy] & Gaussian ($\mu = 0.034$, $\sigma=0.008$) & $0.031^{+0.003}_{-0.003}$ & measurement noise of the VLT observations\\
$\sigma_\mathrm{IRAC}$ [mJy] & Gaussian ($\mu = 0.65$, $\sigma=0.2$) & $0.678^{+0.020}_{-0.019}$ & measurement noise of the IRAC observations\\
\tableline
% \end{tabular}

% \end{center}
% \end{table*}

% \begin{table*}
% \begin{center}
% \caption{Priors and Posteriors of our Bayesian analysis, power-law(K)/log-normal model(M).} \label{results_double_logn}
% \begin{tabular}{llll}
% \tableline
% \tableline
% &&&\\
% & & Mean of &\\
% Parameter & Prior & Posterior & Description\\
% \tableline
% &&&\\
\null\\[-1ex]
\multicolumn{4}{l}{\quad Case~3: $K$ log-normal/$M$ log-normal model \added{+ spectral information from synchronous data}}\\
$\gamma_{1}$ & flat\tablenotemark{a} on [1.2, 3.5] & $2.10^{+0.10}_{-0.09}$ & primary PSD slope\\
$\gamma_{2}$\tablenotemark{b} & flat\tablenotemark{a} on [1.2, 10.0] & $5.8^{+2.8}_{-2.4}$ & secondary PSD slope\\
$f_{b}$ [$10^{-3} \rm{minutes}^{-1}$] & flat\tablenotemark{a} on [1.0, 600.] & $4.11^{+0.76}_{-0.65}$ & primary correlation frequency\\
$f_{b,2}$ [$\rm{minutes}^{-1}$] & flat\tablenotemark{a} on [0.001, 0.6] & $0.31^{+0.19}_{-0.15}$ & secondary break frequency\\
&& $> 0.118$ & $(95\%$ \replaced{confidence level}{credible} level) \\
$\mu_{\logn, {K}}$ & flat on [$-$8.3, 3.7] & $-1.35^{+0.62}_{-0.60}$ &  log-normal mean in $K$-band\\
$\sigma_{\logn, {K}}$ & flat on [0.001, 4.0]  & $0.56^{+0.24}_{-0.21}$ &  log-normal standard deviation in $K$-band \\
$\mu_{\logn, {M}}$ & flat on [$-$6.0, 6.0] & $1.01^{+0.47}_{-0.44}$ & log-normal mean in $M$-band\\
$\sigma_{\logn, {M}}$ & flat on [0.001, 4.0] & $0.39^{+0.15}_{-0.13}$ & log-normal standard deviation in $M$-band\\
$\sigma_\mathrm{Keck}$ [mJy] &Gaussian ($\mu = 0.017$, $\sigma=0.008$)  & $0.013^{+0.005}_{-0.006}$ & measurement noise of the Keck observations\\
$\sigma_\mathrm{VLT}$ [mJy] & Gaussian ($\mu = 0.034$, $\sigma=0.008$) & $0.030^{+0.002}_{-0.003}$ & measurement noise of the VLT observations\\
$\sigma_\mathrm{IRAC}$ [mJy] & Gaussian ($\mu = 0.65$, $\sigma=0.2$) & $0.677^{+0.013}_{-0.013}$ & measurement noise of the IRAC observations\\
\tableline
\end{tabular}
\end{center}
\tablenotetext{a}{The joint prior distributions are flat under the conditions $f_{b,2} > f_{b}$ and $\gamma_2 > \gamma_1$, respectively; see \ref{priors}.}
\tablenotetext{b}{Unconstrained by the data; posterior is a minor alteration of the prior.}

\end{table*}
\FloatBarrier

\noindent increases dramatically without carrying much information about the intrinsic variability. Therefore we chose a logarithmic binning scheme, roughly equally spaced in logarithmic time lags, with a spacing large enough to allow for a similar number of points in the {long-time-lag} bins. We included time lags up to half the size of the observing window, $\sim$700 minutes in the case of the IRAC data. For the NaCo and the NIRC2 data, which have a wide range of observing window durations, we used points of similar variance increase in the structure function, 300 minutes and 40 minutes, respectively. For the ranges of [160, 700] minutes (IRAC), [50, 300] minutes (NaCo), and [10.5, 40] minutes (NIRC2), we used a single large bin with three times the weight in the distance function as the lower bins (see Equation~\ref{distdef}).\footnote{It is not necessary to densely sample the shape of the structure function around the break timescale. Because the mock data are computed as the Fourier transform of the PSD, the break frequency contributes to {\em all} timescales. However, the plateau of the structure function at the longest timescales is directly related to the variance of the process and crucially helps to constrain the PSD parameters.} 
This approach makes conservative use of the complementary but overlapping information provided by each instrument, with IRAC providing  the longest timescales covering the coherence timescale, NaCo at medium timescales between 100 and 10 minutes, and NIRC2 at the shortest timescales to below 1 minute.

The slope of the structure function is related to the slope of the underlying PSD but is also a function of the overall variance of the process and the variance of the measurement noise. In particular, for red noise with quickly decreasing amplitudes toward higher frequencies, the structure function at the shortest timescales close to the data cadence $\tau_{\rm{cad}}$ is 
\begin{equation}\label{lowwn}
V(\tau \approx \tau_{\rm{cad}}) \approx 2 \sigma^2 \quad, 
\end{equation}
with $\sigma$ the measurement noise. If the red-noise process has finite variance, then at timescales much larger than the coherence timescale $ \tau_b$, the structure function is
\begin{equation}\label{highwn}
V(\tau \gg \tau_b) \approx 2 \cdot {\rm Var}[F(t)] + 2 \sigma^2 \quad,
\end{equation}
with ${\rm Var}[F(t)]$ the variance of the variability process.

\cite{2015A&C....13....1I} implemented ABC sampling in Python and gave
a detailed description of the method.  Following their approach,
we developed our own C++ implementation.\footnote{Our C++ 
implementation (\ref{ABC}) is based on FFTW, 
uses an efficient algorithm (\ref{FFT}) for calculating 
structure functions,  and is fully parallelized for large computational 
clusters.} \ref{ABC} gives
a more detailed description of the algorithm and the underlying model. 

We tested three models of the flux density PDFs:
\begin{itemize}
\item Case 1 \added{(exploratory)}: independent power-law parametrizations of the flux-density 
PDFs in $K$-band and $M$-band
\item Case 2 \added{(exploratory)}: a power-law parametrization of the 
flux\discretionary{-}{}{-}density PDF in $K$-band
and a log\discretionary{-}{}{-}normal parametrization in $M$-band
\item Case 3 \added{(main result)}: independent log-normal parameter\-izations of the flux-density PDFs
in $K$- and $M$-band while including $\Re(M/K)=12.4\pm0.5$ from the synchronous 
$K$- and $M$-band data (Section~\ref{simobs})
\end{itemize}

All of the above  parametrizations describe the data in the limited flux-density range 
observed, and at least in the $K$-band, they are equally valid.
The choices were informed by the analyses of \cite{2011ApJ...728...37D}
and \cite{2012ApJS..203...18W}. While
a log-normal distribution can be expected from accretion variability processes (e.g., \citealt{2005MNRAS.359..345U}), and indeed a log-normal distribution can also describe the 
observed $K$-band flux densities,  the log-normal parameters derived are related to the location of the mode of the PDF. For the NaCo data, which constitute the majority of the $K$-band data, the mode is \replaced{within}{close to} the white-noise-dominated
part of the distribution. This makes both parameters difficult to 
determine with precision. In contrast, power-law parameters---slope and 
normalization---describe mainly the tail, which is well above the white noise. 
For $K$-band, 
\cite{2012ApJS..203...18W} showed that the power-law description is 
advantageous, but it makes the simplifying (and possibly unphysical) assumption \added{that the PDF increases monotonically toward smaller flux densities until hitting}
 a sharp cutoff at zero flux density.  Nevertheless, the baseline Case~1 
fit uses a power law for both bands. Because we do not have, a priori, a 
detailed understanding of the $M$-band distribution, and also motivated by \added{(but not explicitly using) the}
additional information drawn from synchronous data\deleted{ (described below)}, Case~2
investigates a log-normal distribution for $M$-band. Finally, adding constraints 
from simultaneous $K+M$ data lets even the double log-normal parametrization
give well-constrained parameters, and Case~3\replaced{explores this possibility}{, our preferred model, gives results for this possibility.}

To simultaneously fit the structure functions of the three datasets, 
the model parameters (Table~\ref{results}) are as follows: 
\begin{itemize}
\item in all cases the respective instrumental
measurement uncertainties $\sigma$ and four PSD parameters: slopes $\gamma_1$
and $\gamma_2$ and break frequencies $f_{b}$ and $f_{b,2}$; 
\item for Case~1, flux-density PDF parameters $F_0$ (pole), $\beta_K$ and 
$\beta_M$ (power-law slopes), and the $M$- to $K$-band ratio factor $s$;\footnote{This choice of parametrization was motivated by the reports that $\Re(M/K)$ is invariant within uncertainties, 
at least over a wide range of
timescales (except for very minor short-timescale fluctuations) and 
flux-density levels (\citealt{2007ApJ...667..900H,2014IAUS..303..274W} but 
disputed by, e.g., \citealt{2017MNRAS.468.2447P}). However, this parametrization permits  a flux-density-dependent $\Re(M/K)$ if $\beta_K \neq \beta_M$. In this case $s$ loses its meaning as the $M$- to $K$-band ratio factor (see \ref{ratioform}).} 
\item for Case~2, $K$ power-law parameters $F_0$ (pole) and $\beta_K$ and 
$M$ log-normal
parameters $\mu_{\logn,{M}}$ and $\sigma_{\logn,{M}}$.
\item for Case~3, two pairs of log-normal parameters $\mu_{\logn,{K}}$,
$\sigma_{\logn,{K}}$, $\mu_{\logn,{M}}$, and $\sigma_{\logn,{M}}$. The Case~3 analysis is additionally based on a modified distance function (Equation~\ref{dist_log_n}) to select combinations of log-normal PDFs that result in \replaced{$\Re[M/K, F(K) = 0.07~\rm{mJy}]\approx 12.4$}{$\Re[M/K, F(K) = 0.15~\rm{mJy}]\approx 12.4$} (see Section~\ref{alpha_disc} for details).
%\item Noise levels $\sigma_{\rm{IRAC}}$, $\sigma_{\rm{NIRC2}}$, and
%$\sigma_{\rm{NaCo}}$.
\end{itemize} 
Table~\ref{results} %, \ref{results_logn}, and \ref{results_double_logn} 
lists the priors for each of the parameters \added{(see also \ref{priors}).
We used informative Gaussian priors for the measurement noise levels, which are independently determined, and for the power-law parameters in exploratory Cases~1 and~2. The reasons are further discussed in Section~\ref{quality}. For Case~3, we used flat priors for the unknown parameters in order to let the data {dominate} the posteriors.}

\begin{figure*}
\begin{center}
\includegraphics[scale=0.27, angle=0]{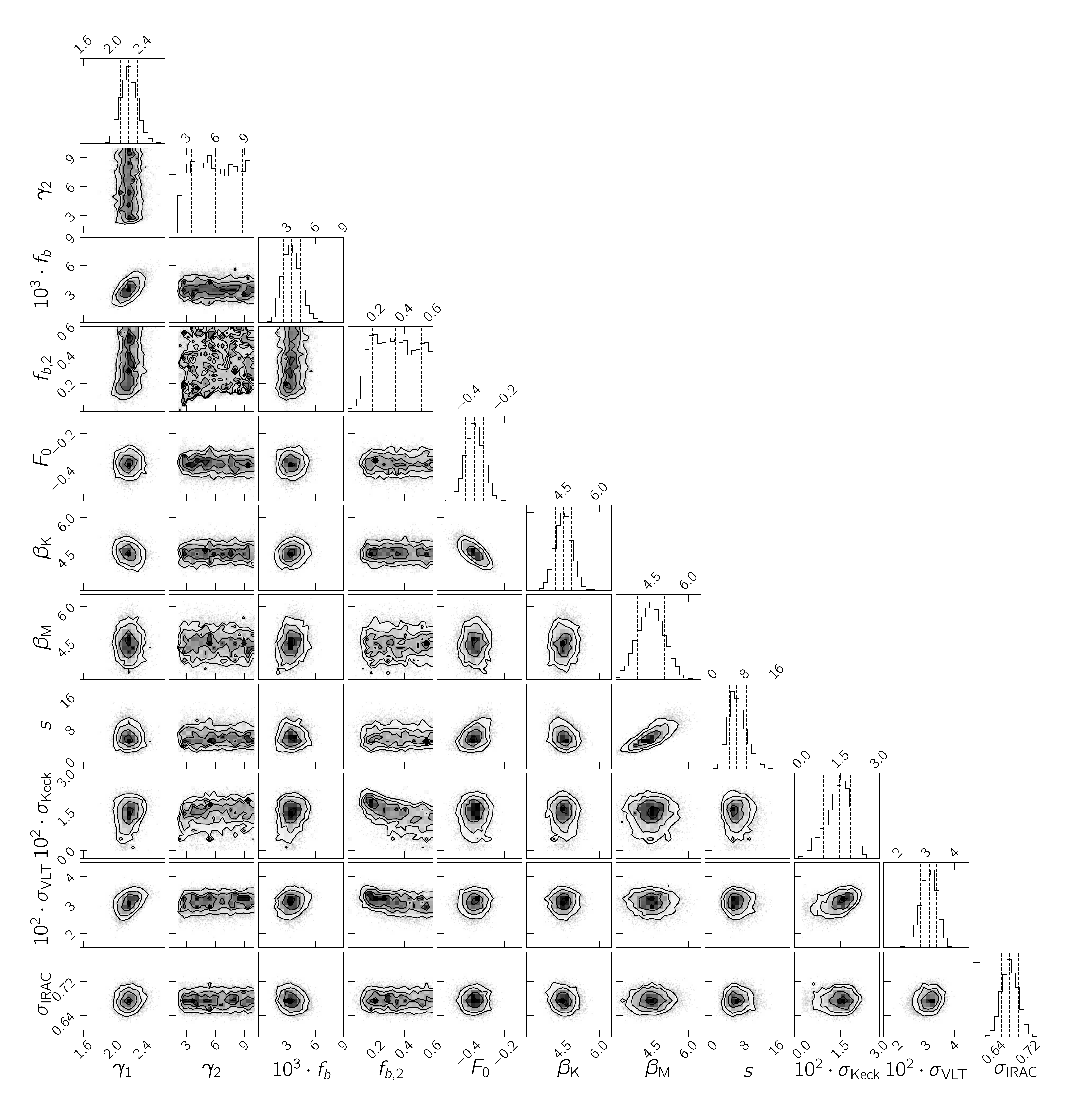}
\end{center}
\setlength{\abovecaptionskip}{-20pt}
\caption{Results of the Bayesian structure function fit for Case~1 (power-law/power-law; see \S\ref{meth}). Contours show the joint (posterior) probability density for each parameter pair, and panels along the upper right edge show histograms of the marginalized posterior of each parameter defined in Table~\ref{results}. \added{For each histogram, the dashed lines mark the 16\%, 50\%, and 84\% quantiles.} \replaced{The 95\%-confidence level upper limit for $1/f_{b,2}$ is 8.3~minutes}{The upper limit for $1/f_{b,2}$  with a probability of 95\% is 8.3~minutes.}}\label{corner_pl}
\end{figure*}

\begin{figure*}
\begin{center}
\includegraphics[scale=0.27, angle=0]{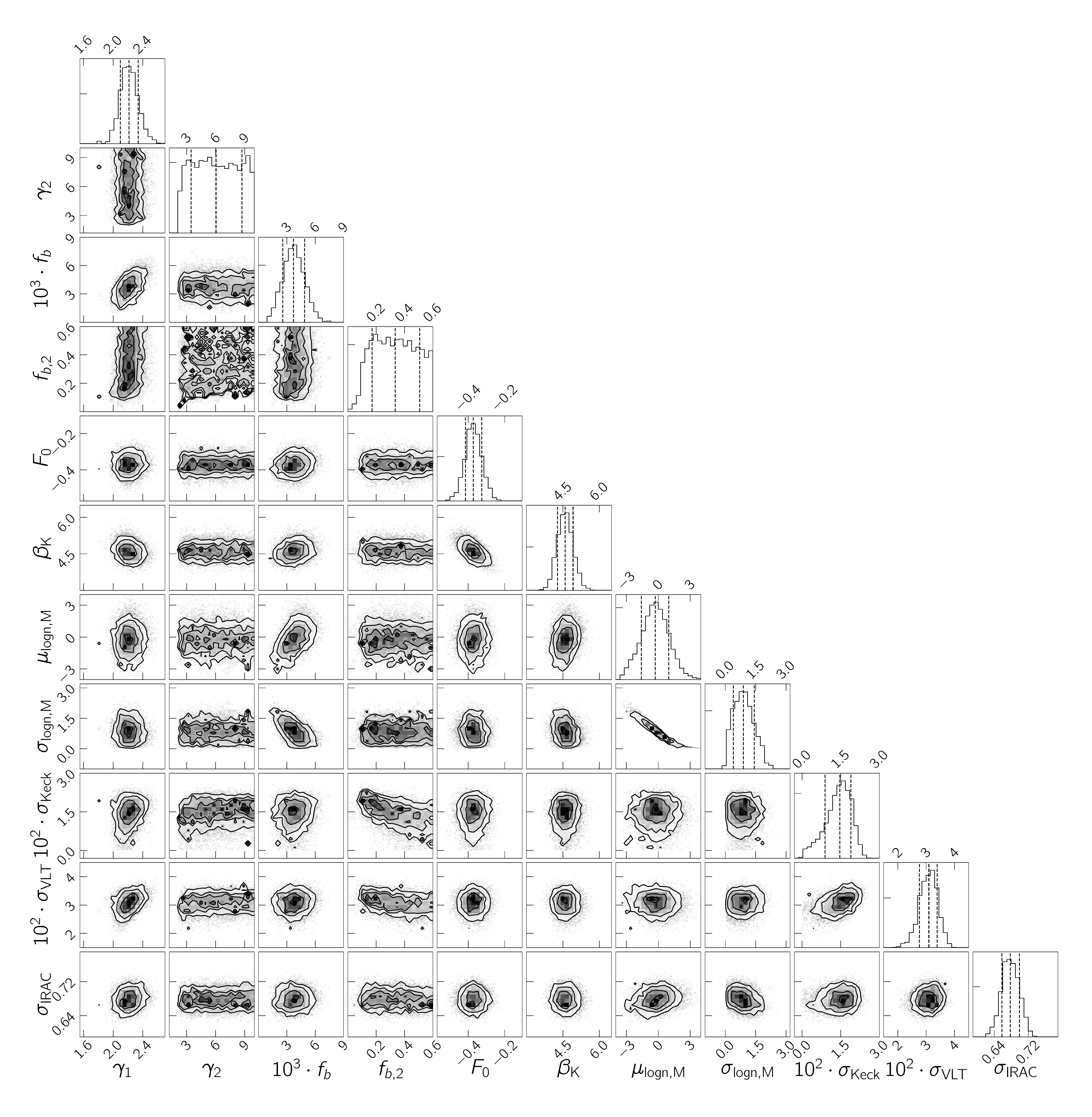}
\end{center}
\setlength{\abovecaptionskip}{-20pt}
\caption{Results of the Bayesian structure function fit for Case~2 (power-law/log normal: see \S\ref{meth}). Contours show the joint (posterior) probability density for each parameter pair, and panels along the upper right edge show histograms of the marginalized posterior of each parameter defined in Table~\ref{results}. \added{For each histogram, the dashed lines mark the 16\%, 50\%, and 84\% quantiles.}  \replaced{The 95\%-confidence level upper limit for $1/f_{b,2}$ is 8.6~min}{The upper limit for $1/f_{b,2}$  with a probability of 95\% is 8.6~minutes}, nearly the same as Case~1. The strong correlation between the $M$ log-normal parameters is expected when the mode of the intrinsic log-normal distribution below the white noise level.}\label{corner_logn}
\end{figure*}

\begin{figure*}
\begin{center}
\includegraphics[scale=0.27, angle=0]{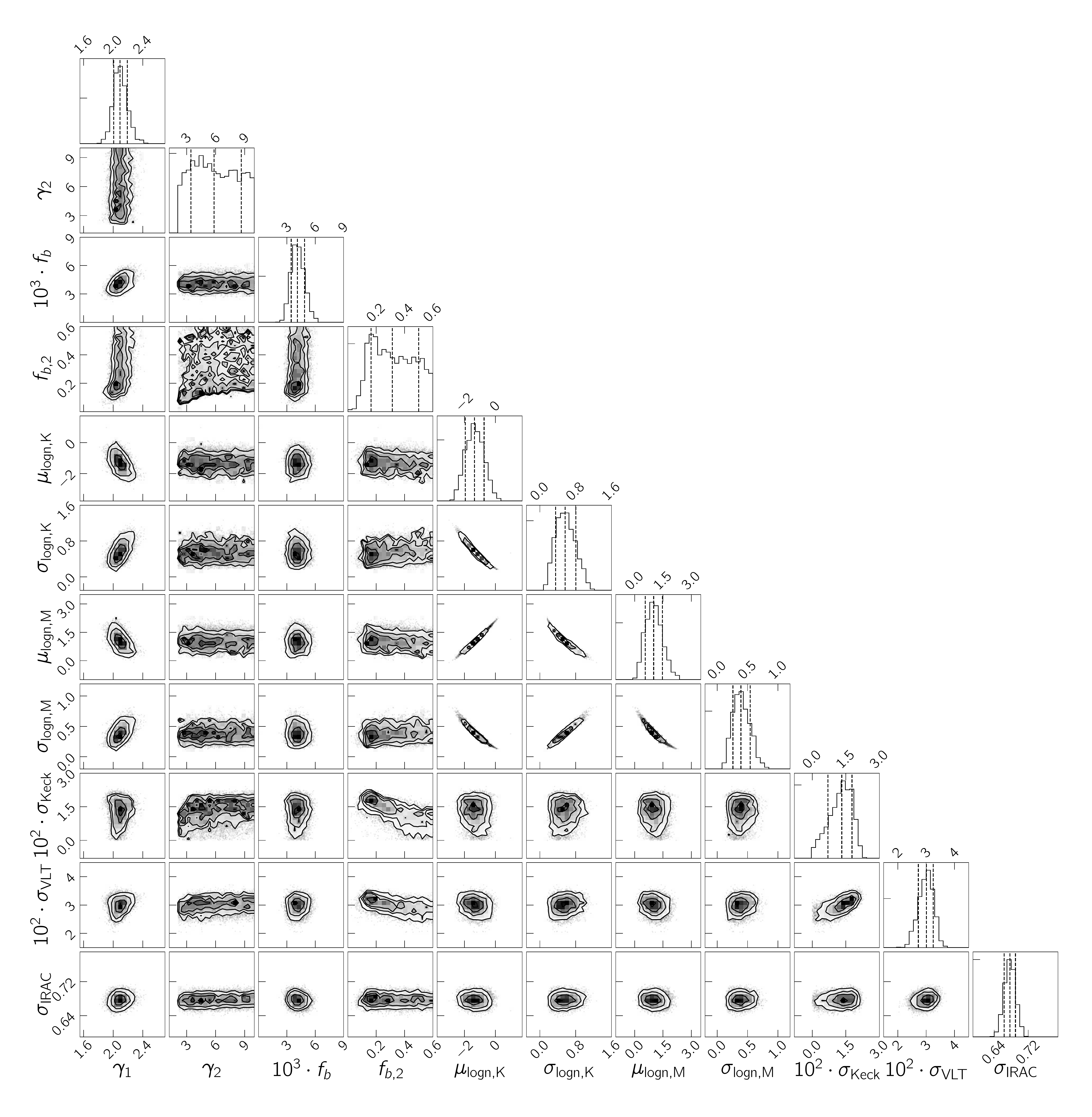}
\end{center}
\setlength{\abovecaptionskip}{-20pt}
\caption{Results of the Bayesian structure function fit for Case~3 (log-normal/log-normal; see \S\ref{meth}). Contours show the joint (posterior) probability density for each parameter pair, and panels along the upper right edge show histograms of the marginalized posterior of each parameter defined in Table~\ref{results}. \added{For each histogram, the dashed lines mark the 16\%, 50\%, and 84\% quantiles.}  \replaced{The 95\%-confidence level upper limit for $1/f_{b,2}$ is 8.3~min}{The upper limit for $1/f_{b,2}$  with a probability of 95\% is 8.5~minutes}, about the same as Cases~1 and~2.}\label{corner}
\end{figure*}

Developing and running the ABC algorithm required an extensive effort in optimization of code and adaptation of the distance function to the problem to achieve the results presented here. The large number of calculations involved in the massive iterative generation and evaluation of light curves---including both test and final analysis runs---required in total about 60,000 CPU hours on our UCLA Hoffman cluster node and 250,000 CPU hours on the XSEDE super clusters Stampede1, Comet, and Bridges (\citealt{6866038}). Each of the runs reported here took 2 days on 24 cores, and the last iteration with 10,000 parameter sets took about 1 day each on 800--1200 cores executing ${\sim} 2\times10^{10}$ FFTs. The results of our Bayesian analyses are shown in Figures~\ref{corner_pl}, \ref{corner_logn}, and~\ref{corner}, and the weighted averages and standard deviations are listed in Table~\ref{results}. %--\ref{results_double_logn}. 

For Case~1 (power-law/power-law), all parameters are well constrained with the exception of the secondary break frequency $f_{b,2}$ and slope $\gamma_2$. The secondary break frequency has a lower limit $f_{b,2} > 0.120~\rm{minutes}^{-1}$ or equivalently an upper limit for the secondary break timescale of 8.3~minutes at \replaced{95\% confidence}{the 95\% credible level}. The main break timescale  $\tau_{b} = 286^{+191}_{-94}$~minutes (90\% \replaced{confidence}{credible level}).

For Case~2 (power-law/log-normal), all parameters are similarly well constrained, again with the exception of the secondary break frequency  and slope. The limit is $f_{b,2} > 0.112~\rm{minutes}^{-1}$ or equivalently an upper limit for the secondary break timescale of 9.0 minutes (95\% \replaced{confidence}{credible level}). The main break timescale  
$\tau_{b} = 270^{+261}_{-92}$~minutes ($90\%$ \replaced{confidence}{credible level}).

{For Case~3 (log-normal/log-normal), again all parameters but the secondary break frequency $f_{b,2}$ and slope $\gamma_2$ are well constrained. The limit is $f_{b,2} > \replaced{0.114}{0.118}$~\rm{minutes}$^{-1}$ or equivalently an upper limit for the secondary break timescale of \replaced{8.8 minutes (95\% confidence)}{8.5~minutes (95\% credible level}). The main break timescale \replaced{$\tau_{b} = 245^{+88}_{-61}$~minutes (90\% confidence}{$\tau_{b} = 243^{+82}_{-57}$~minutes (90\% credible level}).}

% {For Case~3 (log-normal/log-normal), again all parameters but the secondary break frequency $f_{b,2}$ and slope $\gamma_2$ are well constrained. The limit is $f_{b,2} > 0.118~\rm{min}^{-1}$ or equivalently an upper limit for the secondary break timescale of 8.5 min (95\% \replaced{confidence}{credible level}). The main break timescale $\tau_{b} = 243^{+82}_{-57}$~min ($90\%$ \replaced{confidence}{credible level}).}

\deleted{The non-detection of secondary breaks is not an artifact of the method or the data.
We tested our algorithm on mock data sets {(with the same cadence and flux-density PDFs as the real data)} having $\tau_{b}=270$~minutes and $\gamma_1=2.25$.  For the secondary break timescale and slope, we explored two cases:  one with  $\tau_{b,2}=70$~minutes and $\gamma_2=4.5$ and another with $\tau_{b,2}=15$~minutes and $\gamma_2=5.5$.  For both cases we were able to recover the secondary break frequency and all other input parameters except the secondary slope. Inability to constrain $\gamma_2$
is a result of the transition from the red noise to the white noise regime that dominates the shape of the structure function at the lower timescales.  {In other words, while the secondary break frequency with a slope
$\gamma_2$ distinctly steeper than $\gamma_1$ changes the variance at higher timescales enough to be detected in the mock data structure function, the difference between different values for the secondary slope are dominated by the the white noise variance.  As a result a precise measurement of $\gamma_2$ is impossible.}}

\section{Discussion}\label{disc}

\subsection{Validation of the distance function}

\added{The posterior distributions derived from our analysis depend on the choice of distance function.  The ABC posterior will only approach the actual  distribution if the distance correctly encapsulates all information relevant to parameter estimation.  Without an analytic likelihood function, determining the validity of the distance function is difficult.  However, given a mock data set derived from a set of assumed parameters, we can determine whether our analysis and distance function recover the known parameters.  We tested our algorithm on mock data sets constructed with $\tau_{b}=270$~minutes and $\gamma_1=2.25$ and  with the same cadence and flux-density PDFs as the real data.  For the secondary break timescale and slope, we explored two cases:  one with  $\tau_{b,2}=70$~minutes and $\gamma_2=4.5$ and another with $\tau_{b,2}=15$~minutes and $\gamma_2=5.5$.  For both cases, we were able to recover the secondary break frequency and all other input parameters except  $\gamma_2$. Inability to constrain $\gamma_2$
is a result of the data being dominated by instrumental white noise at the shorter timescales.   In other words, while a secondary break to a slope
$\gamma_2$ distinctly steeper than $\gamma_1$ changes the variance at short timescales enough to be detected in the mock data structure function, the actual value for the secondary slope is dominated by the white noise variance.  As a result a precise measurement of $\gamma_2$ is impossible, but the data can reveal a break if one is present.}

\subsection{Quality of the statistical analysis}\label{quality}

\begin{figure}
\begin{center}
\includegraphics[scale=0.58, angle=0]{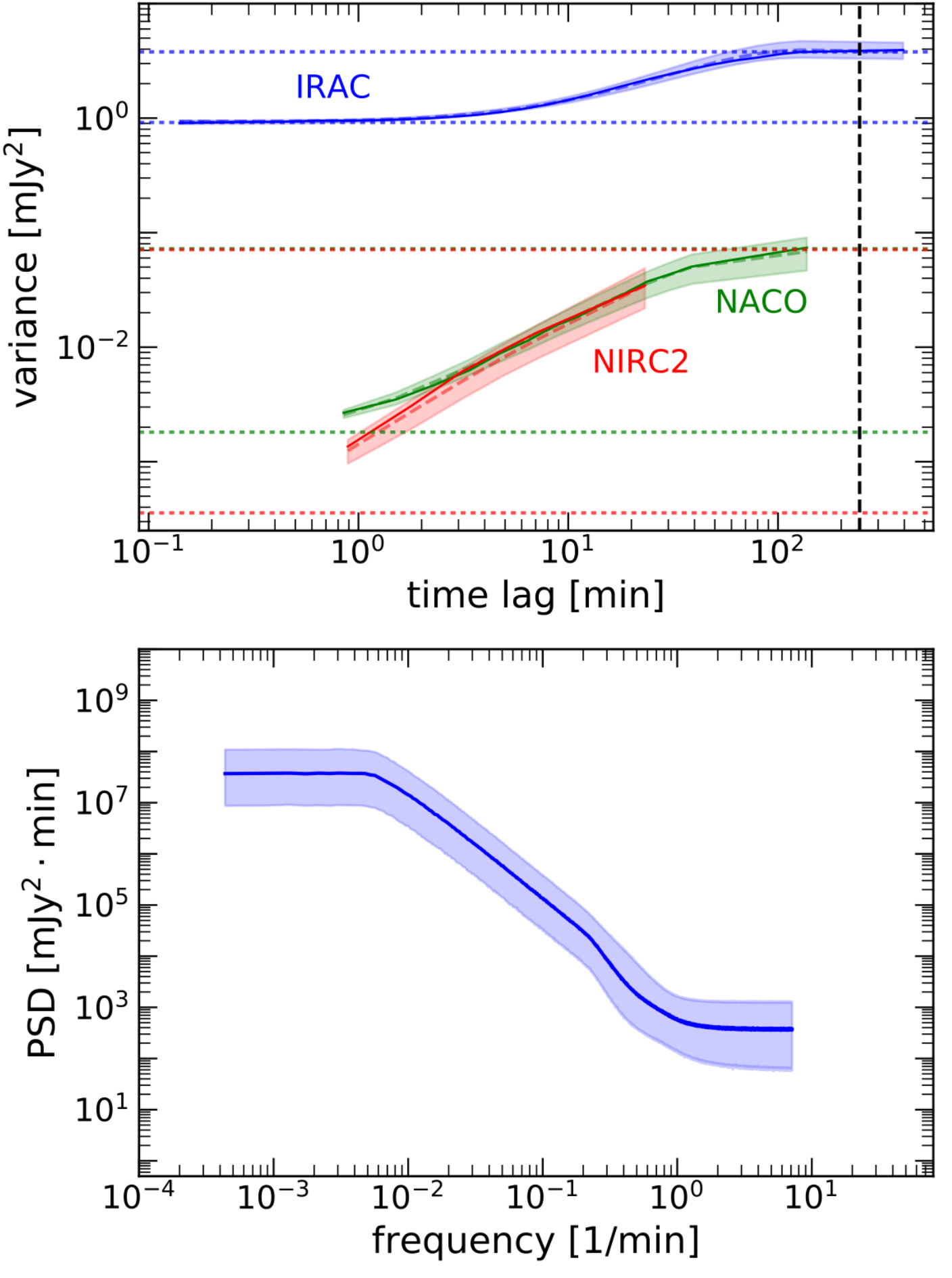}
\end{center}
\setlength{\abovecaptionskip}{-5pt}
\caption{Structure functions and power spectral density.  The upper panel shows
structure functions (Eq.~\ref{structurefunction}) for the three instruments.
Solid lines show the observed data (as presented in Figure~\ref{structplot}), and corresponding dashed curves show the median of 10,000 \replaced{Case~1}{Case~3} (see \S\ref{meth}) model structure functions for the respective instruments. The shaded
envelopes denote the model $68\%$ \replaced{Bayesian confidence}{credible} intervals for each time lag.
The vertical dashed line marks the derived correlation timescale \replaced{${\gg} \tau_{b}$}{$1/f_{b}$}. Pairs of horizontal short-dashed lines, color-coded for each instrument,
mark the two noise levels of each measurement. The lower line of each pair indicates the measurement noise (Eq.~\ref{lowwn}), and the upper line the intrinsic red noise of the \Sg\ variability when sampled at timescales ${\gg} \tau_{b}$ combined with measurement noise (Eq.~\ref{highwn}).  (The upper lines for NaCo and NIRC2 are nearly indistinguishable.)
The details of generating the structure functions, including the choice of time lag ranges, are described in \S\ref{meth} and~\ref{dist}. The slope of the structure function relates to the slope of the PSD but also depends on the underlying white noise level and is therefore different for each observatory despite the common PSD.   The lower panel shows power spectral densities
of 10,000 mock IRAC light curves derived from the final \replaced{Case~1}{Case~3} parameters. The mock light curves have the same cadence as the \deleted{first 23 hours of} IRAC data but the lower white noise of the NIRC2 data. The solid line shows the median for each frequency, and the shaded areas show the $68\%$ \replaced{confidence}{credible} intervals.  {Because the PSD is a function of frequency, short time lags are to the right. The units of the PSD are $\rm{mJy}^2 \cdot \rm{minutes}$, but the scaling of PSD values shown here is arbitrary.} The slight break in slope around 0.2~minutes$^{-1}$  is \replaced{not significant}{well within the 1$\sigma$ envelope}. It arises from the condition $\gamma_{2} > \gamma_{1}$ and the lack of sensitivity to structure below 9~minutes, close to the white noise level.}\label{synthPSD}
\end{figure}

\replaced{The posterior distributions (Table~\ref{results}) converge to excellent agreement between the derived models and observed data.}{The mock structure functions resulting from the derived posterior distributions (Table~\ref{results}) are in excellent agreement with the observed structure functions. Based on the final Case~3 iteration, we created 10,000 structure functions for each instrument, and these closely resemble the measured structure functions as shown in the upper panel of Figure~\ref{synthPSD}. This figure additionally shows the short- and long-timescale white noise levels of the processes (see Equations~ \ref{lowwn} and \ref{highwn}). The latter were directly derived from the Case~3 log-normal parameters. The measured structure functions asymptotically approach the calculated levels.} 

\begin{figure}
\begin{center}
\includegraphics[scale=0.70, angle=0]{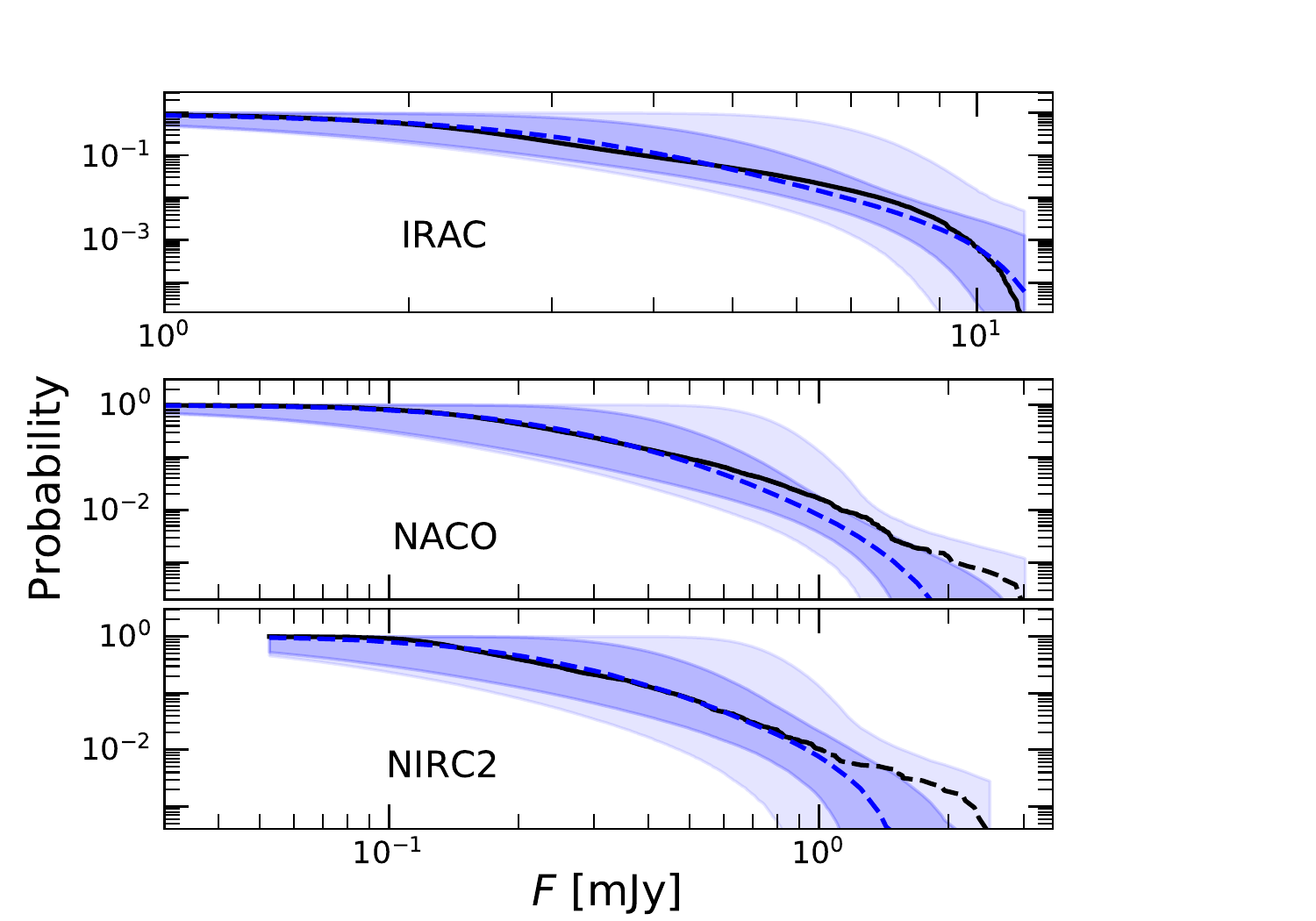}
\end{center}
\setlength{\abovecaptionskip}{-7pt}
\caption{\added{Cumulative distribution functions of Sgr~A* 4.5~\micron, 2.18~\micron, and 2.12~\micron\ flux densities (top to bottom). The black lines show the CDFs observed by the respective instruments. For the VLT and Keck, the dashed sections of the black lines indicate flux densities that stem from the single brightest flux density excursion observed with that instrument (discussed in \S~\ref{alpha_disc}). The dashed blue lines show the median
CDFs from the Case~3 model, and shaded areas show $68\%$ and $95\%$ credible intervals derived from 10,000 light curves drawn from the  Case~3  parameters (\S\ref{meth} and Table~\ref{results}).}}\label{sgradistrcdf}
\end{figure}

\added{Figure~\ref{sgradistrcdf} shows the excellent agreement of the cumulative distribution functions (CDFs) of the $M$- and $K$-band data with the Case~3 posteriors. Light curves derived for Cases~1 and~2 show agreement between mock and observed data similar to Case~3. However, for the power-law parametrization in these cases, we could not use wide, flat priors because the resulting parameters for the $K$-band CDF did not describe the observed distribution. The reason seems to be 
%the lower representation of the $K$-band data in the distance function in combination with the fact 
that a power law is simply not the correct model for the lowest flux densities. In order to force the proper description of the $K$-band CDF, we used informative priors based on earlier analysis  \citep{2012ApJS..203...18W}. In Case~3 this  reliance on informative priors is not needed.
% We will come back to this point in Section~\ref{alpha_disc}.
} 

\deleted{The Case~1 power-law 
parameters for the $K$-band flux-density PDF agree with those derived by \cite{2012ApJS..203...18W} and \cite{2014ApJ...793..120H}, who used different
analysis methods than ours. The derived measurement noise for each observatory agrees with the independently estimated values mentioned in Section~\ref{meth}. The power-law index in $M$-band is identical within uncertainties to the one in $K$-band, and Figure~\ref{sgradistrcdf} shows the excellent agreement of the cumulative distribution function (CDF) of the $M$-band data with a CDF drawn from uncorrelated power-law data plus white noise with the determined median values of $\beta_{{M}}$,  $s \cdot F_{0}$, and $\sigma_{\rm IRAC}$. Additionally, the final Case~1 iteration created 10,000 structure functions for each observatory, and these closely resemble the measured structure functions as shown in the upper panel of Figure~\ref{synthPSD}. This figure additionally shows the short- and long-timescale white noise levels of the processes (see Equations~ \ref{lowwn} and \ref{highwn}). The latter were directly derived from the variance of the power laws, i.e., from the power-law parameters. The measured structure functions asymptotically approach the calculated levels.  There is a similar agreement between observed data and the light curves derived for Cases~2 and~3. The derived measurement noise for each observatory agrees with the independently estimated values mentioned in Section~\ref{meth}.}

\subsection{Power spectral density of NIR variability}

Based on our combined modeling of the PSD  and the flux-density PDFs (and in Case~3 the additional constraints from $K$- to $M$-band spectral properties), we can derive a well-constrained estimate of the PSD of the Sgr~A* NIR variability. The lower panel of Figure~\ref{synthPSD} shows 
%a simple Fourier transform power spectrum of the IRAC light curves together with 
a PSD synthesized from the final Case~3 parameters. This synthesized PSD shows a well-constrained shape over three orders of magnitude in frequency.
The IRAC data fully cover the coherence timescale of the variability process (as expected), and there is no significant evidence for a second break timescale below 20~minutes. However, FFT periodograms on real data with white noise and irregular sampling are not statistically consistent estimators and not well suited for precision measurements of the PSD parameters, motivating our use of the ABC sampler.
%The PSD is precisely measured over three orders of magnitude and featureless.
The coherence timescale for Case~1 is $\tau_{b} \equiv 1/f_{(b)} = 286^{+191}_{-94}$~minutes at the $90\%$ \replaced{confidence}{credible} level. Case~2  gives much the same timescale $\tau_{b} = 270^{+261}_{-92}$~minutes but with a larger uncertainty because of the uncertainty in the log-normal parameters. Case~3 shows a slightly different (but consistent within 1$\sigma$) and more precise $\tau_{b} = 243^{+82}_{-57}$~minutes. The validity of the smaller error bars is dependent on whether or not one considers $\Re({M}/{K})$ derived from the synchronous data as representative of the true ratio at that flux density. All three cases give the most precise determination of the PSD parameters so far, and all are consistent with the earlier estimate $\tau_{b} = 128^{+329}_{-77}$~minutes (\citealt{2009ApJ...694L..87M}).  Figure~\ref{conf} compares the \replaced{confidence}{credible} contours of the respective analyses.

 {Break timescales of several hours are consistent with viscous timescales rather than with dynamical timescales (e.g., orbital modulations due to inhomogeneities in the accretion flow; \citealt{2014MNRAS.442.2797D}).  
 \citeauthor{2014MNRAS.442.2797D} analyzed the characteristic timescale of \Sg\ from 230, 345, and 690 GHz submm data and found $\tau_{b,\rm{submm}} = 480^{+180}_{-240}$~minutes at the 95\% \replaced{confidence}{credible} level. 
The authors pointed out that the timescale of $\sim$8~hr in the submm is more than $3\sigma$ larger than the \added{former} NIR timescale of $\sim$2.5~hr \citep{2009ApJ...694L..87M}. 
\citealt{2014MNRAS.442.2797D} discussed the possibility of the NIR emission originating from the same process as the submm but at smaller radii. 
The dependence of the viscous timescale on the radius is $t_{\rm{visc}} \propto R^{3/2}$. 
Therefore the timescales above suggest the NIR radius to be \replaced{1/6}{$\sim$0.5} of the submm radius. 
For a canonical size $R_{\rm{submm}}=3R_S$ of the submm emission region (with $R_S$ the Schwarzschild radius), this puts the \added{entire} NIR emitting process \replaced{right at}{very close to} the ISCO (which is unlikely). The authors concluded that a difference in radius is likely not the reason for the different timescales and suggested that adiabatically expanding plasma with delayed submm emission at larger sizes could be a natural explanation of the timescales. 

Our findings change \replaced{this picture}{the interpretation of the relative timescales}. 
%On one hand 
$\tau_{b,\rm{NIR}}= 243^{+82}_{-57}$~minutes is statistically consistent with the submm values. 
%On the other hand, the NIR radius derived from our value for $\tau_{b,\rm{NIR}}$ is $\sim$\replaced{1/3}{0.6} of the submm radius. Both facts 
This suggests a more direct relation between the NIR and submm emission  (e.g., both wavelengths stemming from the same optically thin synchrotron source). A detailed analysis of a larger submm dataset with similar statistical tools as used here and further simultaneous observations are needed to refine this relation\deleted{ in detail}.

\begin{figure}
\begin{center}
\includegraphics[scale=0.58, angle=0]{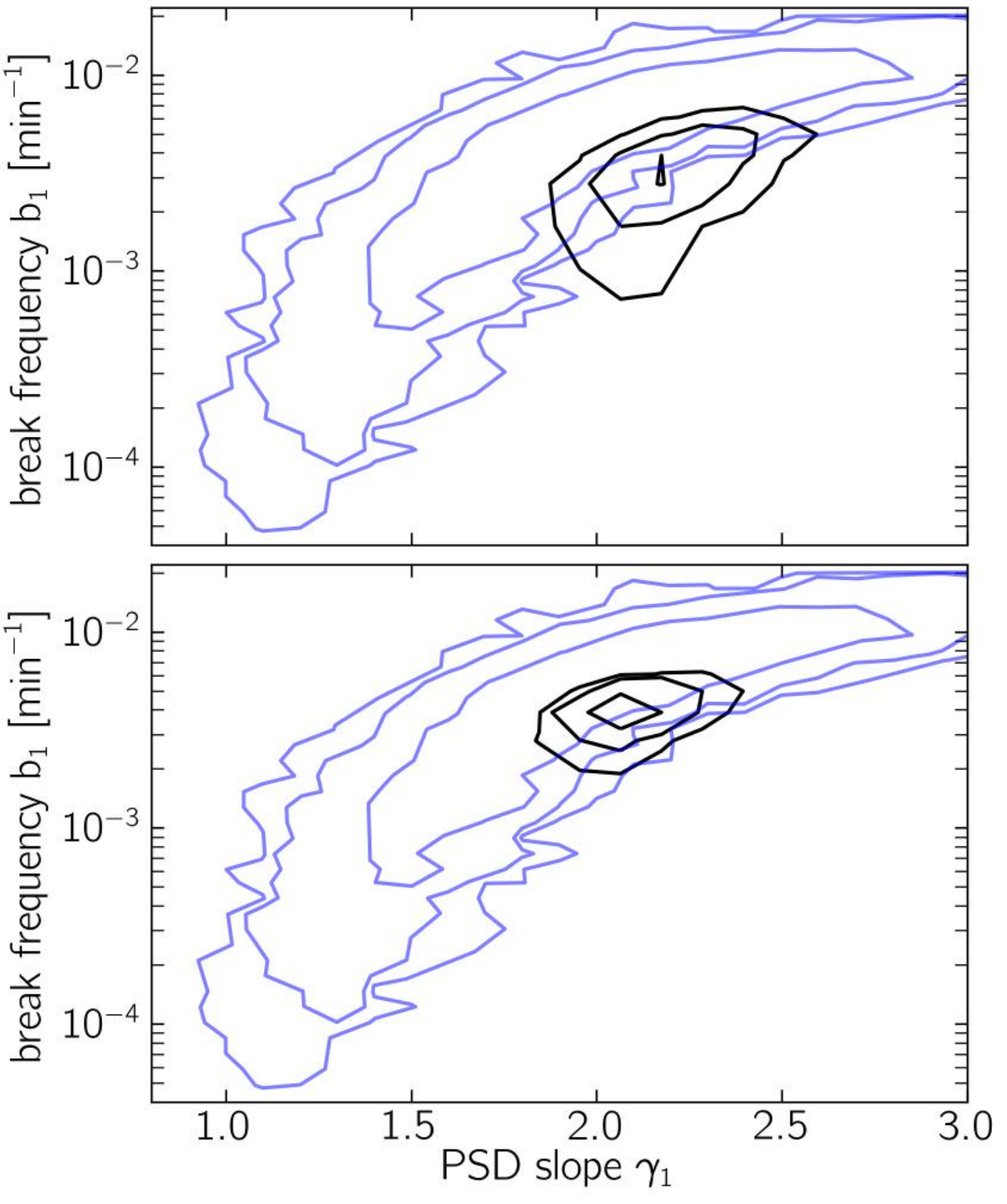}
\end{center}
\setlength{\abovecaptionskip}{-5pt}
\caption{\replaced{Bayesian confidence}{Credible} contours ($68\%,95\%,99\%$) for the parameters $\gamma_{1}$ and $f_{b}$ (Table~\ref{results}). The upper panel shows Case~1 and the lower panel shows Case~3. The blue contours are from Figure~3 of \cite{2009ApJ...694L..87M}, and the black contours show results of the present analysis.
The posteriors have been
marginalized over all other parameters.}\label{conf}
\end{figure}

Despite the ability of the ABC algorithm to detect secondary timescales in mock data, there is little indication of a second break in the real data, regardless of the choice of parametrization. Indeed, a second break can be restricted to timescales $<$9~minutes. Only Case~3 has even a small peak in the posterior  \replaced{around 6~min, but it is not statistically significant.}{with $1/f_{b,2} \approx
6$~minutes. (See the $f_{b,2}$ histogram in Figure~\ref{corner}.)  Shorter break times are consistent with the data, and the secondary break slope $\gamma_2$ is unconstrained. The existing data therefore do not require a second break at all.}

Several models predict modulation of the NIR light at frequencies related to motion at the innermost stable circular orbit (ISCO) of the black hole, either as a QPO \citep{2006A&A...460...15M,2010A&A...510A...3Z,2012ApJ...746L..10D} or a loss of PSD power below the ISCO timescale. Either would create a second break (\citealt{2012ApJ...746L..10D}). 
%The ISCO timescale for a Kerr black hole is dependent on the spin:
%
%\begin{equation}
%\frac{P}{\rm{min}} =0.328 \cdot \frac{2\pi}{\omega_{\rm{p}}}\quad.
%\end{equation}
%
%$\omega_{\rm{p}}$ is the prograde angular velocity:
%
%\begin{equation}
%\omega_{\rm{p}} = \frac{1}{r_{\rm{p}}}\quad,
%\end{equation}
%
%with $r_{\rm{p}}$ the prograde ISCO radius:
%
%\begin{eqnarray}
%r_{\rm{p}} & = & 3 + p - \sqrt{(3-q)(3+q+2p)}\\
%q & = & \Big \{ 1 + (1-a^2)^{\frac{1}{3}} \cdot \big[(1+a)^{\frac{1}{3}} + (1-a)^{\frac{1}{3}} \big] \Big \} \\
%p & = & \sqrt{3a^2 + q^2}\quad,
%\end{eqnarray}
%with $a$ the black hole spin.
%
%In Figure~\ref{spin} we show this dependence.
If these or other processes near the ISCO modulate the light curve of \Sg,
the absence of a secondary break in the PSD implies a lower limit on the black hole spin.
The orbital period for a direct-rotation, equatorial orbit at the ISCO is
\begin{equation}\label{iscop}
P = 2\pi (x_{\rm{ISCO}}^{3/2} + a) \frac{G M_{bh}}{c^3}
\end{equation}
where $0 \le a < 1$ is the dimensionless black hole spin, and
$x_{\rm{ISCO}}$, the radius of the ISCO in units of $G M_{bh}/c^2$, is
given by
\begin{equation}
x_{\rm ISCO} = 3 + Z_2 - [(3 - Z_1)(3 + Z_1 + 2 Z_2)]^{1/2}.
\end{equation}
Here $Z_1 \equiv 1 + (1 - a^2)^{1/3} [ (1 + a)^{1/3} + (1 - a)^{1/3} ]$
and $Z_2 \equiv (3 a^2 + Z_1^2)^{1/2}$ (\citealt{1972ApJ...178..347B}).
Figure~\ref{spin} shows $P(a)$ for $M_{bh} = 4\times 10^6\ \rm{M}_{\odot}$.
Only ISCO modulation periods shorter than the 9~minute upper limit and therefore black hole spins $a>0.9$ are consistent with the light curve data, unless there are no NIR flux variations at the frequency of the ISCO. The hint of a posterior peak for case~3 at about 6~minutes would, if taken seriously, point to maximum spin {\it if} the power is generated at the ISCO. The models as presented by, for example, \cite{2006A&A...460...15M,2007A&A...473..707M} and \cite{2010A&A...510A...3Z} can be ruled out because they predict NIR variability with typical timescales of 15--20~minutes.

\begin{figure}
\begin{center}
\includegraphics[scale=0.58, angle=0]{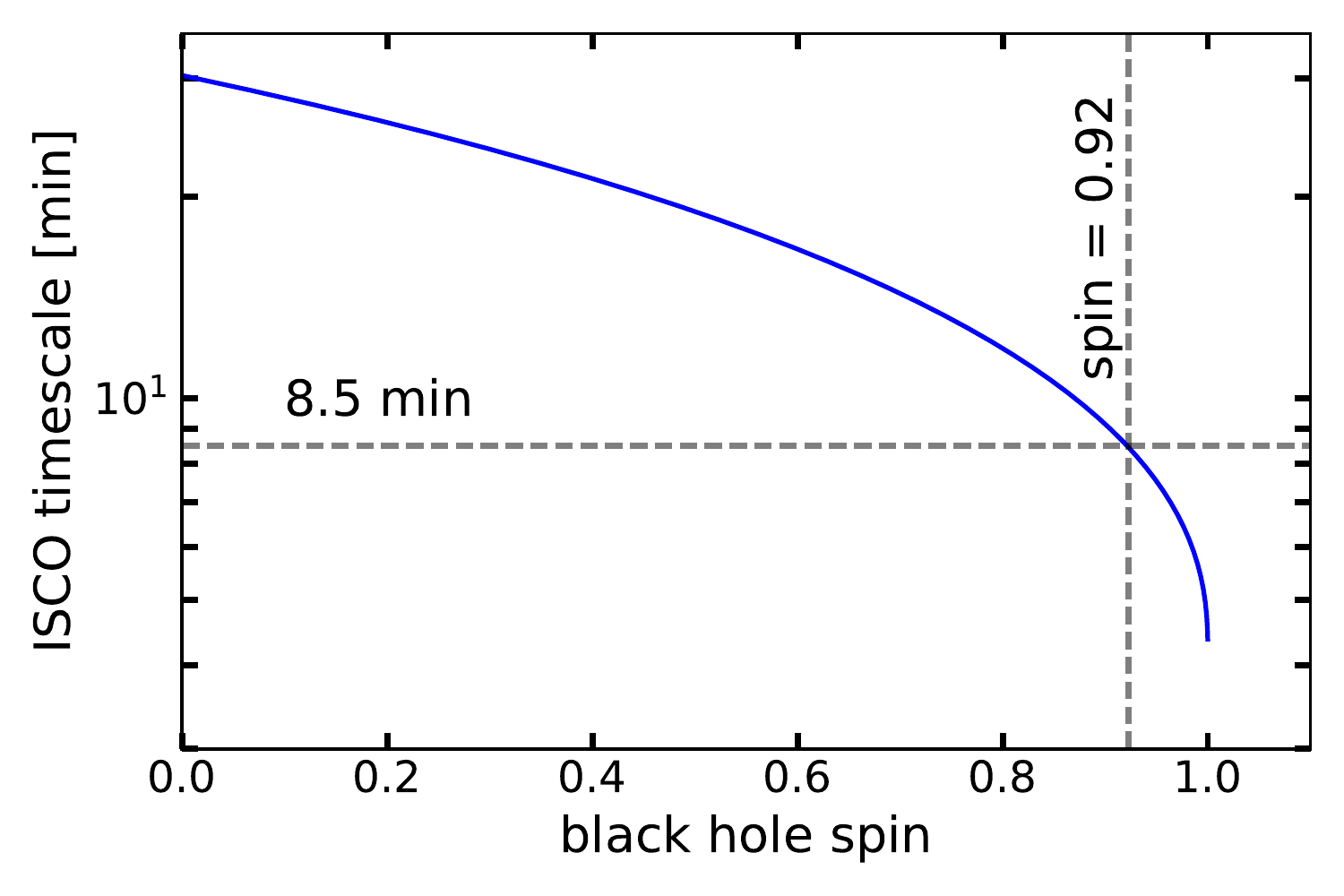}
\end{center}
\setlength{\abovecaptionskip}{-5pt}
\caption{ISCO orbital period (Eq.~\ref{iscop}) as a function of black hole spin in the Kerr metric for a black hole mass $4\cdot 10^6$~${M}_{\odot}$. The horizontal dashed line indicates a period of  \replaced{8.8}{8.5}~minutes, the upper limit on a secondary break timescale.  The vertical dashed line shows the corresponding dimensionless spin $a$.}\label{spin}
\end{figure}

\subsection{{Sgr~A*}'s NIR spectral index}\label{alpha_disc}

The $K$- to $M$-band ratio derived from the Case~1 (power-law/power-law) ABC fit $s = 5.9^{+2.5}_{-1.9}$ ($1\sigma$) is in excellent agreement with the value $s = 5.8^{+4.8}_{−2.9}$ calculated from the published NIR source spectral index $\alpha_s = 0.6 \pm 0.2$.  That index was derived from synchronous 1.6~$\mu$m to 3.7~$\mu$m measurements (\citealt{2007ApJ...667..900H,2014IAUS..303..274W}).  
However, $s\approx6$ is in striking disagreement with $\Re({M}/{K}) \approx 12$ derived from the simultaneous $K$ and $M$ data during its particularly dim flux-density level with a median of \replaced{$F(K)=0.07$}{$F(K)=0.15$}~mJy (Section~\ref{simobs}). 

In order to test how $s\approx6$ is related to our choice of prior, we attempted to alter the prior such that a higher value of $s$ was preferred. In all tests with Gaussian priors centered around $s > 6.0$, the ABC sampler consistently found a posterior about $1\sigma$ below the mean value of the prior to approach $s = 6.0$. Altering the prior for $s$ to exclude $s = 6.0$ and prefer higher values led to significantly different power-law indices $\beta$ for the flux-density PDFs in the two bands (and thus to a flux-density-dependent spectral index; see \ref{ratioform}). In the case of flat priors wide enough to encompass $s = 6.0$, the ABC code always reverted to a posterior $s \approx 6.0$ (Figure~\ref{corner_pl}). 
This  behavior shows that, integrated over the entire datasets
%at flux densities levels significantly higher than the noise levels (where the ABC sampler is most sensitive for measuring $\Re({M}/{K})$) 
and in the absence of simultaneous data, $s = 6.0$ describes the data well enough to match the total variance in both bands (i.e., the levels and shapes of the structure functions at longer time lags). \added{This result, however, requires use of informative priors for the power-law PDF parameters. Flat priors produced flatter, but still equal, power-law slopes for the $K$- and $M$-band PDFs but gave a poor fit to the $K$-band PDF. The ratio $s$  preferred higher values but remained only loosely constrained.}

The tension\added{ in Case~1 with informative priors} between the statistically derived ratio $s$ and the \deleted{real time} observed \added{(Figure~\ref{simdata})} ratio suggests a variable spectral index, in particular a \deleted{possible} trend of $\alpha_s$ with flux-density level. All three parametrizations allow the NIR spectral index to be a function of flux-density level. Based on the fact that the light curves at different wavelengths within the NIR are almost identical in shape (ignoring the minor short-timescale fluctuations discussed in Section~\ref{simobs} and \citealt{2014IAUS..303..274W}), the basic assumption is that if one NIR band rises or falls, the other rises or falls too. As a consequence, the quantiles of the flux-density PDFs must be equal for corresponding flux densities, and it is possible to derive the flux-density ratio between two bands as a function of flux density in one of the bands and the PDF parameters. These dependencies are calculated in \ref{ratioform} for our three different combinations of power-law and log-normal PDFs. In Case~1 our posterior distributions for $F_0$, $\beta_{K}$, $\beta_{M}$, and $s$ result in an almost perfectly constant $\Re({M}/{K})$ independent of $F_K$. This is expected because the posteriors of the power-law slopes $\beta_{K}$ and $\beta_{M}$ are almost identical, and the PDFs in both bands are the same except for a factor $\Re({M}/{K}) \approx s$. 

In the context of matching quantiles, larger values for $\Re({M}/{K})$ at low flux-density levels imply different distributions for $K$ and $M$-band flux densities, in particular a flattening of the $M$-band flux-density PDF toward low flux densities relative to the $K$-band PDF. The IRAC dataset is competitive with the S/N of the ground-based telescopes (\citealt{2014ApJ...793..120H}). The measured large value for $\Re({M}/{K})$ is an indicator that, in contrast to $K$-band, in $M$-band we start to discern the intrinsic turnover at the mode of the flux-density PDF despite measurement noise. \cite{2011ApJ...728...37D} originally suggested a log-normal flux-density PDF parametrization for \Sg. Parameterizing the $M$-band PDF as a log-normal while keeping the power-law parametrization for $K$-band (as a well-constrained reference) is one way to test for the presence of an intrinsic turnover in the $M$-band PDF. Case~2 analyzes this possibility.

Figure~\ref{three_plot} illustrates how the different $K$ and $M$ PDFs lead to a variable flux-density ratio that naturally reaches $\Re({M}/{K}) = 12.4$ at the average offset-corrected flux density $F_{\rm{avg}} = 0.15$~mJy measured for the 2016 data. Unfortunately, because the log-normal parameters cannot be well constrained from non-synchronous data only, the marginalized distribution of the flux density ratios is much wider in Case~2 than in Case~1 (Figure~\ref{three_plot}). At low flux densities, the 1$\sigma$ and 2$\sigma$ contours cover a huge range of possible flux-density ratios. However, the distributions peak at about the same ratio, and the flux-density ratio at high flux densities is about the same in both cases. This suggests that the power-law/log-normal parametrization of Case~2 can naturally explain both the redder spectral indices observed for low phases of \Sg\ (\citealt{2005ApJ...628..246E,2006ApJ...640L.163G,2006ApJ...642L.145K,2011AA...532A..26B,2017MNRAS.468.2447P}) and the bluer spectral indices during brighter phases (\citealt{2005ApJ...635.1087G,2007ApJ...667..900H,2011AA...532A..26B,2014IAUS..303..274W}).

This discussion takes $\Re({M}/{K}) \approx 12$ at face value despite evidence for short-timescale fluctuations. However, this value is integrated over $\sim$3~hr during which the source fluctuated around the low level of $F_K\sim0.15$~mJy with a maximal variation amplitude of $\Delta F_K \approx 0.1$~mJy. In the following, we assume that this ratio is representative for $F_K \approx 0.15$~mJy. 

In Case~3 we assumed a log-normal parametrization for both bands. (It would be surprising for the $K$-band PDF to have a fundamentally different form than the $M$-band PDF.) This case exploits the additional information from the synchronous data in our statistical analysis of the non-synchronous datasets. This is achieved by a modification of the distance function, as given by Equation~\ref{dist_log_n}. This approach has immense constraining power and allows us to derive tight posteriors for the log-normal parameters of both bands. 
Equation~\ref{log_n_ratio} gives $\Re({M}/{K})$ as a function of $F(K)$ as derived from the posteriors.
Figure~\ref{three_plot_2} shows the drastic improvement of the 1$\sigma$ and 2$\sigma$ envelopes. Interestingly, the flux density distributions derived from the posteriors predict $F(K)\le 0.15$~mJy to occur with a probability of only $\sim$23\% (this flux density is located left of the peak of most distributions \added{in the particle system}), and the flux-density-ratio histogram in the range directly observed is peaked around \replaced{$\Re({M}/{K}) \approx 8.0$}{$\Re({M}/{K}) \approx 9.0$}, close to the value derived in Case~1.

\begin{figure*}
\begin{center}
\includegraphics[scale=0.55, angle=0]{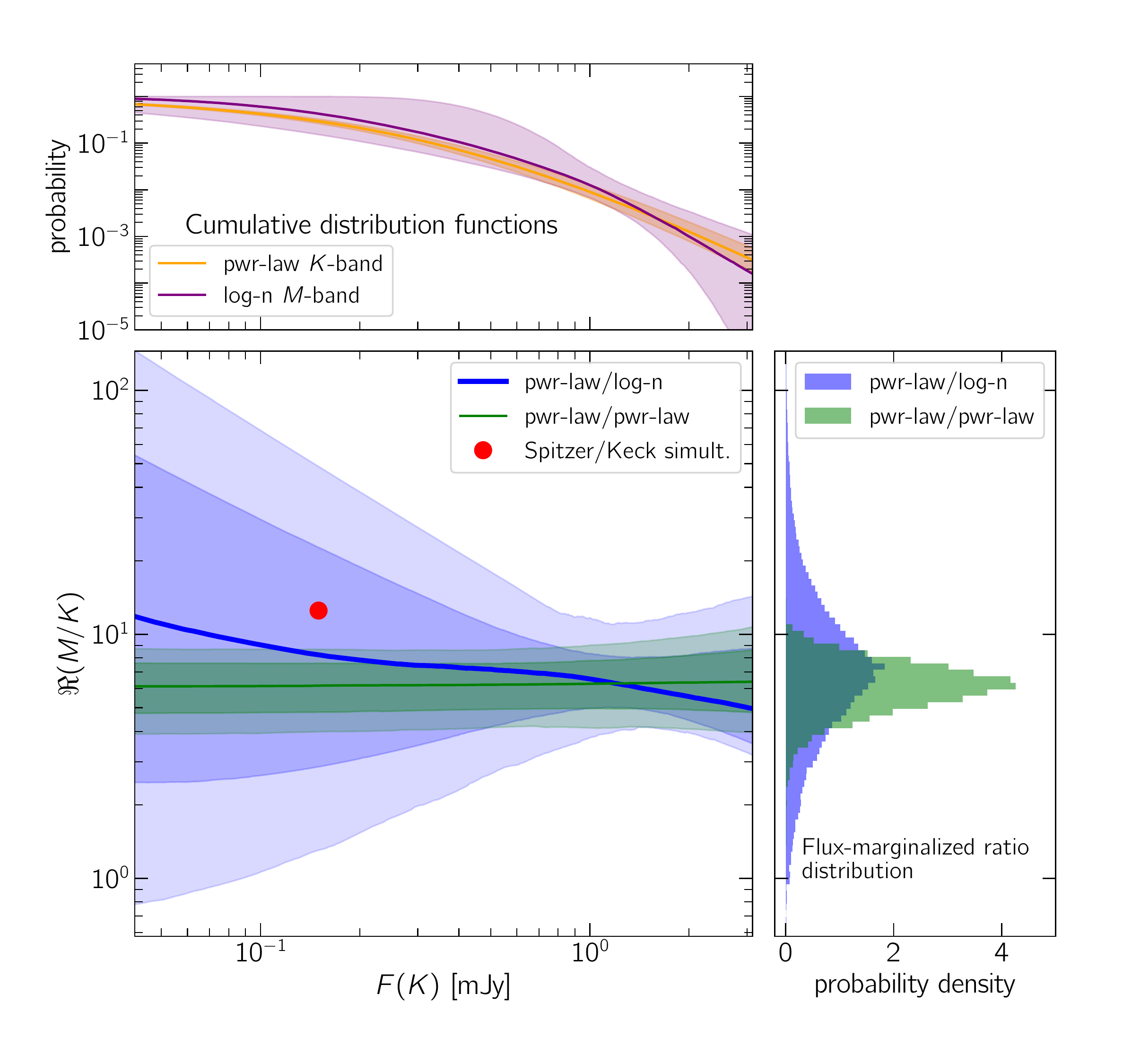}
\end{center}
\setlength{\abovecaptionskip}{-20pt}
\caption{Flux-density ratio $\Re(M/K)$ as a function of $K$-band flux density, as derived from our posteriors for Cases~1 and~2. The central panel shows the median and 68\% and 95\% \replaced{Bayesian confidence}{credible} contours for Case~1 in green and for Case~2 in blue. The red point denotes the flux-density ratio derived from the simultaneous observations (\S\ref{simobs}). The upper panel shows the CDFs for 
Case~2: $M$-band in purple and  $K$-band in orange. Shading indicates the limits at the 68\% \replaced{confidence}{credible} level. {In order to make the CDFs comparable, the abscissa of the $M$-band CDF (i.e., the $M$-band flux densities) has been scaled by a factor 1/5.9 (with 5.9 being the average $\Re(M/K)$ in Case~1) to place them on the same scale as the $K$-band flux densities}.  The right panel shows histograms of \replaced{$M$-band flux density}{$\Re(M/K)$} marginalized over the actually observed flux-density range.  Case~1 is in green and Case~2 in blue.}\label{three_plot}
\end{figure*}

\begin{figure*}
\begin{center}
\includegraphics[scale=0.55, angle=0]{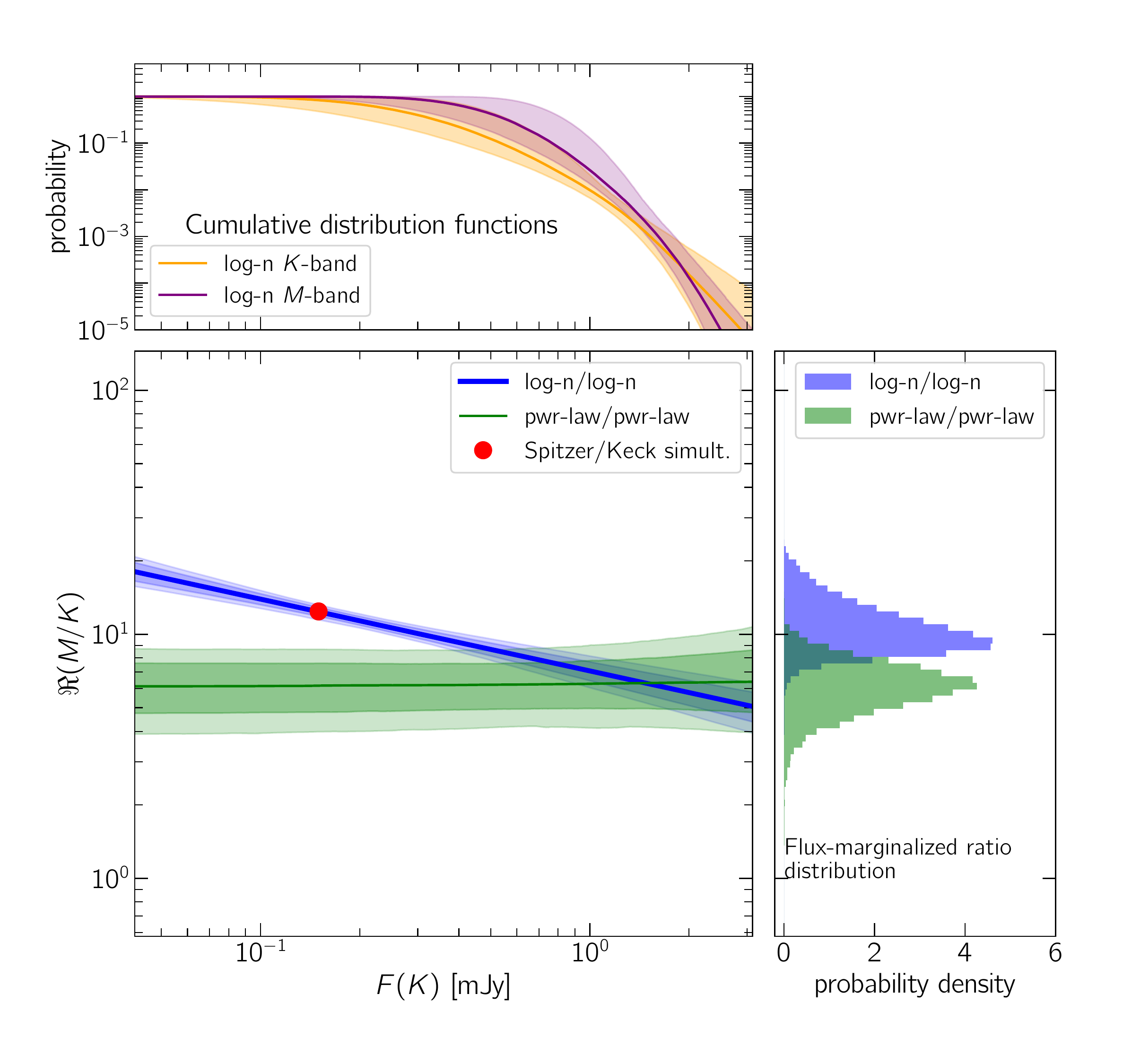}
\end{center}
\setlength{\abovecaptionskip}{-20pt}
\caption{Flux-density ratio $\Re(M/K)$ as a function of $K$-band flux density for Case~3. The central panel shows the median and 68\% and 95\% \replaced{confidence}{credible} contours for Case~1 in green and for Case~3 in blue. The red point denotes the flux-density ratio derived from the simultaneous observations (\S\ref{simobs}). The upper panel shows the CDFs for 
Case~3: $M$-band CDF in purple and  $K$-band CDF in orange. Shading indicates  the 68\% \replaced{confidence}{credible} intervals. {In order to make the CDFs comparable, the abscissa of the $M$-band CDF (i.e., the $M$-band flux densities) has been scaled by a factor 1/5.9 (with 5.9 being the average $\Re(M/K)$ in Case~1) to place them on the same scale as the $K$-band flux densities}. The right panel shows histograms of \replaced{$M$-band flux density}{$\Re(M/K)$} marginalized over the actually observed flux-density range.  Case~1 is in green and Case~3 in blue.}\label{three_plot_2}
\end{figure*}

\begin{figure}[h!]
\begin{center}
\includegraphics[scale=0.40, angle=0]{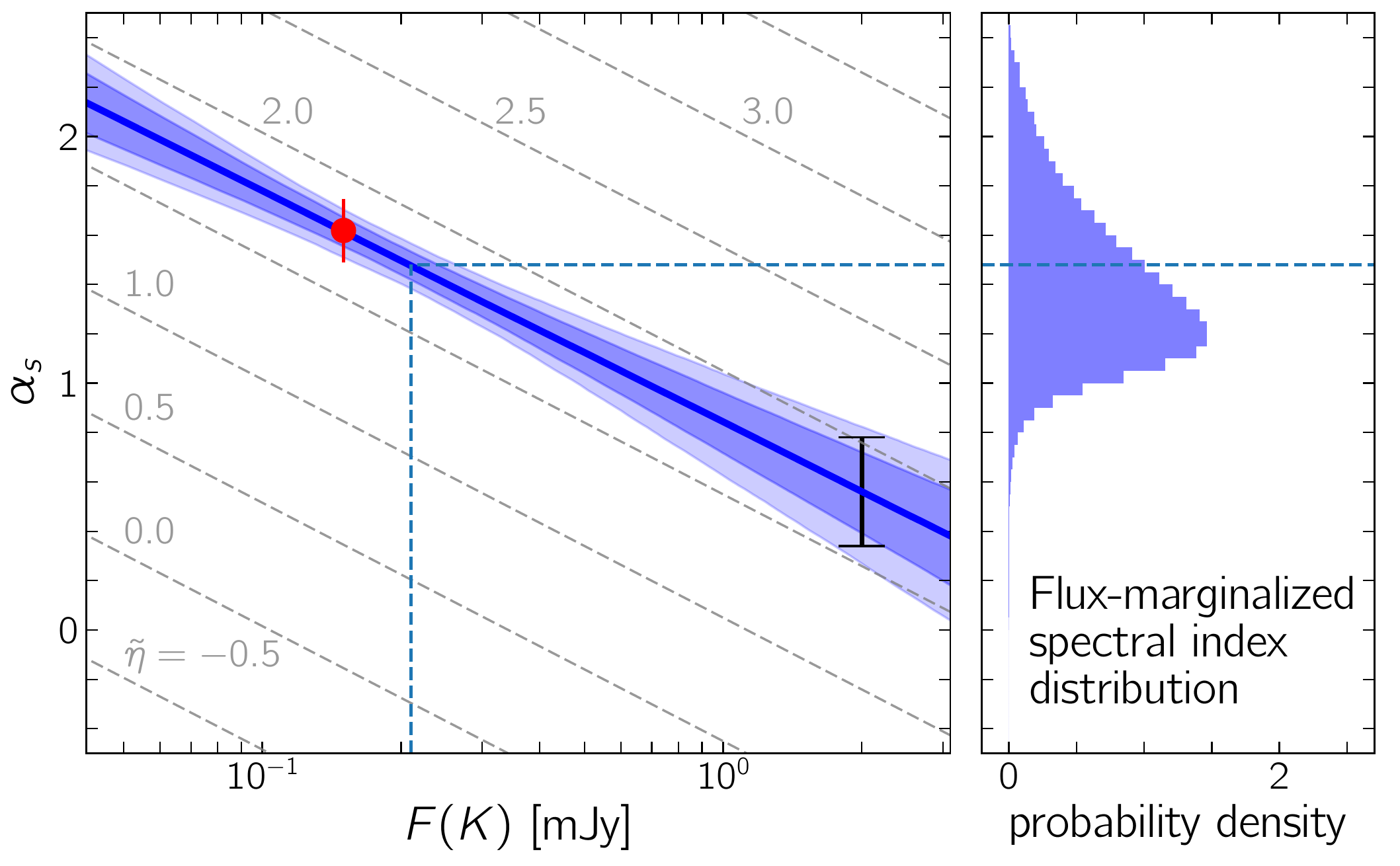}
\end{center}
\setlength{\abovecaptionskip}{0pt}
\caption{\replaced{NIR}{$K$-band to $M$-band} spectral index as a function of  $K$-band flux density. 
The filled red circle in the left panel shows the result of the simultaneous data (Fig~\ref{simdata}),
and the solid line shows the relation from Equation~\ref{eqalpha} with parameter posteriors of Case~3. Shaded areas show
the  68\% and 95\% \replaced{confidence}{credible} contours.
Absolute values of $\alpha_s$ are based on extinctions $A_K = 2.46$~mag and $A_M = 1.0$~mag, \added{and the black error bar in the lower right indicates the uncertainty in the spectral index due to  uncertainties in these reddening values}. The right panel shows the histogram of $\alpha_s$ marginalized over the actually observed flux-density range. The dashed lines show (vertical) the typical flux-density level and (horizontal) the corresponding $\alpha_s$. Previous studies (e.g., the \citealt{2007ApJ...667..900H} determination of $\alpha_{K-L}$)  became noise dominated below \replaced{the indicated flux density}{$\sim$0.2~mJy}.
Gray curves show \added{values of $\alpha_s$ predicted by a simple synchrotron model} (Equation~\ref{eq7}) for several values of $\tilde{\eta}$ as labeled.
}\label{alpha_f}
\end{figure}

In summary, in all three cases $\Re({M}/{K})$ at high flux density is consistent with $\alpha_s \approx 0.6$ (e.g., \citealt{2007ApJ...667..900H} and reddening values from Section~\ref{intro}). At $F(K)= 0.15$~mJy, 
$\alpha_s = 1.64 \pm 0.06$. This is the most precise determination of a spectral index change with flux density in the existing literature. This value is consistent with $\alpha_s = 1.7$ determined by \cite{2006ApJ...640L.163G} for their off-state-subtracted dim state. The combined data are consistent with well-constrained log-normal parameters for both $M$ and $K$
and require $\alpha_s$ to depend on the flux-density level.  
For Case~3, an empirical equation for $\alpha_s$ as a function of observed $F(K)$ is
\begin{equation}\label{eqalpha}
\begin{split}
%\alpha_s = (-0.71 \pm 0.12) \cdot \log{\bigg[\frac{F(K)}{\rm{mJy}} \bigg]} + 2.65^{+0.13}_{-0.14} \\
\alpha_s = \xi \cdot \log{\bigg[\frac{F(K)}{\rm{mJy}} \bigg]} + \eta + 1.2708 \cdot (A_M-A_K)\quad,
\end{split}
\end{equation}
with $\xi = -0.93 \pm 0.16$ and $\eta = 2.7 \pm 0.1$ (1$\sigma$ uncertainties).
%and the cross-covariance $C(\zeta, \eta) = 0.0078$. 
%The given uncertainties are the standard deviations. 
Equation~\ref{si} shows how Equation~\ref{eqalpha} was derived, and   Figure~\ref{alpha_f} illustrates the resulting $\alpha_s$ dependence on flux density. The correlation between $\xi$ and $\eta$ is $\rho(\xi, \eta) = C(\xi, \eta)/(\sigma_{\xi}\sigma_{\eta}) = 0.87$ (with $C(\xi, \eta) = 0.0115$ the cross-covariance) -- that is, the two parameters are strongly correlated\footnote{When explaining results from Equation~\ref{eqalpha}, we will use the variables $\xi$ and $\eta$ instead of their numerical values. With the uncertainties of $\xi$ and $\eta$ being strongly correlated, numerical values with uncertainties could be misinterpreted as independent.}. 
%Figure~\ref{alpha_f} shows the median and 68\% and 95\% confidence contours.
For $F(K)< 0.35$~mJy, Case~3 predicts a deviation of more than $2\sigma$ \added{from the constant spectral index of Case~1}. 
%Case~2 also shows $\alpha_s$ deviating from 0.6 for low flux densities.

The change of $\alpha_s$ with flux density is in the same direction but
 less extreme than found by \cite{2005ApJ...628..246E}, \cite{2006ApJ...642L.145K}, or \cite{2017MNRAS.468.2447P}. A direct comparison between the studies is difficult because of the following: 
\begin{itemize}
\item Different $S/N$ from  the various instruments; for Gaussian white measurement noise, $\alpha_s$ at low, noise-dominated flux densities becomes the logarithm of a Cauchy-distributed random variable (i.e., a distribution with extreme tails in both directions)
\item Different levels of background contamination and different methods of background subtraction 
\item Intrinsic momentary variations outside the general trend (which we determined here with integral methods; e.g., the simultaneous $K$ and $M$ data presented here show an extreme value of $\alpha_s \geq 1.9$ in one brief time interval)
\end{itemize}
The present analysis benefits from two advantages: (1) The comparably high S/N in both bands thanks to the IRAC $M$-band data. (2) The determination of $\alpha_s$ from flux-density PDFs, which themselves are derived from structure functions (i.e., from flux-density differences rather than from absolute flux-density levels). Background contamination is only an issue for the simultaneous dataset, which is one of the longest Keck light curves available and which has a distinct shape. That makes the determination of the \added{relative} offset and the \deleted{relative} flux-density ratio very accurate.

The intrinsic short-timescale variations of $\alpha_s$ seem to be based on small flux-density deviations of one band relative to the shape of the other. As a consequence, they significantly change the spectral index only at low flux-density levels. As the flux-density levels rise, the spectral index should follow the trend of Equation~\ref{eqalpha} with increasing precision. A simple comparison with the \cite{2007ApJ...667..900H} and \cite{2014IAUS..303..274W} data suggests consistency with such a mild trend. An in-depth analysis of additional data will be published separately.

\added{
While the Case~3 log-normal parametrization is consistent with most of the data, the $K$-band CDFs (Figure~\ref{sgradistrcdf}) show tails at high flux density outside the 
$68\%$ credible level (but within the 95\% level).  These are caused by a single particularly bright flux density excursion in the NaCo data and one similarly bright in the NIRC2 data. While these tails were one reason  \cite{2012ApJS..203...18W} chose a power-law approach, the spectral properties discussed in section~\ref{alpha_disc} are a strong indication that for the majority of flux densities, a log-normal parametrization is preferred over a power-law. The need for informative priors in Case~1 is another hint that a power-law parametrization is not an appropriate model. \cite{2011ApJ...728...37D} interpreted the tail in the VLT data as an indication for a second population of power-law-distributed flux densities.  It is intriguing that the independent Keck dataset shows a similar tail as the VLT data. However, both tails constitute about 1\% of the $K$-band observed time. If $M$-band (which does not show any indication of a higher rate of occurrence at higher flux densities) is included,  they constitute only about 0.6\%. Whether the extreme flux densities are extraordinary or not crucially depends on the baseline model. One way to  integrate the spectral properties at the lowest flux densities with the extended tails without adding more parameters would be a log-log-normal (double logarithmic) parametrization as proposed and used by \cite{2014ApJ...791...24M}. In the context of accretion processes, however, a log-normal distribution is the better established choice (e.g., \citealt{1538-4357-697-2-L167}). More data are needed to properly test whether a different PDF or a second population of extreme events is needed, but the Case~3 log-normal model is adequate for the existing data.
}

\subsection{Implications of the flux-density-dependent spectral index for a radiative model}
%with exponential cutoff of the electron energy distribution}

{
For the parametrization of Case~3, we can provide a physical context why the spectral index is a linear function of $\log{\left[F(K)\right]}$. In the following we analytically compare our results to the submm/NIR variability model discussed by \cite{2006A&A...450..535E} and \cite{2011AA...532A..26B}. \cite{2006A&A...450..535E} argued that the submm ($>$1~THz) to NIR emission is pure synchrotron radiation or synchrotron radiation with an additional contribution from synchrotron self-Compton emission. During brighter phases of Sgr~A*, $\alpha_s = 0.6$ is close to the canonical value for optically thin synchrotron radiation ($\alpha_s = 0.7$; e.g., \citealt{1975gaun.book..211M}). The turnover of the synchrotron spectrum from optically thick to optically thin is assumed to be at frequencies $\la$1~THz. The steeper spectral indices during dim phases discussed in the literature (\citealt{2005ApJ...620..744G,2005ApJ...628..246E,2006ApJ...640L.163G,2006ApJ...642L.145K}) are interpreted as the result of a changing electron energy distribution with a changing exponential cutoff at high energies due to synchrotron losses. As derived in \ref{cutoff_alpha}, the dependence of $\alpha_s$ on $S(\nu)$ is :
 \begin{eqnarray}\label{eq13}
%\alpha_s & = & \zeta \cdot \log{\left[\frac{S(\nu)}{S(\tilde{\nu})}\right]}+  \delta \cdot \left[1 + \zeta \cdot \log{\left(\frac{\nu}{\tilde{\nu}}\right)}\right] \\
\alpha_s & = & \tilde{\xi} \cdot \log{\left[\frac{S(\nu)}{\rm{mJy}}\right]} + \tilde{\eta}\quad,
\end{eqnarray}
with $\tilde{\xi}$ and $\tilde{\eta}$ being parameters related to the observing frequencies and submm spectral index \added{and flux density} and defined in \ref{cutoff_alpha}.
Equation~\ref{eq13} has the same form as Equation~\ref{eqalpha}, and one can see this as a motivation to use the log-normal/log-normal parametrization in the context of this model. \replaced{However, with}{With} $\nu_1 = 6.66 \times 10^{13}$~Hz ($M$-band) and $\nu = \nu_2 = 1.375 \times 10^{14}$~Hz ($K$-band), $\tilde{\xi} = -0.96576 \approx \xi$, i.e., the spectral index slope for this model is \replaced{significantly different from}{in excellent agreement with} the empirical slope determined in Case~3. This is illustrated in Figure~\ref{alpha_f}. 
\deleted{In order to accommodate a shallower slope $\xi$, $\tilde{\eta}$ itself must be a function of $F(K)$. Combining Equations~\ref{eq13} and \ref{eqalpha}, and changing from $S(\nu)$ to $F(K)$, the empirical dependence is
% \begin{eqnarray}\label{eq10}
% \tilde{\eta} & = & (\xi + 0.96576) \cdot \log{\bigg[\frac{F(K)}{\rm{mJy}} \bigg]} + \eta \nonumber \\ 
% && +1.2708 A_M - 0.8845 A_K \quad,
% \end{eqnarray}
and with Equations~\ref{eq9} and \ref{eq10} we can express the submm flux density $S(\tilde{\nu})$ ($\tilde{\nu} \approx 1$~THz) as an empirical function of $F(K)$ and the submm spectral index $\tilde{\alpha}_{s}$:}
%\begin{eqnarray}\label{eq12}
%\log{\left[ \frac{S(\tilde{\nu})}{\rm{mJy}}\right]} & = & \frac{1}{0.96576} \cdot \left\{\left(\xi + 0.96576\right) \cdot \log{\bigg[\frac{F(K)}{\rm{mJy}} \bigg]} + \eta \right\}  \nonumber \\
%&& +1.3159 A_M - 0.91586 A_K \nonumber \\
%&& + (13.102 - \log{\tilde{\nu}})  \cdot \alpha_{\rm{submm}} \quad.
%\end{eqnarray}
% \begin{eqnarray}\label{eq12}
% \log{\left[ \frac{S(\tilde{\nu})}{\rm{mJy}}\right]} & = & -\frac{1}{\xi} \cdot \Bigg(\left(\xi + 0.96576\right) \cdot \log{\bigg[\frac{F(K)}{\rm{mJy}} \bigg]} + \eta  \nonumber \\
% && +1.2708 A_M - 0.8845 A_K \nonumber \\
% && - \bigg\{1 + \xi \cdot  \bigg[14.1383 - \log{\bigg(\frac{\tilde{\nu}}{\rm{Hz}}\bigg)}\bigg]\bigg\}  \cdot \tilde{\alpha}_{s}\Bigg) \quad. \nonumber \\
% \end{eqnarray}
\deleted{Our findings in Case~3 therefore imply that, in addition to variations of the cutoff frequency, variations of the submm flux density and/or $\tilde{\alpha}_{s}$  correlated with changes in $\nu_0$ are required to explain the observed spectral index changes with flux density in the context of this model. In other words, our analysis implies a strict correlation of all NIR fluctuations with an optically thin submm component. Similarly,}  
\added{Our findings in Case~3 therefore imply that variations of the cutoff frequency $\nu_0$ are sufficient to explain the observed flux density and spectral index variations. Equation~\ref{eq9} then implies a linear relation between the submm spectral index $\tilde{\alpha}_{s}$ and the submm flux density $ \log{\left[ \frac{S(\tilde{\nu})}{\rm{mJy}}\right]}$ independent of $F(K)$.} Equations~\ref{eq5} and  \ref{eqalpha} \replaced{imply}{give} for the break-off frequency $\nu_0$ 
\begin{eqnarray}\label{eq11}
\frac{\nu_0}{\rm{THz}} & = & 24.1921 \cdot \Bigg\{ \tilde{\alpha}_{s} - \xi \cdot \log{\bigg[\frac{F(K)}{\rm{mJy}} \bigg]} \nonumber \\
&& - \eta - 1.2708 \cdot (A_M-A_K)\Bigg\}^{-2}\quad,
\end{eqnarray}
with the condition
\begin{eqnarray}\label{eq14}
\log{\bigg[\frac{F(K)}{\rm{mJy}} \bigg]} < \left[\frac{\tilde{\alpha}_{s}-\eta-1.2708\cdot(A_M-A_K)}{\xi}\right].
\end{eqnarray}
This last inequality states that $\tilde{\alpha}_{s}$, \added{and consequently $ \log{\left[ \frac{S(\tilde{\nu})}{\rm{mJy}}\right]}$,} can only be constant for a certain flux range (e.g., $0~\rm{mJy} < F(K) < 3.0~\rm{mJy}$ for $\tilde{\alpha}_{s} = 0.4$).  For flux densities higher than this range, $\tilde{\alpha}_{s}$ needs to become smaller. \added{However, no flux densities higher than this range have been observed.}

Figure~\ref{S_nu_F} shows $\nu_0$ \replaced{and $S(\tilde{\nu})$ as functions}{as a function} of $F(K)$ in the range of 0.4--1.5~mJy for a constant, optically thin spectral index $\tilde{\alpha}_{s} = 0.4$ and at a submm frequency $\tilde{\nu} = 1$~THz. The required flux-density of the optically thin submm component is $\sim$2.3~Jy for \replaced{an observed $K$-band flux density of ${\sim} 1.0$~mJy}{$\tilde{\alpha}_{s} = 0.4$}.\replaced{K-band fluctuations of 2~mJy observed correspond to submm changes $\sim1.2$~Jy.These value are}{ This value is similar to, but somewhat smaller than,} the typical submm levels,\replaced{and variability amplitudes}{ which indicates that such an optically thin submm component might not account for all submm radiation}. The predicted cutoff frequencies for moderately bright phases in the NIR are between $50$~THz and $200$~THz. \added{Even if a constant combination of $\tilde{\alpha}_{s}$ and $ \log{\left[ \frac{S(\tilde{\nu})}{\rm{mJy}}\right]}$ seems sufficient to explain the NIR statistics,} Equations~\ref{eq9} and~\ref{eq14} leave open the possibility for a rich interdependence of $F(K)$, $S(\tilde{\nu})$, and $\tilde{\alpha}_{s}$ that is testable with synchronous observations. Indeed, a close correlation between submm fluctuations seen with SMA and the 2014 June 18 IRAC light curve has been observed \citep{Fazio2018}. However, other studies have found evidence for optically thick synchrotron radiation at submm wavelengths (e.g., \citealt{2008ApJ...682..361Y}). The simple model \added{presented here} only begins to address the question of how the NIR and an optically thin submm component might be related. It does not provide any explanation about the origin of the variability in the submm, the origin of the non-thermal electrons, the acceleration mechanisms, or the link to the X-rays, which are crucial for  understanding  the high energy end of the electron distribution (see, e.g., \citealt{2017MNRAS.468.2447P}).

\begin{figure}
\begin{center}
\includegraphics[scale=0.55, angle=0]{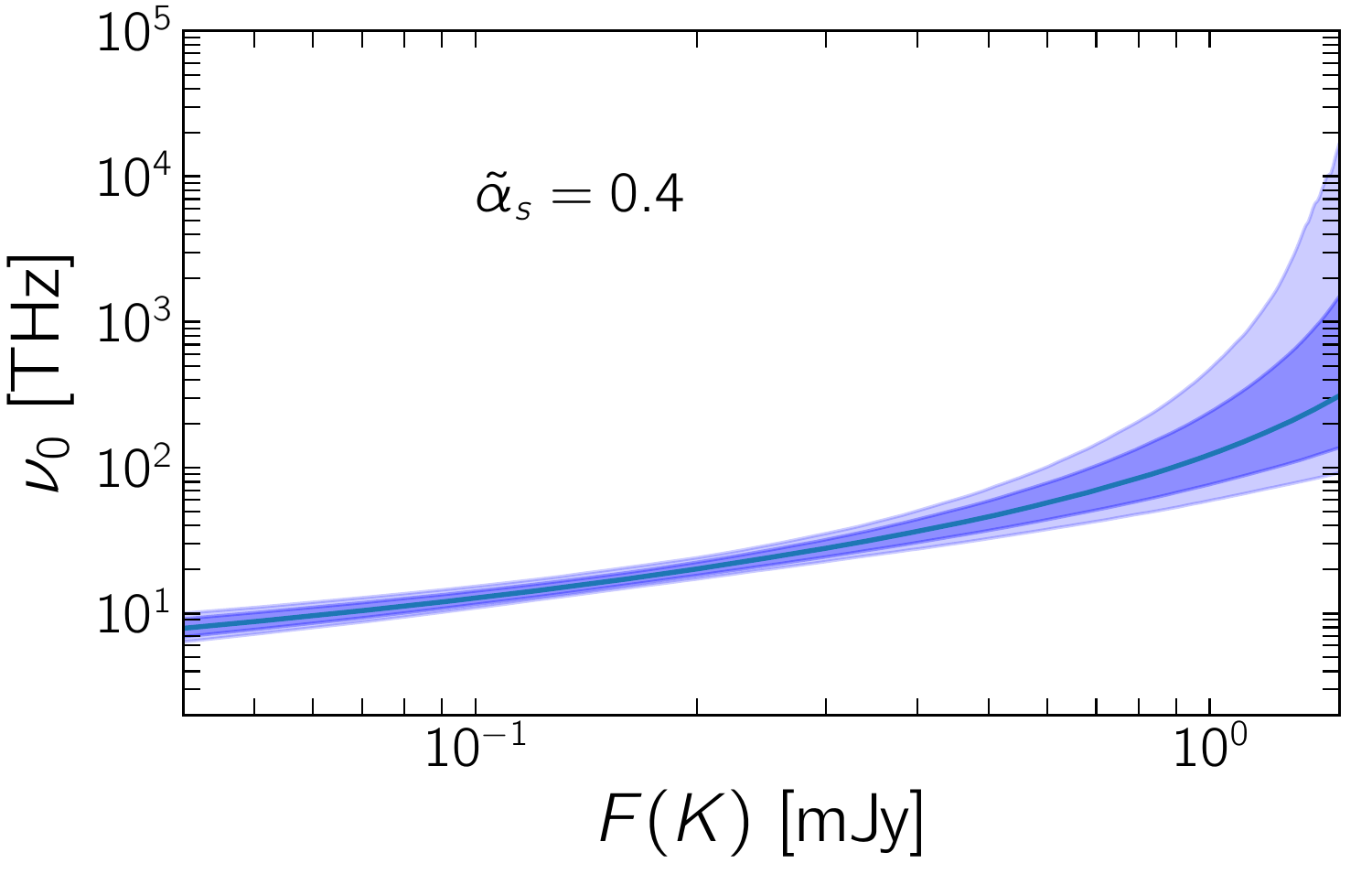}
\end{center}
\setlength{\abovecaptionskip}{-5pt}
\caption{\replaced{Upper panel:}{Synchrotron} cutoff frequency as a function of observed $K$-band flux density according to Eq.~\ref{eq11} for $\tilde{\alpha}_{s} = 0.4$. \replaced{Lower panel: predicted submm flux density at 1~THz according to Eq.~[DELETED] for $\tilde{\alpha}_{s} = 0.4$. In both panels solid}{Solid} lines show the median and shaded areas the 68\% and 95\% \replaced{confidence}{credible} contours.}\label{S_nu_F}
\end{figure}
}

\subsection{The \Sg\ spectral energy distribution}

In Case~3, the inferred log-normal parameters allow us to derive the mode and the {\it expected} flux-density PDF for each band. These quantities provide information on the lower limits of NIR flux densities. 
The modes of the log-normal distributions are  
\begin{eqnarray}
\mathcal{M}[F(M)] & = & 2.34^{+1.75}_{-1.04}~\rm{mJy}~~\rm{and} \nonumber \\ 
\mathcal{M}[F(K)] & = & 0.19^{+0.24}_{-0.11}~\rm{mJy}
\end{eqnarray}
for the $M$- and $K$-band, respectively. With a Galactic center distance of 8.3~kpc (and  extinctions given in Section~\ref{intro}), these flux densities correspond to
\begin{eqnarray}
\mathcal{M}(\nu_{M} L_{\nu_{M}}) & = & (3.2^{+2.4}_{-1.4})~10^{34}~\rm{erg}~\rm{sec}^{-1}~~\rm{and} \nonumber \\
\mathcal{M}(\nu_{K} L_{\nu_{K}}) & = & (2.6^{+2.0}_{-1.2})~10^{34}~\rm{erg}~\rm{sec}^{-1}\quad.
\end{eqnarray}
The error bars do not include uncertainties in the extinction or distance. These values are in full agreement with previously published upper limits (\citealt{2003Natur.425..934G} and references therein). 

In order to put the NIR flux densities in context, it is important to understand how the SED was estimated in the radio regime. The radio levels were obtained as average flux densities of multiple observations (e.g., \citealt{1998ApJ...499..731F}). Because of the symmetry of the intrinsic flux-density PDFs in the radio regime, the average is identical with the mode. The NIR modal values, being the most probable flux densities of Sgr~A* during its least variable moments,  are the natural counterparts to these radio flux density levels and can be interpreted as characteristic flux densities of \Sg\ within their bands. In this picture, a distinction between a quiescent (or steady) and a variable NIR state, as often proposed in the literature, is unnecessary. The modal values are merely particular flux densities within the distributions of variable flux densities.

Despite its attractive simplicity, representing the variable flux densities of \Sg\ by a single value is misleading. A full characterization of  flux densities is provided by the {\it expected} flux-density PDF.
This PDF incorporates information on both the intrinsic variability and the uncertainty in the parameters of the log-normal distributions given our data, and therefore is the proper tool for comparing SED models with our findings. The expected PDF is defined as 
\begin{eqnarray}\label{exp_flux}
\mathcal{P}(F \mid  \mathscr{D}) = \int{\mathcal{P}(F \mid \theta)\widetilde{\mathcal{P}}(\theta \mid \mathscr{D})d\theta}\quad,
\end{eqnarray}
 with $\mathcal{P}(F \mid \theta)$ the log-normal PDF defined in Equation~\ref{eq:fd2} and $\widetilde{\mathcal{P}}(\theta \mid \mathscr{D})$ the approximate posterior defined in Equation~\ref{post_def}. 
To estimate these, for each mock parameter set, we drew 100 flux density values  from the corresponding log-normal distribution and assigned each the weight corresponding to the parameter set. We then derived weighted quantiles from the resulting $10^6$ values. The results are presented in Table~\ref{percentiles} and Figure~\ref{sed}.
\begin{table}[ht!]
\begin{center}

\caption{Percentiles of the {\it expected} flux-density PDFs}
\begin{tabular}{cccrr}
\tableline
\tableline
&&&&\\[-1ex]
Percentile  & $F(K)$    &         $\nu_{K} L{\nu_{K}}$    &     $F(M)$   &   $\nu_{M} L{\nu_{M}}$ \\
 & (mJy) & ($10^{34}~\rm{erg}~ \rm{s}^{-1}$)& (mJy) & ($10^{34}~\rm{erg}~ \rm{s}^{-1}$)\\
\tableline
&&&&\\[-1ex]
 5th & 0.055 &  0.60 & 0.94 & 1.30 \\
15th & 0.110 &  1.19 & 1.49 & 2.06 \\
25th & 0.158 &  1.71 & 1.92 & 2.65 \\
35th & 0.208 &  2.26 & 2.33 & 3.21 \\
45th & 0.263 &  2.85 & 2.75 & 3.79 \\
55th & 0.325 &  3.52 & 3.19 & 4.40 \\
65th & 0.398 &  4.31 & 3.70 & 5.10 \\
75th & 0.489 &  5.29 & 4.33 & 5.98 \\
85th & 0.618 &  6.69 & 5.22 & 7.20 \\
95th & 0.877 & 9.49 & 6.94 & 9.57 \\
99th & 1.190 & 12.88 & 9.12 & 12.58 \\
\tableline
\end{tabular}
\end{center}
\label{percentiles}
\tablecomments{Percentile flux densities for Case~3 (log-normal/log-normal parametrization). The luminosities were derived assuming a distance of the Galactic center of 8.3~kpc and extinctions $A_K = 2.46$~mag and $A_M = 1.0$~mag. The uncertainties of these quantities are not included in the calculations of expected luminosities.}
\end{table}

Figure~\ref{sed} represents the first systematic characterization of \Sg's SED in the NIR at the lowest flux densities. The lowest quantiles are extrapolations to flux densities that are unobservable because of measurement noise. They are valid under the assumptions of Case~3, which has 5th percentiles  $F_{\rm{5\%}}(K) = 0.055$~mJy and $F_{\rm{5\%}}(M) = 0.94$~mJy. These can serve as lower limits for the typical flux density range. In contrast, the quantiles above 25\%  are above the $3\sigma$ detection levels of NIRC2, and the median level is above the $3\sigma$ for IRAC.  \deleted{Validity of these levels is independent of the choice of log-normal or power-law parametrization.}

Characterization of the dim-phase SED of Sgr~A* constrains the radiative processes at work. For radio wavelengths $>$3~cm, the SED is dominated by synchrotron radiation from non-thermal electrons with a power-law energy distribution (\citealt{1998Natur.394..651M,2000ApJ...541..234O,2003ApJ...598..301Y}). The models predict a significant contribution of this non-thermal electron population to the NIR. Figure~\ref{sed} compares the corresponding luminosities with the \cite{2003ApJ...598..301Y} spectral energy distribution model (as shown in their Figure~1). The NIR flux densities agree remarkably well with the model of synchrotron radiation, which was derived entirely from the radio part of the SED for a slope of the electron energy distribution of 3.5.

\begin{figure*}
\begin{center}
\includegraphics[scale=0.6, angle=0]{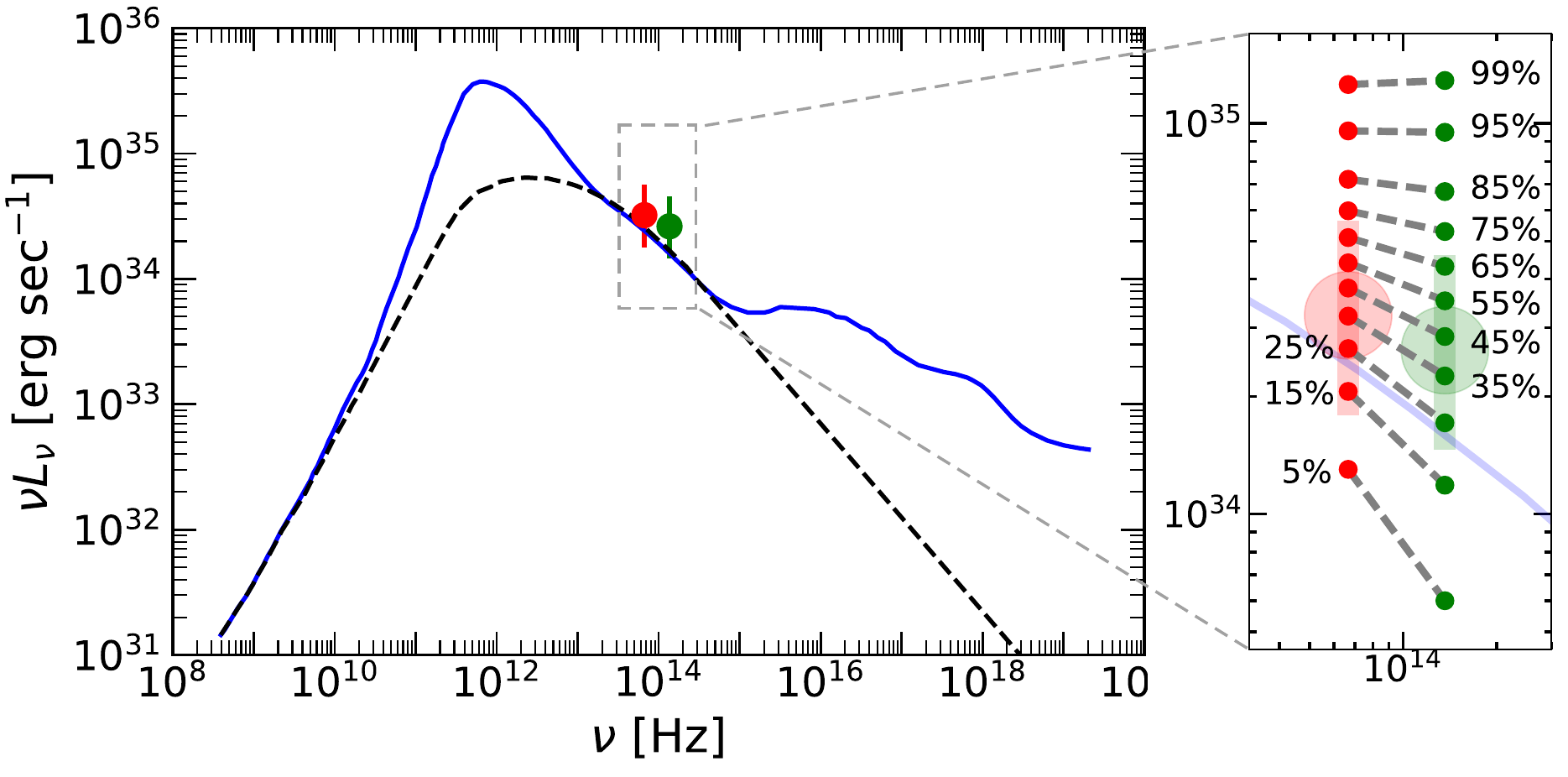}
\end{center}
\setlength{\abovecaptionskip}{-10pt}
\caption{Observed and model spectral energy distributions for \Sg.  
Green and red points show, respectively,
the 2.18 and 4.5~$\mu$m dim-phase luminosity densities of \Sg\, as derived from 
the modes of the Case~3 analysis. The inset on the right shows several percentiles of the {\it expected} flux-density PDF as defined in Equation~\ref{exp_flux}. The gray connection lines indicate the change of the $\nu L_{\nu}$ spectral slope $1-\alpha_s$ with luminosity.
The blue line shows an SED model  \citep{2003ApJ...598..301Y} derived and
normalized entirely
from the radio part of the SED. The SED model assumes synchrotron radiation from electron populations with thermal and non-thermal energy distributions for the radio to NIR, and inverse Compton and bremsstrahlung emission for the higher frequencies. The black, dashed curve shows the non-thermal synchrotron model component \citep{2003ApJ...598..301Y}.}\label{sed}
\end{figure*}

\subsection{Black hole mass, luminosity, and rate of stochastic variability power}\label{s:rsvp}

\cite{2009ApJ...694L..87M} found their \Sg\ break timescale consistent with mass--timescale  relations of AGN in X-rays. However, it can be very difficult to obtain reliable break timescales from AGN light curves \citep{2013ApJ...779..187K}. \citeauthor{2013ApJ...779..187K}\ analyzed X-ray and $0.51~\mu$m light curves of 39~AGN by introducing a parameter called ``rate of stochastic variability power'' (RSVP, designated $\xi^2$). This parameter is defined for damped random walks and quantifies the rate at which stochastic power driving the random walk is inserted. The RSVP is related to the total variance of the Ornstein-Uhlenbeck (OU) variability process (\citealt{2009ApJ...698..895K,2013ApJ...779..187K}) by
 \begin{equation}\label{eqrsvp}
 \zeta^2 = 4\pi f_{b} \cdot {\rm Var}[F(t)]\quad.
 \end{equation}
For the 39~AGN observed by \citeauthor{2013ApJ...779..187K}, $\zeta^2$ measured in X-rays correlates closely with black hole mass.  While $\zeta^2$ as determined from visible light curves also scales with black hole mass, the (anti-)correlation with luminosity is even tighter.

As we have found here,
Sgr~A* is well described by an OU-process with a PSD slope of $\sim$2 with one break timescale (see also \citealt{2014ApJ...791...24M}). Therefore we can use Equation~\ref{eqrsvp} to derive $\zeta^2$ from the variance of the flux-density PDF and the break timescale. 
For \Sg\ at $M$, $\log{\zeta^2} = -2.61^{+0.16}_{-0.17}$. The value predicted by the empirical {\em mass}--RSVP relation  \citep{2013ApJ...779..187K} is  $\log{\zeta^2} \approx -7.4$, about five orders of magnitude smaller. This discrepancy might be expected because  AGN are highly accreting objects, whereas Sgr~A* has a tiny Eddington ratio. However, it is remarkable that the empirical {\em luminosity}--RSVP relation \citep{2013ApJ...779..187K} predicts $\log{\zeta^2} \approx -3.63$, close to the value for Sgr~A*. Figure~\ref{rsvp} compares \Sg\ with the \citeauthor{2013ApJ...779..187K}\ AGN, which have luminosities about nine orders of magnitude larger. While the uncertainties of the empirical relation put Sgr~A* just outside the 1$\sigma$ envelope, the agreement is striking. These findings are even more surprising considering that the \cite{2013ApJ...779..187K} interpretation for the luminosity--RSVP anti-correlation identifies the likely origin of the visible radiation as blackbody radiation from the outer parts of a thick accretion disk, whereas for Sgr~A*, the NIR emission is non-thermal synchrotron radiation from the innermost accretion region.

\begin{figure}
\begin{center}
\includegraphics[scale=0.5, angle=0]{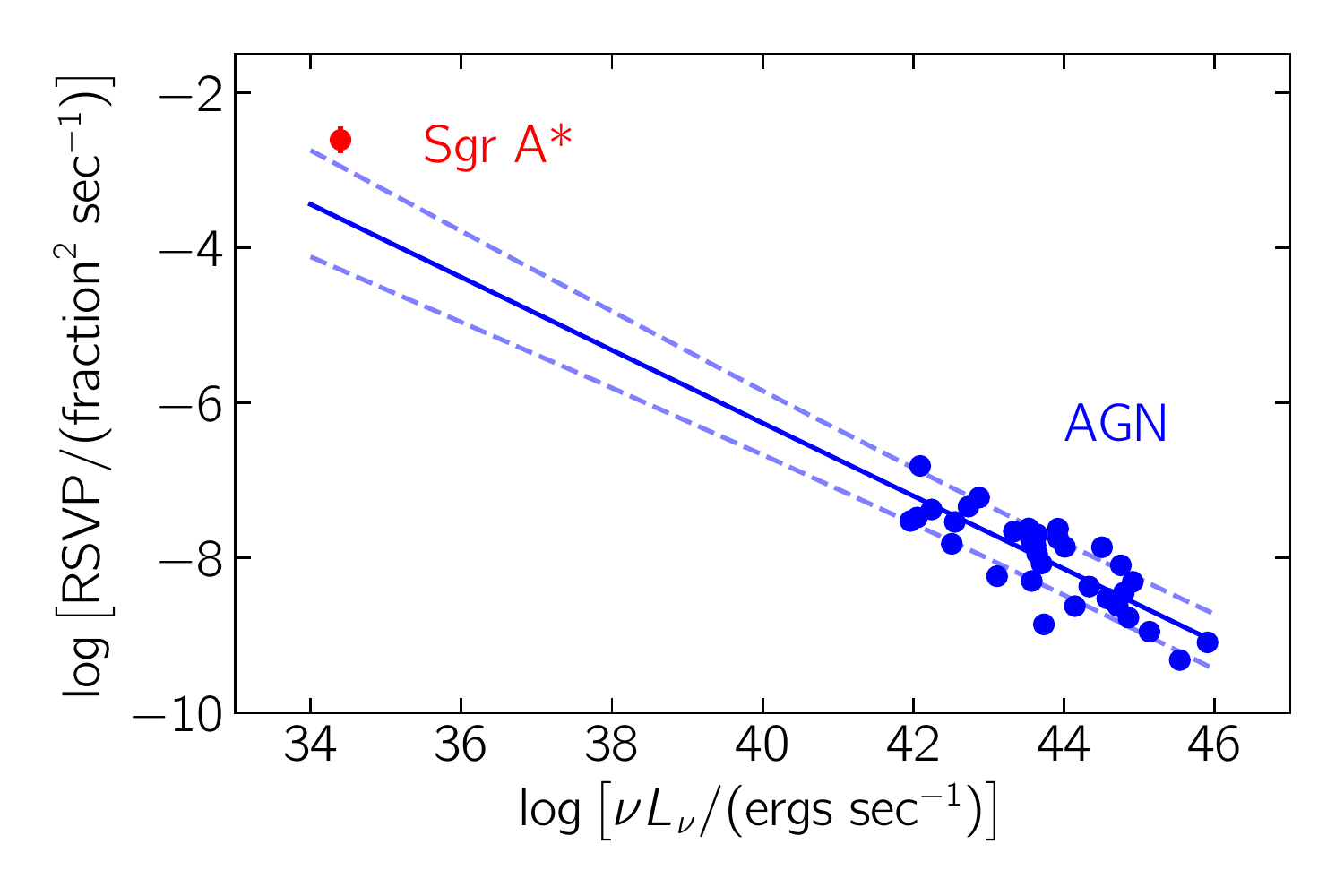}
\end{center}
\setlength{\abovecaptionskip}{-10pt}
\caption{Rate of stochastic variability power (RSVP, \S\ref{s:rsvp})
as a function of luminosity. Blue points denote 39 AGN observed by \citet{2013ApJ...779..187K}, and the solid line shows those authors' Eq.~29. Dashed lines show the corresponding uncertainty envelope. The red point shows the RSVP for Sgr~A* derived here.}\label{rsvp}
\end{figure}

\subsection{Telescope photometric performance}

While the observations from ground-based observatories do not include data after 2010 for the VLT and after 2013 for Keck (except the single 2016 data set), the VLT and Keck data used here constitute the most comprehensive and best characterized datasets available.
They include most of the previously published $K$-band data for \Sg. In particular, they have been used in the statistical analyses of \cite{2012ApJS..203...18W} and \cite{2014ApJ...791...24M} and therefore provide a well understood baseline for the analysis of the 4.5~$\mu$m \Sp\ data. Our analysis of the flux-density PDFs tells us which flux-density level in $M$-band corresponds to which level in $K$-band. This enables us to compare the relative sensitivity of each observatory to a given flux-density excursion. For a representative clock time of $\approx$1~minute (for which the \Sp\ 8.4~s noise scales down by ${\approx}\sqrt{7}$ to $\sigma_{\rm{IRAC}} = 0.05$~mJy) and large flux densities, the S/N proportions IRAC:NaCo:NIRC2 are 1:1.7:3.7. For low flux densities where $\Re({M}/{K}) \approx 12$, S/N proportions become 2:1.7:3.7 (i.e., \Sp/IRAC observing in $M$-band is competitive in S/N with ground-based AO imaging with 8--10~m-class telescopes observing in $K$).  The future {\it James Webb Space Telescope} should be far superior at these wavelengths.

\section{Conclusions}\label{concl}

\added{The existing 2.2 and 4.5~\micron\ variability data of \Sg\ can be explained by a relatively simple model.  The model incorporates log-normal PDFs at both wavelengths and a broken power-law PSD with a single break near 4~hr.  The two brightest observed epochs of \Sg\ hint at but do not require either a separate process for these rare events or a PDF that is not log-normal.}

This paper has aimed to do the following:
\begin{itemize}
\item Presented the most comprehensive available set of NIR light curves of Sgr~A*.  Data were compiled from three observatories: the \SST, the ESO VLT, and the Keck observatory.
\item Demonstrated the value of the new PSF extraction and fitting tool AIROPA on photometry for Sgr~A*
\item Introduced a new Bayesian method to determine the power spectral density of irregularly sampled, red-noise-dominated time series with non-Gaussian flux-density PDFs
\item Determined the power spectral density and characteristic timescale $\tau_{b} = 243^{+82}_{-57}$~minutes of the variability process with unprecedented precision
\item Excluded PSD structure at timescales of 10--100~minutes.  Such timescales  correspond to the innermost stable circular orbit for black hole spin parameter $a<0.9$.
\item Determined the spectral NIR properties of Sgr~A* and the intrinsic flux-density PDFs with unprecedented accuracy. In particular, we confirmed the NIR spectral index of $\alpha_s \approx 0.6$ for flux densities above 0.3~mJy and found a redder spectral index at lower flux densities.
\item Explored the spectral index dependence on flux density within the context of a electron energy distribution with exponential cutoff. We find the predicted submm levels and variability amplitudes to be consistent with the observed submm properties.
\item Determined the dim phase SED in the NIR based on synchronous $K$- and $M$-band data, assuming a log-normal parametrization for the flux-density PDFs
\item Demonstrated that Sgr~A* is in agreement with the anti-correlation between luminosity and rate of stochastic variability power
derived from visible light curves of more luminous AGN
\item Showed that the \SST\ has relative photometric performance at $4.5~\mu$m on Sgr~A* competitive with ground-based AO observations at $2.18~\mu$m 
\end{itemize}

These results are especially of interest for the GRAVITY interferometric experiment at the Galactic Center. One of its goals is to measure the astrometric signature of hot spots moving close to the ISCO of the black hole. We expect that GRAVITY will not detect any such signature at timescales longer than 9~minutes as we do not find any NIR ISCO signature at these timescales. It seems imperative to design GRAVITY to operate at timescales significantly shorter than 10~minutes.

Another interesting result is the indication of the intrinsic turnover of the flux-density PDF at low flux-density levels in $M$-band. This means that $M$-band space-based observations are uniquely suited to explore all relevant timescales {\it and} the low flux-density regime, where the changes in timing and flux-density PDF possibly carry essential information about the physical processes at work. \Sg\ will be an essential target for the much more sensitive {\it James Webb Space Telescope}.  

It is surprising how steady the statistical, spectral, and polarimetric parameters describing the variability of Sgr~A* have been since the beginning of AO observations.  The fact that the PSD parameters can be determined more precisely with a more extensive dataset and a better method implies that the PSD and PDF parameters are indeed self-consistent and nearly stationary over the last $\sim$15 years. (We have, however, not strictly tested stationarity in this analysis.) While one might expect the accretion process to be susceptible to abrupt changes in the supply of material (e.g., material stripped off  G1 or G2), the NIR variability process shows no indication of that. The timescales at which matter travels from the typical periapsis distance of the G sources or S-stars ($>$100--200~au) are not clear. The interaction of infalling matter with the large number of fast-orbiting stars in the S-cluster might prevent larger clumps of gas from coherently finding their way to the innermost accretion region, thus regulating the steady supply of matter. 

\acknowledgments
  
\added{We thank the anonymous referee for helpful comments.
We thank Arno Witzel, Rainer Sch\"odel, Andreas Eckart, Dan Marrone, Stefan Gillessen, Matthew Malkan, Aurelien Hees, Zhiyuan Li, Leo Meyer, and Silke Britzen for fruitful discussions.}
We thank Jean Turner for giving us access to her UNIX server for debugging our C++ code. 
We thank Nick Robertson for his excellent IT support. 
This work is based on observations made with the \SST, which is 
operated by the Jet Propulsion Laboratory,
California Institute of Technology, under a contract with
NASA. We thank the staff of the \Sp\ Science Center for their help in 
planning and executing these demanding observations.
The W. M. Keck Observatory is operated as
a scientific partnership among the California Institute
of Technology, the University of California, and the National
Aeronautics and Space Administration. The authors
wish to recognize that the summit of Mauna Kea
has always held a very significant cultural role for the
indigenous Hawaiian community. We are most fortunate
to have the opportunity to observe from this mountain.
The observatory was made possible by the generous financial support of the W. M. Keck Foundation.
Support for this work was provided by NSF grants
AST-0909218, AST-1412615.
R.N. was supported by the NSF grant AST-1312651.
C.F.G. is supported by NSF grant AST-1333612 and AST-1716327.
This work used the Extreme Science and Engineering Discovery Environment (XSEDE), which is supported by National Science Foundation grant number ACI-1548562.
The XSEDE allocation IDs are TG-AST170006 and TG-AST080026N. The computations were executed on the clusters Stampede, Comet, Bridges, and SuperMIC.
This work used the UCLA Hoffman2 cluster.

\noindent
\software{Astropy \citep{2013A&A...558A..33A},
                              Matplotlib \citep{4160265}, AIROPA \citep{2016SPIE.9909E..1OW}, StarFinder \citep{2000SPIE.4007..879D}, FFTW \citep{FFTW05}, NumPy \citep{numpy} , Stan \citep{stan_ref}, corner.py \citep{corner}}
                              
\noindent
\facility{Spitzer/IRAC, Keck/NIRC2, VLT/NaCo}

\bibliographystyle{apj}
\bibliography{mybib_gc}{}

\appendix
\renewcommand{\thesection}{Appendix~\Alph{section}}

\section{Bayesian estimation of the flux-density ratio from synchronous data}\label{bayratio}
\setcounter{equation}{0}

Figure~\ref{simdata} shows simultaneous $K$ and $M$ light curves that match very closely.
In order to derive the best flux-density ratio from these data, we modeled the $M$ flux-density as $F(M)=s \cdot F(K)+c$, where $s$ and $c$ are constants to be derived from the light curves.  Each data point in the observed light curves can be modeled as
\begin{eqnarray}
F_{i, \rm{real}}(M) & \sim & \mathcal{N}(s \cdot F_{i,\mathrm{real}}({K}) + c,\sigma_{\mathrm{disp}}^2)\\
F_{i, \rm{obs}}(M) & \sim & \mathcal{N}(F_{i,\mathrm{real}}(M),\sigma_{M}^2) \\
F_{i, \rm{obs}}(K) & \sim & \mathcal{N}(F_{i,\mathrm{real}}(K),\sigma_{K}^2) \quad,
\end{eqnarray} 
where $x \sim \mathcal{N}(\mu, \sigma^2)$ denotes \added{a random variable distributed according to} a normal distribution with mean $\mu$ and standard deviation $\sigma$. 
$F_{\mathrm{obs}}(M)$ and $F_\mathrm{obs}(K)$ are
the observed flux densities including measurement white noise; 
$F_{\mathrm{real}}(M)$ and $F_{\mathrm{real}}(K)$ are idealized flux densities without 
measurement noise;  $ \sigma_{\rm{disp}}$ is an additional dispersion to allow 
the ideal ratio of $F_\mathrm{real}(M)$ and $F_\mathrm{real}(K)$ to differ as 
implied by the previously observed low-level and short-timescale spectral index
fluctuations \citep{2014IAUS..303..274W}.
Integrating over the model parameters  $F_\mathrm{real}(M)$ and $F_\mathrm{real}(K)$ gives
\begin{equation}
F_{i,\mathrm{obs}}(M)  \sim  \mathcal{N}(s \cdot F_{i,\mathrm{obs}}(K) +
c,\sigma_{{M}}^2+ s \cdot 
\sigma_{{K}}^2+\sigma_{\mathrm{disp}}^2)\quad.
\end{equation}
We implemented this likelihood function with a MCMC Bayesian sampler in Pystan. With $\sigma_{{M}} = 0.212$ (the $M$-band measurement noise for the rebinned IRAC data) and $\sigma_{{K}} = 0.015$, we obtained the posteriors shown in Figure~\ref{corner_mcmc}.

\section{Bayesian structure function analysis}\label{ABC}
\setcounter{equation}{0}

For time series analyses of non-Gaussian-distributed light curves, there is generally no analytic expression for the likelihood function. That rules out a standard Bayesian approach, and instead we use a population Monte Carlo approximate Bayesian computation (PMC-ABC), which requires no prior knowledge of the likelihood function. The procedure was described by \cite{2015A&C....13....1I}, but we have created our own C++ implementation tailored to the task of time series analysis. Our analysis procedure has four functional components:
\begin{itemize}
\item A method to randomly simulate data that mimic the observations as closely as possible. The method is based on a model parameterized by the quantities one wishes to determine.  The model can be either statistical or deterministic or a combination.
\item A distance function that quantifies how close the simulated data come to the available observations
\item A prior distribution for each parameter
\item The PMC-ABC sampler itself, which calls the three components above in the proper order
\end{itemize}
\replaced{The individual components are described in detail by \cite{2015A&C....13....1I} for the PMC-ABC sampler, \cite{2012ApJS..203...18W} for the data simulation, and \cite{2014ApJ...793..120H} for the data simulation and distance function.}{A previous approach to the PMC-ABC sampler was described by
\cite{2015A&C....13....1I}  and to data simulation by \cite{2012ApJS..203...18W} and \cite{2014ApJ...793..120H}.} Notation and details for this work are explained below.

\subsection{Simulating NIR light curves of Sgr~A*} \label{howtosim}

The power spectral density of the simulated data is a red-noise power-law spectrum with breaks at two frequencies, $f_{b}$ and $f_{b, 2}$ (Fig.~\ref{period}).  The first break transitions between a slope $\gamma_0 = 0$ (for low frequencies corresponding to long time lags)
to slope $\gamma_1$ and the second (at high frequencies corresponding to short time lags)
from $\gamma_1$ to $\gamma_2$:
\begin{equation}\label{psddef}
 \mathrm{PSD}(f) \propto \left\{
 \begin{array}{lcl}
  f^{-\gamma_0} & \mathrm{for} & f < f_{b}\\
  f^{-\gamma_1} & \mathrm{for} & f_{b} \leq f < f_{b,2}\\
  f^{-\gamma_2} & \mathrm{for} & f \geq f_{b,2}\quad.
 \end{array}\right.
\end{equation}
%
%to a slope of zero at a temporal frequency $f_{b}$, the characteristic (or coherence) timescale:
%   \begin{equation}\label{psddef}
%   \begin{split}
%   {\rm PSD}(f) &\propto f^{-\gamma_1},\qquad f>f_{b}\\
%   {\rm PSD}(f) &\propto f_{b}^{\gamma_0-\gamma_1}f^{-\gamma_0},\qquad f\le f_{b} \; \; ,
%   \end{split}
%   \end{equation}
%   with $f$ the temporal frequency.
%   
% Here we use a more complex model including a second break described by equation (\ref{psddef}) and the additional equation:
% \begin{equation}\label{psddef2}
% {\rm PSD}(f) \propto f_{b,2}^{\gamma_2-\gamma_1}f^{-\gamma_2},\qquad f\le f_{b,2}\; \; .
% \end{equation} 

For long time intervals $t\gg \tau_{b} \equiv 1/ f_{b}$, the flux-density PDF
(denoted $\mathcal{P}(F)$) is well described by a power law:
\begin{equation}\label{eq:fd}
   \mathcal{P}[F \mid (\beta, F_0)] = \left\{
  \begin{array}{lll}
  \left[\frac{(\beta-1)}{-F_{0}}\right]
      \left[\frac{(F-F_{0})}{(-F_{0})}\right]^{-\beta} & 
    \mathrm{for} & F \ge 0\quad,\\
  0 & \mathrm{for} & F < 0\quad,
  \end{array}\quad
\right.
\end{equation}
where $\beta$ is the power law index and $F_{0} < 0$ is the pole of the power-law.
The cumulative distribution function is then
\begin{equation}\label{eq:cdf1}
{\rm{CDF}}[F \mid (\beta, F_0)] \propto\bigg(\frac{F-F_{0}}{-F_{0}}\bigg)^{-\beta+1}\quad.
\end{equation}

An alternative distribution that describes the data in the observed range is a log-normal:
\begin{equation}\label{eq:fd2}
  \mathcal{P}[F \mid (\mu_{\logn} , \sigma_{\logn})] =
  \begin{array}{lcc}
{{(\sqrt{2 \pi} F \sigma_{\logn})}}^{-1} \cdot \exp{\left({-\frac{\left[{\ln{\left(\frac{F(K)}{\rm{mJy}} \right)}-\mu_{\logn}}\right]^2}{{\sqrt{2} \sigma_{\logn}^2}}}\right)} \quad ,
  \end{array}
\end{equation}
with 
\begin{equation}\label{eq:cdf2}
{\rm{CDF}}[F \mid (\mu_{\logn} , \sigma_{\logn})] = \frac{1}{2} - \frac{1}{2} {\rm{erf}}\left[{\frac{\ln{\left(\frac{F(K)}{\rm{mJy}}\right)} - \mu_{\logn}}{\sqrt{2}\sigma_{\logn}}}\right]\quad,
\end{equation}
where $F \in [0,\inf]$, $\mu_{\logn} \in [-\inf,+\inf]$, and $\sigma_{\logn} \in [0,\inf]$.

To create light curves from the PSD in Eq.~\ref{psddef} that show the flux-density PDF of Equation~\ref{eq:fd}, we used the 
\cite{1995A&A...300..707T} method as further developed by \cite{2012ApJS..203...18W}:
\begin{itemize}
\item Draw Fourier coefficients for each frequency from a Gaussian distribution
with a variance proportional to the value of the PSD at that
frequency. 
\item Fourier transform to the time domain giving normal-distributed random 
variable $y$, and normalize $y$ to unit variance. 

\item Sample $y$ to the cadence of the observed light curve.  

\item Transform $y$ into a power-law distributed random variable $T(y)$ 
that takes on values $0 < T(y) < \infty$:
\begin{equation}
T(y) =   F_{{0}} - F_{{0}} \cdot \left\lbrace \frac{1}{2}\left[ 1 +
    {\mathrm{erf}}\left(\frac{y}{\sqrt{2}}\right)\right]\right\rbrace^{
    {\left(1-\beta\right)}^{-1}}\quad,
\label{transf}
\end{equation}
where `erf' is  the Gaussian error function, $\beta$ is the power-law index, and $F_{0}$ is the power-law pole of Equation~\ref{eq:fd}.
For the alternative log-normal distribution, use instead
\begin{equation}
T(y) =  \exp{\big(\sigma_{\logn} \cdot y + \mu_{\logn}\big)} \quad.
\label{transf2}
\end{equation}

\item Draw Gaussian noise (independent for each point), and add it to each point to account for the measurement errors. 
 \end{itemize} 
 
The above method allows generating light curves according to any calibrated PSD of the form of Eq.~\ref{psddef} -- that is, distributed as the observed data on all timescales. In particular, it enables comparison of the absolute values of the structure functions
of simulated light curves with the observed structure function. The
transformation of Equation~\ref{transf} or~\ref{transf2} changes the PSD of the 
generated light curve slightly
(Figure~\ref{period}), but the break timescales
are invariant under this transformation.   

\begin{figure}
\begin{center}
\includegraphics[scale=0.35, angle=0]{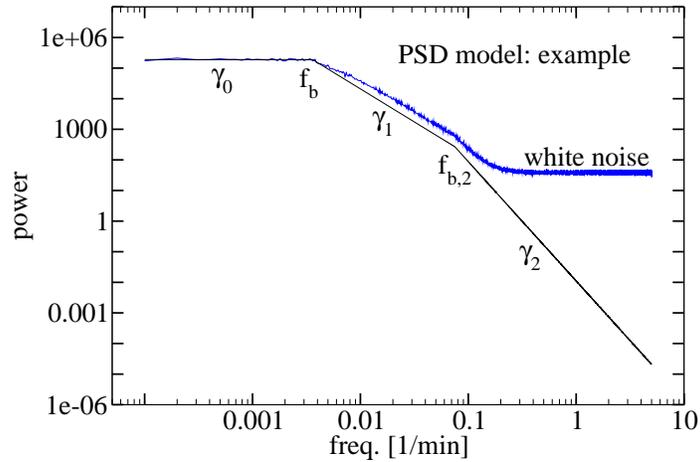}
\end{center}
\setlength{\abovecaptionskip}{-10pt}
\caption{The PSD model before and after nonlinear transformation. The black lines show an example PSD model with two break timescales. Because the abscissa is in frequency space, short timescales are to the right.  In the notation used here, the first break frequency at ${\sim}3\times10^{-3}$~minutes$^{-1}$ is $f_{b}$,
and the second break frequency at $\sim$0.1~minutes$^{-1}$ is $f_{b,2}$. Slopes of the three segments are from left to right $\gamma_0 = 0$, $\gamma_1$, and $\gamma_2$.  
The last is indeterminate in the actual data, because no second break can be found. The blue curves
show a simulated measurement of the example PSD (\ref{howtosim}) based on an  
average FFT periodogram of 1000 equally sampled light curves with 6~s cadence and a duration of $10^4$ minutes.}\label{period}
\end{figure}
 
\subsection{The distance function}\label{dist}

The distance function $\phi$ quantifies the difference between two data sets.  This function is used in the PMC-ABC algorithm to compare randomly drawn mock data sets to the measured data (see section \ref{sec:pmcabc} below).  We based our distance function on the first-order structure function defined by
\begin{equation}
\label{eq:strfundef}
V(\tau_i) = \frac{1}{n_i}\sum_{t_j, t_k}\left[F(t_j) - F(t_k) \right]^{2} \quad\mathrm{for}\quad \tau_i \leq (t_j - t_k) < \tau_{i+1}\quad,
\end{equation}
that is, the sum of $\left[F(t_j) - F(t_k)\right]$ over all $n_i$ existing pairs whose time lags $(t_j - t_k)$ fall within the bin $[\tau_i, \tau_{i+1}]$.  
%
%\begin{equation}
%\Delta_i = [\tau_i-\Delta_{(i,-)}, \tau_i+\Delta_{(i,+)}]\quad.
%\end{equation}
%
We defined the distance between two light curves as the weighted L2 norm of
the difference between the logarithms of the respective structure function's binned values:
\begin{equation}\label{distdef}
\phi(V_1, V_2) = \sum_{i} w_{i} 
  (\log\left[V_1(\tau_i)/V_2(\tau_i) \right])^{2}\quad
\end{equation}
with $w_{i}$ the weights for the chosen binning. These weights control the relative influence of each data set on the results, not the accuracy to which the structure function is approximated. For a sufficient number of iterations, any weighting scheme converges to the same result\added{, but speed of convergence depends on the weights. For this work, weights were chosen by trial and error to give consistent and equal convergence in all bins of all three structure functions}. 
The relative weights \added{adopted were} unity for each structure function bin, except for the single wide bin at large time lags (see \S\ref{meth} \added{and Figure~\ref{structplot}}), which had a weight of 3. \added{This last bin had to be wide in order to compensate for the intrinsic variance of the structure function at high time lags.  However, using a wide bin lowers the effective weight of high time lags, which are essential for determining the characteristic time scale. Using weight three let the mock structure function converge onto the last bin as fast as onto all the others.} Additionally, we weighted the structure function from the NIRC2 data with 0.67 relative to the IRAC and NaCo data\added{ because the relatively short durations of NIRC2 observations led to higher variance in the structure function. Even with its higher variance, the NIRC2 structure function carries most of the information about the shortest timescales. The adopted weights enable the ABC algorithm to first determine the posteriors of the well-determined parameters before finding the best fits for the second break timescale and slope. Searching for $f_{b,2}$ and $\gamma_2$ before settling on good values for the other parameters would have been hopelessly inefficient.}

\subsection{Prior distributions}\label{priors}

We used a combination of flat and Gaussian priors.  The latter are appropriate
for the few cases of independently well-determined parameters such as the measurement noise. In order to guarantee a monotonically decreasing function for the PSD, we applied the conditions
\begin{equation}\label{cond}
f_{b,2} > f_{b}~~ \mathrm{and} ~~
\gamma_2 > \gamma_1
\end{equation}
to the flat joint prior distributions for the PSD, $\mathcal{P}(\gamma_{1}, \gamma_{2})$ and $\mathcal{P}(f_{b}, f_{b,2})$. While the joint prior distributions of  $\gamma_{1}$, $\gamma_{2}$,  $f_{b}$, and $f_{b,2}$ are uniform, their marginalized distributions are not because of the conditions in Eq.~\ref{cond}.  In this case, the cumulative distribution functions of the marginalized probabilities are quadratic in their respective parameters.  Therefore drawing from a joint uniform probability distribution subject to the constraints in Eq.~\ref{cond} can be obtained by
\begin{eqnarray}
 f_{b}  & = & f_{max} - (f_{max}-f_{min}) \cdot \sqrt{ \mathfrak{u}_{f, 1}} \\
 f_{b,2} & = & (f_{max,2}-f_{b}) \cdot \mathfrak{u}_{f, 2} +f_{b}
\end{eqnarray}
and
\begin{eqnarray}
 \gamma_1 & = & \gamma_{max} - (\gamma_{max}-\gamma_{min}) \cdot \sqrt{\mathfrak{u}_{\gamma, 1}} \\
 \gamma_2 & = & (\gamma_{max,2}-\gamma_1) \cdot \mathfrak{u}_{\gamma, 2} +\gamma_1
\end{eqnarray}
where $\mathfrak{u}_{f, i}$ and $\mathfrak{u}_{\gamma, i}$ are random variables uniformly distributed on $[0,1]$.

\subsection{The PMC-ABC sampler}\label{sec:pmcabc}

Approximate Bayesian Computation (ABC) is a useful computational algorithm for Bayesian parameter space exploration where explicit likelihood evaluations are either impossible or not feasible \citep{CameronPettitt12, Marjoram03, Sisson07}.  {The goal is to estimate the posterior by finding a set of mock light curves ($\sim$10,000 in our case) that agree with the actual data within specified limits. The input parameter sets from which these light curves were generated provide an approximation of the source parameters' posterior if these variability parameters were drawn from distributions that are statistically consistent with the priors. In principle, one could simply draw from the prior as often as it takes to find 10,000 accepted light curves. However, this approach becomes computationally impossible for tight limits. Instead, the PMC-ABC algorithm implemented here is iterative and informed by the posterior estimate of the previous iteration, focusing its search on regions of parameter space in which acceptable light curves are most likely to be found.

At its core, the basic ABC algorithm consists} of two Monte Carlo sampling steps and an acceptance step.  Each iteration starts by selecting a random model parameter set $\mathscr{\theta}$ from a predefined probability distribution $\mathcal{P}(\mathscr{\theta})$. 
%(For the first iteration, this distribution is given by the priors.)  
Given the parameter set, a random mock data set $\widetilde{\mathscr{D}}$ is drawn from the likelihood $\mathcal{P}(\widetilde{\mathscr{D}} \mid \mathscr{\theta})$.  This mock data set is compared to the actual data set through the distance function $\phi({\mathscr{D}},\widetilde{\mathscr{D}})$.  If ($|\phi({\mathscr{D}}, \widetilde{\mathscr{D}})| < \epsilon$), where $\epsilon$ is a chosen limit, the drawn parameter set is accepted.  These steps repeat until enough mock data sets are accepted.

The ABC algorithm explores the approximate posterior probability distribution:
\begin{equation}\label{post_def}
 \widetilde{\mathcal{P}}(\mathscr{\theta} \mid \mathscr{D}) \equiv \mathcal{P}(\mathscr{\theta} \mid |\phi({\mathscr{D}}, \widetilde{\mathscr{D}})| < \epsilon) \propto \int \mathcal{W}(\phi(\mathscr{D}, \widetilde{\mathscr{D}}) \mid \epsilon) \mathcal{P}(\widetilde{\mathscr{D}} \mid \mathscr{\theta})\mathcal{P}(\mathscr{\theta})\ d \widetilde{\mathscr{D}}\quad,
\end{equation}
where $\mathcal{W}(\cdot \mid \epsilon)$ is a top-hat window function with width $\epsilon$. \replaced{As $\epsilon \rightarrow 0$, the approximate posterior becomes exact, but the chance of accepting a drawn parameter set becomes vanishingly small.}{Given an adequate distance function, the approximate posterior becomes exact as $\epsilon \rightarrow 0$, but the chance of accepting a drawn parameter set becomes vanishingly small.}  {Therefore, as mentioned, a naive application of the ABC} algorithm would be computationally infeasible.  The PMC-ABC algorithm is a variant of the normal ABC algorithm that attempts to improve ABC performance by iteratively applying a population Monte Carlo technique to ``learn'' the important regions of parameter space \citep{2015A&C....13....1I, DrovandiPettitt11}.  {The PMC-ABC algorithm iteratively modifies a population of parameter combinations (a ``particle system'' $\mathscr{C}$) and a list of weights corresponding to each parameter combination.  With each iteration, the weighted particle system asymptotically approaches the target posterior distribution.}  Each iteration's parameter combinations are drawn from a distribution $\overline{\mathcal{P}}(\mathscr{\theta})$ inferred from the previous iteration's particle system.  In our implementation, we smooth over the previous iteration's particle system using a Gaussian kernel whose dispersion is the dispersion of the \replaced{particle system}{previous particle system ($\{\theta_{n-1, i}\}$), thereby selecting from a distribution of
\begin{equation}
\overline{\mathcal{P}}(\theta) \propto \sum_i \mathcal{K}(\theta \mid \theta_{n-1, i}, \Sigma)
\end{equation}
where $\mathcal{K}(\cdot \mid \cdot, \Sigma)$ is a Gaussian kernel with dispersion $\Sigma$. $\overline{\mathcal{P}}(\theta)$ is truncated to be within the prior range.} The mock data set is then drawn, and the resulting parameter combination is accepted if the distance from the real data is below $\epsilon$.   $\epsilon$  is decreased for each iteration by selecting the 45th percentile largest distance value from the previous \replaced{iteration.}{iteration's accepted parameter sets.} {Because the parameter values at stages after the first are not selected from the prior but from a ``proposal distribution'' based on the previous posterior estimate, the resultant parameter points must be reweighted by a factor $\mathcal{P}(\mathscr{\theta})/\overline{\mathcal{P}}(\mathscr{\theta})$.  This reweighting makes the proposal distribution statistically consistent with the prior}. The algorithm is iterated until the acceptance ratio decreases to a user-specified value.  The diagram Algorithm~\ref{alg:abc} summarizes the required steps.

\begin{figure}[ht]
\centering
\normalsize
\begin{minipage}{.6\linewidth}
\begin{algorithm} [H]
\label{alg:abc}
 \caption{PMC-ABC algorithm.}
 %\KwData{$\mathcal{D} \longrightarrow$ observed catalogue.}
 %\KwResult{ABC-posteriors distributions over the model parameters.}
 \While{\rm{size of} $\mathscr{C}_{\rm{start}} < M$}
 {
   Draw $\mathscr{\theta}$, from the prior, $\mathcal{P}(\mathscr{\theta})$.\\
   Use  $\mathscr{\theta}$ to generate $\widetilde{\mathscr{D}}$.\\
   Save $\mathscr{\theta}$ as part of next generation particle system, $\mathscr{C}_{\rm{start}}$.
 }
 Save N points in $\mathscr{C}_{\rm{start}}$ with the lowest distance values to $\mathscr{C}$.\\
 Set weights to $1/N$\\
 \While{\rm{Acceptance ratio is above} $\alpha$}
 {
   Infer $\overline{\mathcal{P}}(\mathscr{\theta})$ from $\mathscr{C}$.\\
   Set $\epi$ to the 45\% quantile of the ordered distances in $\mathscr{C}$.\\ 
   \While{\rm{size of} $\mathscr{C}_{\rm{next}} < N$}
   {
     Draw $\mathscr{\theta}$ from $\overline{\mathcal{P}}(\mathscr{\theta})$.\\
     Use  $\mathscr{\theta}$ to generate $\widetilde{\mathscr{D}}$.\\
     \If{$|\phi(\widetilde{\mathscr{D}}, \mathscr{D})| < \epi$}
     {
       Save $\mathscr{\theta}$ as part of next generation particle system, $\mathscr{C}_{\rm{next}}$.\\
       Set weight to $\mathcal{P}(\mathscr{\theta})/\overline{\mathcal{P}}(\mathscr{\theta})$
     }
   }
   $\mathscr{C} \leftarrow \mathscr{C}_{\rm{next}}$
 }
\end{algorithm}
\end{minipage}
\end{figure}

\section{Efficient calculation of the first-order structure function}\label{FFT}
\setcounter{equation}{0}

The time-series analysis technique used here requires multiple structure function calculations that share identical observing cadences.  This can be resource-intensive because the number of operations needed to calculate each structure function (Eq.~\ref{eq:strfundef}) is proportional to the square of the number $N$ of light curve data points.  Direct calculation made this step the primary computational bottleneck.  To reduce computation time, we developed a more efficient algorithm for calculating structure functions that share identical observing cadences.  Central to our algorithm is the comparison of a ``perfect'' structure function defined by
\begin{eqnarray}
\label{eq:perfstruct}
 \widetilde{V}_i & = & \frac{1}{N} \sum_{k=0}^{N-i-1} (F(t_{k+i}) - F(t_k))^2\\
 & = & \frac{1}{N} \sum_{k=0}^{N-i-1} (F(t_{k+i})^2 + F(t_k)^2) - \frac{2}{N} \sum_{k=0}^{N-i-1} F(t_{k+i})F(t_k)\\
 & = & \widetilde{\Phi}_i^0 + 2\widetilde{\Phi}_i^1 \label{eq:pfcom}~.
\end{eqnarray}
to the actual structure function calculated from Eq.~\ref{eq:strfundef}, which needs to be evaluated only once.  This is advantageous because the number of operations for the second term $\widetilde{\Phi}_i^1$ of Eq.~\ref{eq:pfcom} goes as $N\log N$ when calculated via fast Fourier transforms.  The first term $\widetilde{\Phi}_i^0$ can be calculated recursively ($\widetilde{\Phi}_i^0 = \widetilde{\Phi}_{i+1}^0 + F(t_{N-1-i})^2 + F(t_i)^2; \widetilde{\Phi}_{N}^0 = 0$), which is linear in $N$.  Replacing flux pairs in the actual binned structure function (Eq.~\ref{eq:strfundef}) with the corresponding perfect structure function values therefore shortens computation time. Unfortunately, some perfect structure function values will be shared between multiple bins, and these values will need to be split between these bins.  In such cases, light curve pairs must be explicitly calculated and added to or subtracted from the affected bins.  In practice, for observing cadences that are fairly even, few light curve pairs need to be calculated directly, and these have negligible effect on computational performance.  For observing cadences that are very uneven, up to 30\% of the light curve pairs require direct calculation.  Even in such cases, though, the above algorithm still offers significant performance improvements for multiple structure function calculations that share the same observing cadence.

\section{Ratio between $M$- and $K$-band derived from power-law and log-normal distributions}\label{ratioform}
\setcounter{equation}{0}

Any model that allows different flux-density PDFs for $M$- and $K$-band predicts a varying $F(M)/F(K)$ ratio as a function of $F(K)$. This is true even when both distributions are power laws if the power-law index $\beta$ differs for the two bands. A simple way to calculate the ratio function -- that is, the ratio  $\Re[M/K, F(K)] \equiv F(M)/F(K)$ as a function of $F(K)$ -- is the assumption that  the cumulative distribution functions (as defined in Eq.~\ref{eq:cdf1} or~\ref{eq:cdf2}) are equal for all corresponding pairs $[F(K),(F(M)]$. This is simply asking for a match of the lowest $5\%$ in $K$-band with the lowest $5\%$ in $M$-band, the lowest $10\%$ with the lowest $10\%$, and so on. In other words, it assumes that when $K$-band rises, $M$-band rises, and when $K$-band falls, $M$-band falls. The simultaneous NIR (1.65--3.8~\micron) light curves in the literature indeed demonstrate this behavior.
Under this assumption, we get for Case~1 (power-law/power-law):
\begin{equation}
\Re[M/K,F(K)] =\frac{s \cdot F_{0}(M)}{F(K)} \bigg\{ 1 - \bigg[ \frac{F(K) - F_{0}(K)}{-F_{0}(K)}\bigg]^\frac{\beta_{{K}} - 1}{\beta_{{M}} - 1} \bigg\} \quad,
\end{equation}
with $F_{0,{M}},F_{0,{K}} < 0$. For Case~2 (power-law/log-normal),
\begin{equation}
\Re[M/K,F(K)] = \bigg[\frac{F(K)}{\rm{mJy}} \bigg]^{-1} \cdot \exp{\bigg[{\mu_{\operatorname{logn,M}} \sqrt{2} \sigma_{\operatorname{logn,M}} \cdot \mathrm{erf}^{-1}\big(1 - 2 \kappa \big)}\bigg]}\quad,
\end{equation}
with
\begin{equation}
\kappa = \bigg(\frac{F_{{K}} - F_{0,{K}}}{-F_{0,{K}}}\bigg)^{1-\beta_{{K}}} \quad,
\end{equation}
and with $F_{0,{K}} < 0$. For Case~3 (log-normal/log-normal),
\begin{equation}\label{log_n_ratio}
\Re[M/K,F(K)] = 
\left[\frac{F(K)}{\rm{mJy}} \right]^{-1} \cdot
\exp{\left\{\left(\ln{\left[\frac{F(K)}{\rm{mJy}} \right]} - \mu_{\operatorname{logn},K}\right)
\cdot \frac{\sigma_{\operatorname{logn},M}}{\sigma_{\operatorname{logn},K}} + \mu_{\operatorname{logn},M} \right\}}\quad.
\end{equation}

In order to use the information about the flux-density ratio determined from the synchronous data ($\Re[M/K,0.15~{\rm mJy}] = 12.4 \pm 0.5$), we can use this equation and extend the Equation~\ref{distdef} distance function to
\begin{equation}\label{dist_log_n}
\phi(\theta) = \phi(V_{\theta}, V_{\rm{obs.}}) + w \cdot \bigg\{
 \big[X-\Re(M/K,0.15)\big] / 0.5\bigg\}^2\quad,
\end{equation}
with $X \sim \mathcal{N}(12.4, 0.5^2)$ and $w$ a chosen weight. {(Here, $w = 0.002$ relative to the weights defined in \ref{dist}.)}
{The corresponding NIR spectral index
\begin{equation}\label{log_n_alpha}
\alpha_s = 
\left\{{\log{\big(\frac{\lambda_M}{\lambda_K} \big)}}\right\}^{-1} 
\cdot \bigg\{ \log{\big[\Re(M/K)\big]} + 0.4 \cdot (A_M-A_K) \bigg\}\quad,
\end{equation}
where $A_M$ and $A_K$ are the adopted interstellar extinctions in magnitudes.
For our case with
\begin{equation}
\log\big\{\Re[M/K, F(K)]\big\} = \bigg(\frac{\sigma_{\operatorname{logn},M}}{\sigma_{\operatorname{logn},K}}-1\bigg) \cdot \log{\bigg[\frac{F(K)}{\rm{mJy}} \bigg]} + 0.4343 \cdot \bigg(\mu_{\operatorname{logn},M} -  \frac{\sigma_{\operatorname{logn},M}}{\sigma_{\operatorname{logn},K}} \cdot \mu_{\operatorname{logn},K} \bigg)
\end{equation}
}
and with $\lambda_M = 4.5$~\micron\ and $\lambda_K = 2.18$~\micron,
\begin{eqnarray}
\alpha_s & = & 3.1771 \cdot \bigg(\frac{\sigma_{\operatorname{logn},M}}{\sigma_{\operatorname{logn},K}}-1\bigg) \cdot \log{\bigg[\frac{F(K)}{\rm{mJy}} \bigg]} + 1.3798 \cdot \bigg(\mu_{\operatorname{logn},M} -  \frac{\sigma_{\operatorname{logn},M}}{\sigma_{\operatorname{logn},K}} \cdot \mu_{\operatorname{logn},K} \bigg) + 1.2708 \cdot (A_M-A_K) \label{si} \\ 
& = & \xi \cdot \log{\bigg[\frac{F(K)}{\rm{mJy}} \bigg]} + \eta + 1.2708 \cdot (A_M-A_K)\quad.
\end{eqnarray}

\section{NIR spectral index as a function of flux density
%in the case of synchrotron emission from an electron energy distribution with exponential high energy cutoff
}\label{cutoff_alpha}
\setcounter{equation}{0}

For synchrotron radiation from an electron energy distribution with an exponential cutoff, the spectrum in the optically thin frequency regime is a power-law with an exponential cutoff at a frequency $\nu_0$ (e.g., \citealt{1985ApJ...288...32B}).
The flux density $S(\nu)$ at a given frequency $\nu$ in the optically thin regime is
\begin{equation}\label{eq1}
S(\nu) = k_0 \cdot \nu^{-\tilde{\alpha}_{s}} \cdot \exp{\big[-\big({\nu}/{\nu_0}\big)^{{1}/{2}}\big]}\quad,
\end{equation}
with $\tilde{\alpha}_{s}$ the spectral index of the optically thin power-law spectrum \added{and $k_0$ a proportionality constant}. For typical electron energy cutoffs,  $\nu_0$ will be located in or slightly above the NIR frequency range. Thus varying flux densities in the NIR and a flux-density-dependent NIR spectral index could be the consequence of a changing energy cutoff in the electron energy distribution.

We want to derive the flux-density dependence of the spectral index for this scenario. For a frequency $\tilde{\nu} \ll \nu_0$ (but still in the optically thin regime; i.e., in the submm regime close to 1~THz), Equation~\ref{eq1} becomes
\begin{equation}\label{eq2}
S(\tilde{\nu}) = k_0 \cdot \tilde{\nu}^{-\tilde{\alpha}_{s}}\quad,
\end{equation}
and we can eliminate the proportionality factor:
\begin{equation}\label{eq3}
S(\nu) = S(\tilde{\nu}) \cdot \big({\nu}/{\tilde{\nu}}\big)^{-\tilde{\alpha}_{s}} \cdot \exp{\big[-\big({\nu}/{\nu_0}\big)^{{1}/{2}}\big]}\quad.
\end{equation}
This equation implies that while the flux density in the NIR overall scales with flux density in the submm, for a given submm flux density the NIR variability is \replaced{dominated}{caused} by the changes in $\nu_0$. The flux-density ratio between two frequencies $\nu_1$ and $\nu_2$ is
\begin{eqnarray}\label{eq4}
{S(\nu_1)}/{S(\nu_2)} &= &\big({\nu_1}/{\nu_2}\big)^{-\tilde{\alpha}_{s}} \cdot \exp{\big[\big({\nu_2}/{\nu_0}\big)^{{1}/{2}} - \big({\nu_1}/{\nu_0}\big)^{{1}/{2}}\big]} \quad,
\end{eqnarray}
and
\begin{equation}\label{eq5}
\alpha_s = \tilde{\alpha}_{s} - 
{\nu_{0}^{-{1}/{2}}} \cdot \frac{\nu_{2}^{\sfrac{1}{2}} - \nu_{1}^{\sfrac{1}{2}}}{\ln{(10)} \cdot \log{\big({\nu_1}/{\nu_2}\big)}}\quad.
\end{equation}
Equation~\ref{eq3} gives
\begin{eqnarray}\label{eq6}
{\nu_{0}^{-\sfrac{1}{2}}} &=& 
{-\nu^{-{1}/{2}}} \cdot {\ln{(10)}}\cdot
\left\{ \log{\left[{S(\nu)}/{S(\tilde{\nu})}\right]}
+ \tilde{\alpha}_{s} \cdot \log{\left({\nu}/{\tilde{\nu}}\right)} \right\}~~\rm{with} \\[2ex]
S(\tilde{\nu}) &<& 10^{\tilde{\alpha}_{s}\cdot\log{(\sfrac{\nu}{\tilde{\nu}})}}\cdot S(\nu) \quad.
\end{eqnarray}
Inserting Equation~\ref{eq6} in Equation~\ref{eq5} gives
\begin{eqnarray}\label{eq7}
%\alpha_s & = & \zeta \cdot \log{\left[\frac{S(\nu)}{S(\tilde{\nu})}\right]}+  \delta \cdot \left[1 + \zeta \cdot \log{\left(\frac{\nu}{\tilde{\nu}}\right)}\right] \\
\alpha_s & = & \tilde{\xi} \cdot \log{\left[\frac{S(\nu)}{\rm{mJy}}\right]} + \tilde{\eta}\quad,
\end{eqnarray}
with
\begin{equation}\label{eq8}
\tilde{\xi} = \frac{\nu_{2}^{\sfrac{1}{2}} - \nu_{1}^{\sfrac{1}{2}}}{\nu^{\sfrac{1}{2}} \cdot \log{\big({\nu_1}/{\nu_2}\big)}}
\end{equation}
and
\begin{equation}\label{eq9}
\tilde{\eta} =  \tilde{\alpha}_{s} \cdot \left[1 + \tilde{\xi} \cdot \log{\left({\nu}/{\tilde{\nu}}\right)}\right] - \tilde{\xi} \cdot \log{\left[\frac{S(\tilde{\nu})}{\rm{mJy}}\right]}\quad.
\end{equation}

%\listofchanges

\end{document}